\documentclass[12pt,preprint,letterpaper,xdvi]{aastex}
\usepackage{amsmath} 
\usepackage{wrapfig} 
\usepackage{array}  
\usepackage{natbib}
\usepackage{brief_bib}
\usepackage{lscape}
\usepackage{verbatim}
		
\newcommand\bvec{{\bf B}}
\newcommand\nhat{{\bf \hat{n}}}
\newcommand\lhat{\textit{\textbf{\^l}}}
\newcommand\lvec{{\bf L}}
\newcommand\rvec{{\bf r}}

\newcommand\half{\frac{1}{2}}
\newcommand\al{$\alpha$}
\newcommand\als{$\alpha\mbox{ }$}

\newcommand{\vol}{{\cal V}}
\newcommand{\llf}{Low \& Lou}
\newcommand{\llfs}{Low \& Lou$\mbox{ }$}

\begin{document}

\title{Reconstructing the Local Twist of Coronal Magnetic Fields and the Three-Dimensional Shape of the Field Lines from Coronal Loops in EUV and X-Ray Images}
\author{A.~Malanushenko, D.W.~Longcope, D.E.~McKenzie}
\affil{Department of Physics, Montana State University\\
Bozeman, MT 59717, USA}

\begin{abstract}
Non-linear force-free fields are the most general case of force-free fields, but the hardest to model as well. There are numerous methods of computing such fields by extrapolating vector magnetograms from the photosphere, but very few attempts have so far made quantitative use of coronal morphology. We present a method to make such quantitative use of X-Ray and EUV images of coronal loops. Each individual loop is fit to a field line of a linear force-free field, allowing the estimation of the field line's twist, three-dimensional geometry and the field strength along it. 

We assess the validity of such a reconstruction since the actual corona is probably not a linear force-free field and that the superposition of linear force-free fields is generally not itself a force-free field. To do so, we perform a series of tests on non-linear force-free fields, described in \citet{LowLou1990}. For model loops we project field lines onto the photosphere. We compare several results of the method with the original field, in particular the three-dimensional loop shapes, local twist (coronal \al), distribution of twist in the model photosphere and strength of the magnetic field. We find that, (i) for these trial fields, the method reconstructs twist with mean absolute deviation of at most 15\% of the range of photospheric twist, (ii) that heights of the loops are reconstructed with mean absolute deviation of at most 5\% of the range of trial heights and (iii) that the magnitude of non-potential contribution to photospheric field is reconstructed with mean absolute deviation of at most 10\% of the maximal value. 
\end{abstract}

\section{Introduction}\label{sec_intro}

Most active region coronal magnetic fields are believed to be in a force-free state, 
\begin{equation}
\nabla\times\bvec=\alpha(\rvec)\bvec, 
\label{eqn_fff}
\end{equation}
where $\alpha$ is a scalar of proportionality \citep[e.g.,][]{Nakagawa1971}. It turns out that \als is closely related to the local twist of magnetic field lines. For example, in a cylindrical uniformly-twisted flux tube, field lines twist about the axis by an angle $\theta=\frac{1}{2}\alpha L$ over axial distance $L$ (Aschwanden, 2006\nocite{Aschwanden_book}). 

If \als varies in space, a solution of Eq.~(\ref{eqn_fff}) is called a non-linear force-free field, as it solves a non-linear system of equations for different components of $\bvec$ and the scalar \al. \citet{Demoulin1997b} described basic problems arising when trying to solve these equations. In particular, the existence and uniqueness of a solution is not entirely clear. Another difficulty is the fact that the only source of boundary conditions available at the moment are vector magnetograms measuring the field within the non-force-free photospheric layer. 

A particular case of a force-free field, called a linear force-free field, or a constant-\als field, occurs when $\nabla\alpha=0$. In this case, using $\nabla\cdot\bvec=0$, Eq.~(\ref{eqn_fff}) is transformed to a Helmholtz equation for $\bvec$. This is much easier to solve and the conditions for existence and uniqueness of solution are known. Nor does the solution require vector magnetogram data, but only a line-of-sight magnetogram and a value of the constant \al.  This boundary condition is affected less by the fact that the photosphere is probably not force-free. There are many methods of solving for linear force-free fields \citep[e.g.,][]{Nakagawa1972, Chiu1977, Altschuler1969, Lothian1995, Alissandrakis1981}. In particular, in this paper we will use the Green's function method, described in \citet{Chiu1977}, as it does not place any restrictions on \als and it generates a field over an entire half-space, without boundaries. 

While they are simpler to generate, linear force-free fields have proven insufficient to model complex geometries of the solar corona. Observations of curvature of H\als structures, visual studies of twist in coronal loops, and estimations of local twist at the photospheric level via vector magnetograms reveal active regions with spatially varying twist, and even varying sign of twist \citep[for example,][]{Burnette2003, Nakagawa1972}. In light of this any constant-\als approximation would appear to be, strictly speaking, incorrect. Over the past decade there have been many attempts to perform extrapolations of non-linear force-free magnetic field into the corona and to assess the quality of the extrapolation by comparing lines of the resulting field to coronal loops \citep[e.g.,][and references therein]{Schrijver2008, DeRosa2009}. 

In the current paper we apply a completely different approach. We use the visible shapes of coronal loops to infer the twist of the magnetic field. Instead of measuring twist in the photosphere, where Eq.~(\ref{eqn_fff}) is not appropriate, we \textit{perform measurements in the region of interest}, in the low-$\beta$ force-free solar corona. The method thus relies solely on EUV or X-Ray images of coronal loops and on line-of-sight magnetograms. 

The basic idea is to try to approximate every visible coronal loop with a field line from a linear force-free field, and allow \als to be different for every loop. Even for non-linear force-free fields, \als must be constant along each field line\footnote{This result is obtained by taking the divergence of both sides of Eq.~(\ref{eqn_fff}) and using $\nabla\cdot\bvec=0$ and the identity $\nabla\cdot(\nabla\times\bvec)=0$, so that $\bvec\cdot\nabla\alpha=0$.}. If \als changes smoothly then it would be reasonable to expect that \als is nearly constant in the vicinity of a given field line. Of course, a superposition of constant-\als fields would not in general be a force-free field and at first sight such a method could not be expected to yield meaningful results. In the next few paragraphs we argue that such a method might work under certain circumstances relevant to the solar corona. Within the core of this work we support the hypothesis with tests first using analytic non-linear force-free fields and then with solar data. 

Our method is similar to the ones proposed by Green et al (2002)\nocite{Green2002} and Lim et al (2007)\nocite{Lim2007}, however, with several important advantages. First, it does not require the full length of a coronal loop to be visible for a successful reconstruction. Second, it does not require either of the footpoints to be visible. Third, it allows the user to draw a smooth curve (a B\'ezier spline) interactively on top of the loop, rather than selecting a few points along the loop. This maximizes the amount of information taken from the coronal image. The fit itself is similar to the one used by van Ballegooijen (2004)\nocite{Ballegooijen2004}; but while van Ballegooijen (2004) fits loops with lines of a particular non-linear force-free field model, we fit loops with lines of many different linear force-free fields, choosing the best \al. 

Consider an imaginary example of two dipoles far apart compared to their sizes. Suppose that they constitute a non-linear equilibrium, having different twist, possibly of opposite signs. Suppose, however, that within each dipole the twist is more or less constant. In such a scenario there would probably be some transition region between the dipoles where \als changed significantly. Provided the dipoles are far apart we may claim that in the vicinity of the footpoints of one of them the current of the other would not significantly perturb the field, and in the close vicinity of each of them the field would be nearly a constant \als field.

To support this reasoning we note that the dipolar term of a magnetic field drops as $1/r^3$ and thus the effect of a distant dipole is in general not very large compared to the nearby dipole. Indeed, this is why studying a magnetic field of an isolated region of the corona is at all meaningful.

We therefore argue that a non-linear force-free field could be considered to be linear in the regions of slowly changing \als (in some sense of the term). Thus the geometry of a field in isolated regions of slowly-changing \als might be approximated by the geometry of a constant-\als field.

What is the limit of applicability of such an assumption? It is quite clear that it could work well for an isolated uniformly-twisted active region. Is it possible to pick a field line in an active region, and to suppose that the field's geometry is not significantly different from that of a linear field in a close vicinity of this field line? We herein conduct several experiments on both synthetic and real data which provide evidence that at least in certain cases of interest such an assumption is reasonable.

The paper is organized in the following manner. In Section~\ref{sec_distance} we explicitly define the function to be minimized in order to obtain a best fit, a ``distance'' between two curves, $d$. In Section~\ref{sec_fit} we describe the minimization process, varying \als and the line-of-sight coordinate $h$, and report that it indeed works in the obvious case, where the loop is a field line from an actual linear force-free field. In Section~\ref{sec_shape} we describe the typical features of the function $d(\alpha, h)$ and attempt to explain their appearance. In Section~\ref{sec_llf} we present the results of applying this procedure to several analytic non-linear force-free fields, described by \citet{LowLou1990}. We also present an additional step proven necessary for the best fit procedure. This step amounts to minimizing $d(\alpha, h)$ in a very specific region of $(\alpha, h)$ parameter space. We demonstrate that this step significantly improves the results for strongly twisted fields. In Section~\ref{sec_realdata} we demonstrate the same method applied to real data: line-of-sight magnetograms from SOHO/MDI and coronal images from Hinode/XRT. In Section~\ref{sec_discuss} we discuss the results and their possible use in studying coronal magnetic fields. 
% An Appendix~1 we make a note on Low \& Lou fields, basically, 
% revising the producing equation and demontrating that the resulting 
% equation has several solutions, some with same sign twist everywhere 
% in space and some others with changing twist of variable sign. 
% We use these different solutions in the paper to study how typical change 
% in twist over a typical length could affect the quality of the reconstruction. 

%It turns out that for a strongly twisted (or strongly non-linear) field, or for a very straight loop in the plane-of-the-sky projection the global minimum of $d(\alpha, h)$ could be misleading. On the other hand, the parameter spaces for many different loops share certain common features, and provided these features could be identified for a specific loop, the local minimum in the specific region of the parameter space has strong correlation with the real

\clearpage
\section{The distance between two lines}\label{sec_distance}
In order to compare two curves, we seek a function quantifying the discrepancy between the curves. The ideal function would yield similar results to those obtained by visual comparison. It would be large when a human observer would consider the two lines to be far apart or unlike one another and small when a human observer would consider them to be similar and close to one another.

We use for this purpose a function first introduced by Green et al (2002)\nocite{Green2002} and later used by Lim et al (2007)\nocite{Lim2007}. We apply it, however, to a different set of objects. While Green et al (2002) and Lim et al (2007) compare \textit{a few points along the loop} to a set of field lines traced from the photosphere, \textit{at the presumed location of the loop's footpoints}, we compare a \textit{smooth curve}, chosen to visually match the loop, to a set of field lines traced from \textit{different locations along the line-of-sight} at some point along the curve. The same method was used by \citet{Ballegooijen2004}. Therefore, unlike method of Lim et al (2007), our method does not require knowledge of the footpoints. In fact, it will work with even a small portion of a loop. %We also do a slight modification of this function to compare lenths of the curves, not only their relative position. 

The \textit{discrepancy} function is defined between two smooth curves in a plane, $\lvec_1=\left\{x_1(l), y_1(l), 0\leq l \leq L_1\right\}$ and $\lvec_2=\left\{x_2(l), y_2(l), 0\leq l \leq L_2\right\}$. For every point $l$ on $\lvec_1$ it is possible to define a minimal distance between that point $\lvec_1(l)$ and $\lvec_2$ \textit{in the classical sense}: the smallest of the distances between the point $\lvec_1(l)$ and every point along $\lvec_2$. It could also be defined as the length of the shortest perpendicular from the point $\lvec_1(l)$ to the curve $\lvec_2$, given sufficient smoothness. We will refer to this distance as $\delta(\lvec_1(l), \lvec_2)$. The discrepancy between the two curves is the average of $\delta(l)$ over curve $\lvec_1$,

\begin{equation}
d(\lvec_1, \lvec_2)=\frac{1}{L_1}\int\limits_{0}^{L_1}{\delta(\lvec_1(l), \lvec_2)dl}.
\label{d_first_def}
\end{equation}

A numerical scheme to compute this integral is quite simple. Sample line $\lvec_1$ into $n_1$ segments with equal lengths $\Delta l_1=L_1/n_1$. Provided the segments are small compared to the local radius of curvature, for every point on $\lvec_1$, 
\begin{equation}
\delta(\lvec_1(l_i), \lvec_2)=\min\left(\left.\sqrt{(x_1(l_i)-x_2)^2+(y_1(l_i)-y_2)^2} \right| (x_2, y_2)\in \lvec_2 \right).
\label{delta_first_def}
\end{equation}
The discrepancy integral can then be approximated by the sum
\begin{equation}
d(\lvec_1, \lvec_2)=\frac{\Delta l_1}{L_1}\sum\limits_{i=0}^{n_1}{\delta(\lvec_1(l_i), \lvec_2)},
\label{d_second_def}
\end{equation}
representing the mean distance between points of one curve and the whole second curve; it has the units of length. 

\begin{figure}[!hc]
 \begin{center}
 \begin{tabular}{cc}
  \includegraphics[width=6cm]{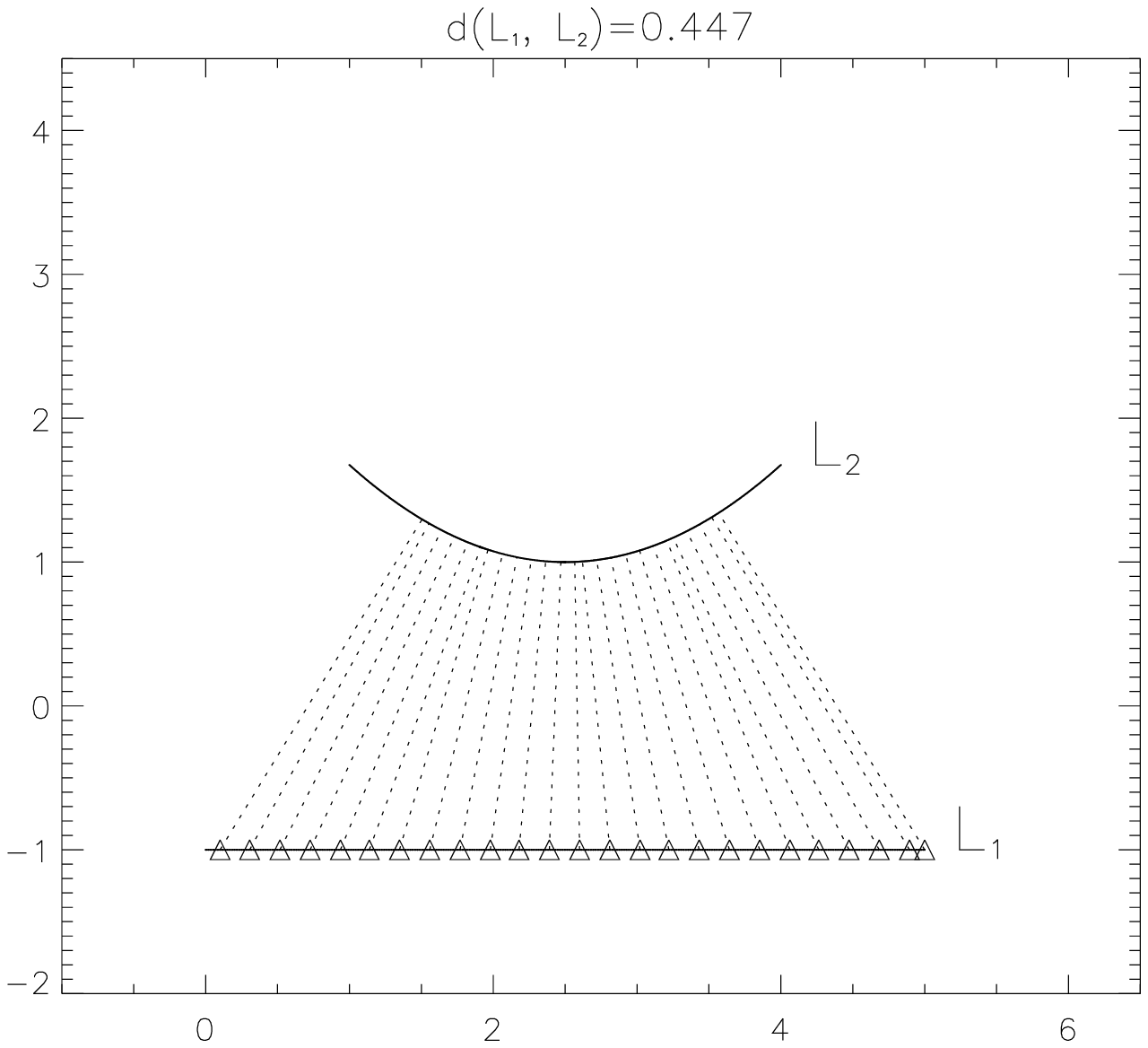} 
  \includegraphics[width=6cm]{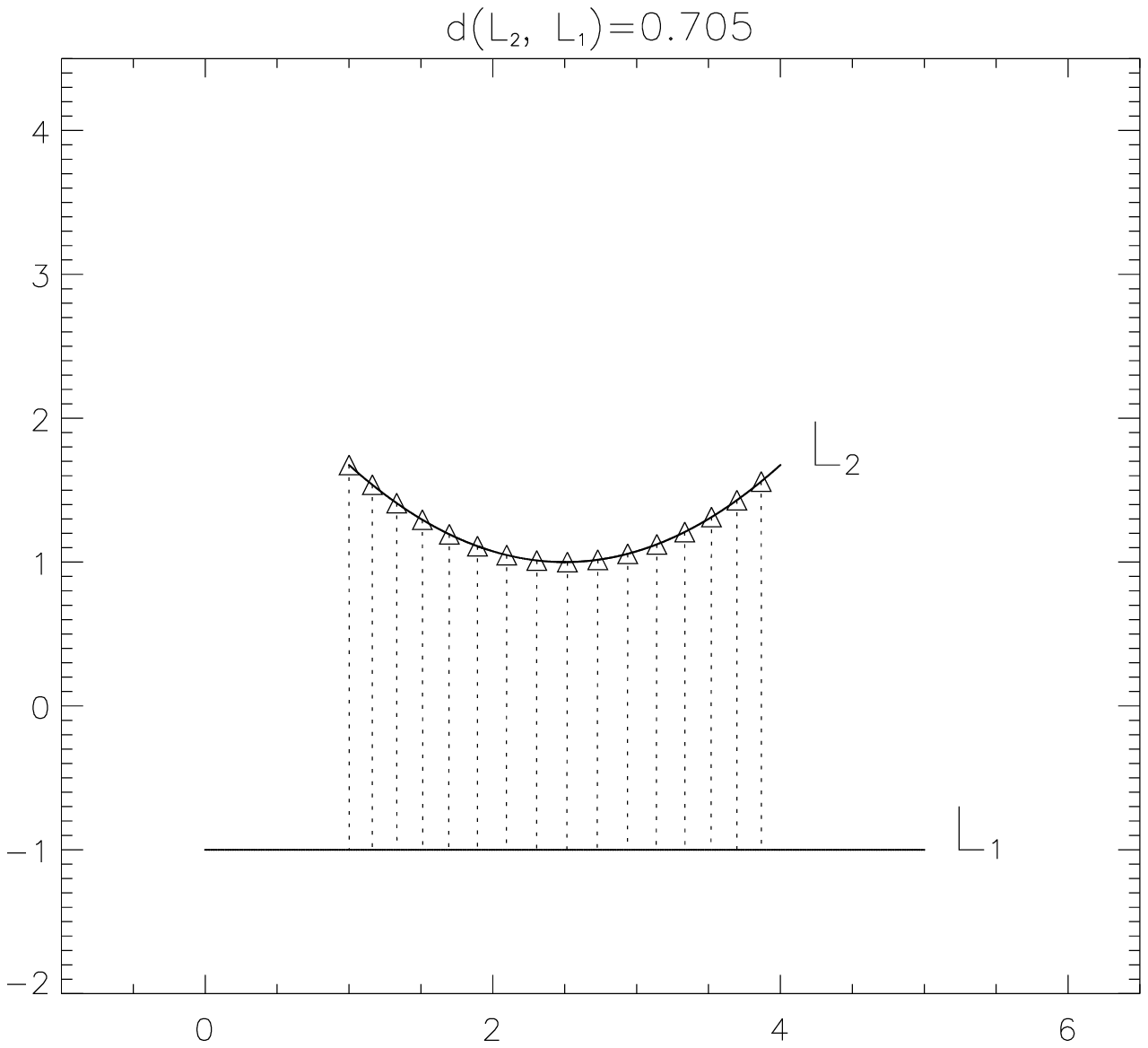} 
 \end{tabular}
 \end{center} 
 \caption{\small{The discrepancy between the two lines. The left panel illustrates the calculation of $d(\lvec_1, \lvec_2)$ and the right shows $d(\lvec_2, \lvec_1)$. For each point on the first curve (the first argument of $d$), denoted by a triangle, one finds the closest point on the second curve (the second argument). The closest distance is shown as a dashed line. The net discrepancy is the average of all such distances. Comparison of the two panels illustrates that the discrepancy is non-commutative.}}
% \caption{\small{The discrepancy between the two lines is obviously noncommutative, since it is defined as an average of distances from \textit{a set of points}, forming one curve (here $\lvec_1$) to the curve (here $\lvec_2$). In this illustration the distances (i.e., $\delta(\lvec_1(l_i), \lvec_2)$) are shown as dashed lines and the points $\lvec_1(l_i)$ are shown as triangles. }}
 \label{fig_noncommutative} 
 \end{figure}

The trivial properties of the discrepancy function $d(\lvec_1, \lvec_2)$ are, first, that it is non-negative and second, it is non-commmutative, meaning $d(\lvec_1, \lvec_2)$ is different in general from $d(\lvec_2, \lvec_1)$, as illustrated in Fig.~\ref{fig_noncommutative}.

\section{\al-h-fit}\label{sec_fit}

For a visible coronal loop it is possible to construct a smooth two-dimensional curve $\lvec_0$ in the plane of the sky visually approximating the loop, or some portion of the loop. The loop is really a three-dimensional structure and for every point on $\lvec_0$, the third coordinate, i.e.\ along the line of sight (LOS), is unknown. If the loop is at disk center, then the LOS coordinate is the height above the photosphere. For simplicity in notation, we will thereafter refer to this coordinate as ``height'', denoted $h$, even when it is not vertical.

The main idea of what we call an \textit{h-fit} is to choose a point $l_0$ on the loop $\lvec_0$ and \textit{prescribe} a certain height. Then, if the magnetic model is known, trace a field line from the three-dimensional location $(L_{0x}(l_0), L_{0y}(l_0), h)$ and compare its plane of the sky projection $\lvec(h)$ to the original loop by calculating $d(h)=d(\lvec_0, \lvec(h))$. Finally, we vary $h$ to find the minimum of $d(h)$.

\begin{figure}[!hc]
 \begin{center}
% \begin{tabular}{cc}
  \includegraphics[width=9cm]{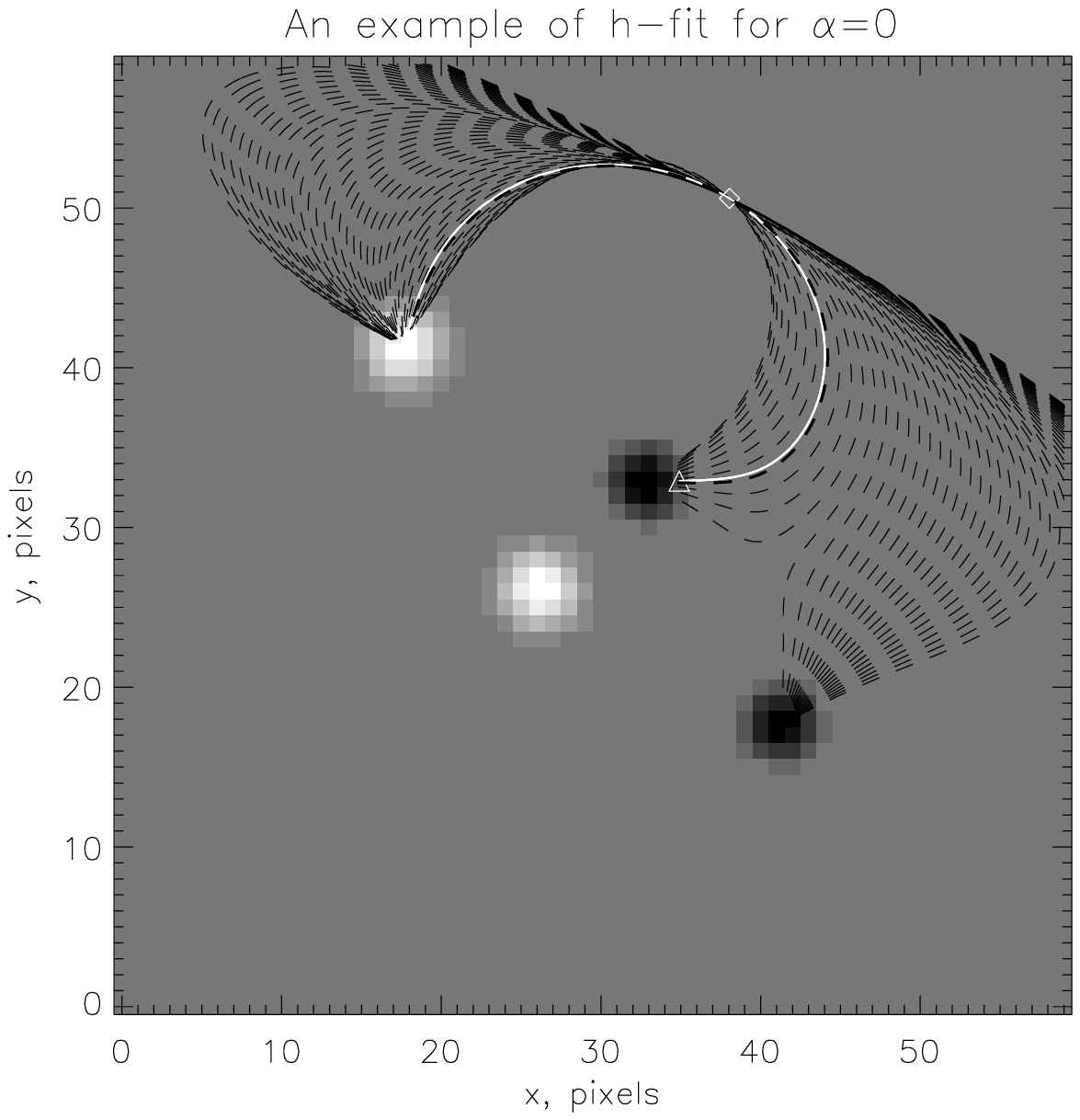} \\
  \includegraphics[width=9cm]{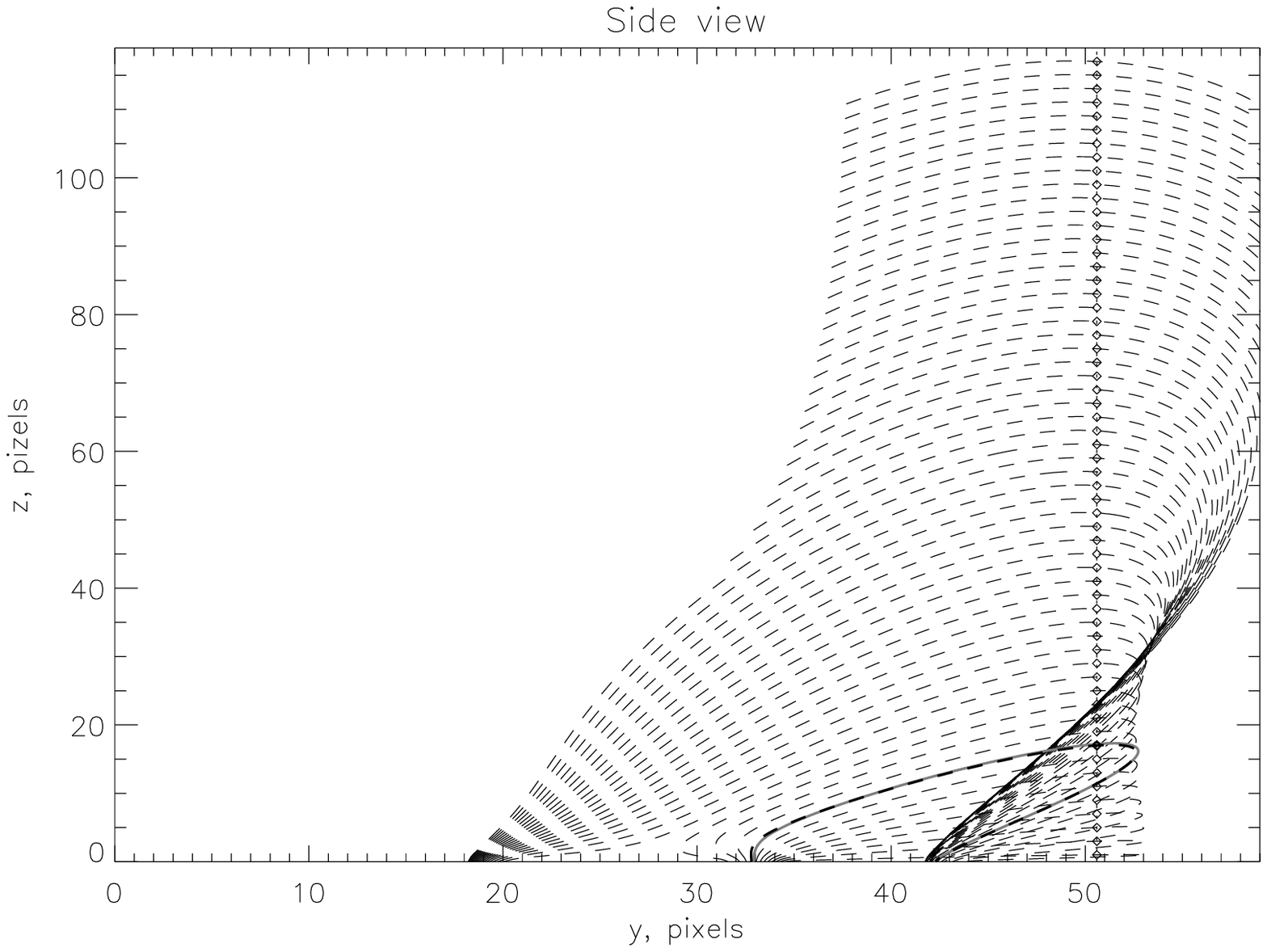} 
% \end{tabular}
 \end{center} 
 \caption{\small{The synthetic example used to illustrate the \textit{h-fitting routine}. \textit{Top}: The synthetic magnetogram (in gray-scale) was used to generate the potential field in $Z_+$. One field line was selected to represent the synthetic loop (in white; its starting point is shown as the triangle). It was projected onto $x$--$y$ plane and compared with field lines, traced from the points with the same $(x_0, y_0)$ (in this case $(x_0, y_0)\approx(38, 51)$), but at different heights. These trial field lines are shown as dashed lines; they are traced for every pixel of the column $(x_0, y_0, z\in[1,n_z-1])$. To make the plot clearer, only every second one is drawn. The point $(x_0, y_0)$ was chosen as the mid-point of the projection of the ``loop'' (shown in diamond). \textit{Bottom}: The same example, viewed in $x$--$z$ projection. The initial ``loop'' is shown as gray, the traced field lines are shown as dashed lines, and their starting points are shown as diamonds. The thick dashed line shows the \textit{best h-fit}.}} 
 \label{fig_qpole_h_fit} 
 \end{figure}

\begin{figure}[!hc]
 \begin{center}
% \begin{tabular}{cc}
  \includegraphics[width=9cm]{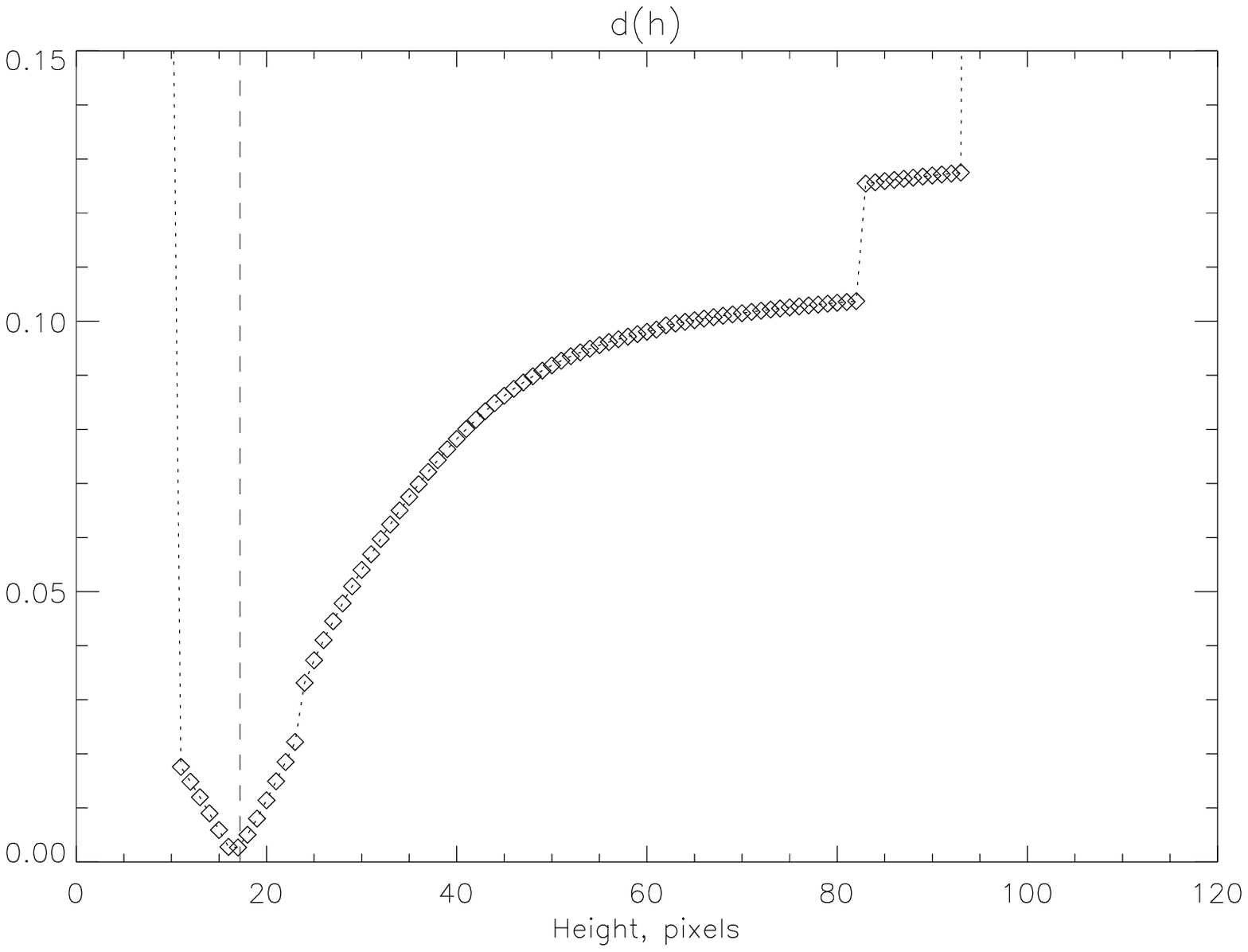} %\\
%  \includegraphics[width=9cm]{../illus/qpole_h_fit_cos_th.eps} 
% \end{tabular}
 \end{center} 
 \caption{\small{The function $d(h)$ has its minumum at a height $h\approx 17$ within the numerical error of the real height of the field line it is modeling. Note that for $h<11$ pix and $h>93$ pix the length of the projection of the traced lines is smaller than the length of the ``coronal loop''. Such lines could be automatically discarded from consideration, since the ``loop'' should be a part of a field line, and the length of a part of a curve cannot be greater than the length of the whole curve. We discard them by making $d(h)$ artificially large if the length of $\lvec_0$ is smaller than the length of $\lvec_1$.}}
 \label{fig_qpole_h_fit_d_th} 
 \end{figure}

To illustrate this method we construct a synthetic magnetogram and generate a magnetic field (the potential field of a magnetic quadrupole in half-space $Z_+$). As a model of the `loop' we take an actual field line, projected onto the $x$--$y$ plane. We then take the mid-point of the projected loop, $\lvec_0(\half L_0)$, and trace field lines at different heights (see Fig.~\ref{fig_qpole_h_fit}). Fig.~\ref{fig_qpole_h_fit_d_th} shows the function $d(h)$ with one minimum at the actual height, to within one step of the $h$ search.

%\clearpage
%\section{\al-h fit: finding \als for constant-\als fields}

Suppose now that the coronal magnetic field is not known, but belongs to a known family of magnetic fields, described by certain parameters. It is then possible to do fitting not only in height, but also in the space of these magnetic field parameters. For example, if there is a reason to believe that the actual field is constant-\als, but with unknown \al, then the discrepancy is a function of both \als and $h$, $d=d(\alpha, h)$ and the minimization must be done over $(\alpha, h)$ space.

To illustrate such a minimization we perform the following experiment. From the quadrupolar magnetogram of Fig.~\ref{fig_qpole_h_fit} we extrapolated a constant-\als field into the corona ($Z_+$). We produced several such fields with different values of \al. For each of these fields we selected several magnetic field lines at random, which we projected onto $x$--$y$ plane. We then treated these projections as synthetic loops and performed an $\alpha-h$ fit. Provided the method works, the best-match \als should have one-to-one correspondence with the real \als for each field line.

The constant-\als fields were generated using Green's function \citep{Chiu1977} for a field in a half-space. This has the advantage that it places no limitations on \al, whereas Fourier methods using periodic images separated by $L$, require $|\alpha|<{2\pi\over L}$ \citep{Nakagawa1971}. In Section~\ref{sec_llf} we demonstrate that our method can reconstruct values of $|\alpha|$ up to ${\pi\over h}$, where the height $h$ of \textit{a point} (not necessarily the highest one) along the loop could be much less than the linear size of the computational domain $L$. For the method we propose, to perform $\alpha-h$ fit within the full region of interest, including $h<L/2$, \als values larger than the maximal allowed by the Fourier method are needed.

%The results of the fit are shown in fig.~\ref{fig_qpole_alpha_h_fit}. 
The results of the fit show that there is indeed a strong correlation between best-match \als and the real \al. However, in some cases (46 points out of 689, about $7\%$ of all field lines) the fit seems to be off by more than one step of \al. We attribute these errors of the fit to several factors. One factor is the gridded search algorithm whereby we calculate $d(\alpha, h)$ for each point on a grid with fixed steps in both \als and h (it is clear that a better algorithm could be implemented, however, in this paper we concentrate on the theoretical possibility of the method, rather than on programming tasks). A second is that the fit is poorly constrained when a field line's shape is hidden by the projection. Finally, there are numerical errors associated with numerical integration of a field line from a field represented only on discrete grid points. %We will further address each issure in details. 

 \begin{figure}[!hc]
 \begin{center}
 \begin{tabular}{cc}
  \includegraphics[width=8cm]{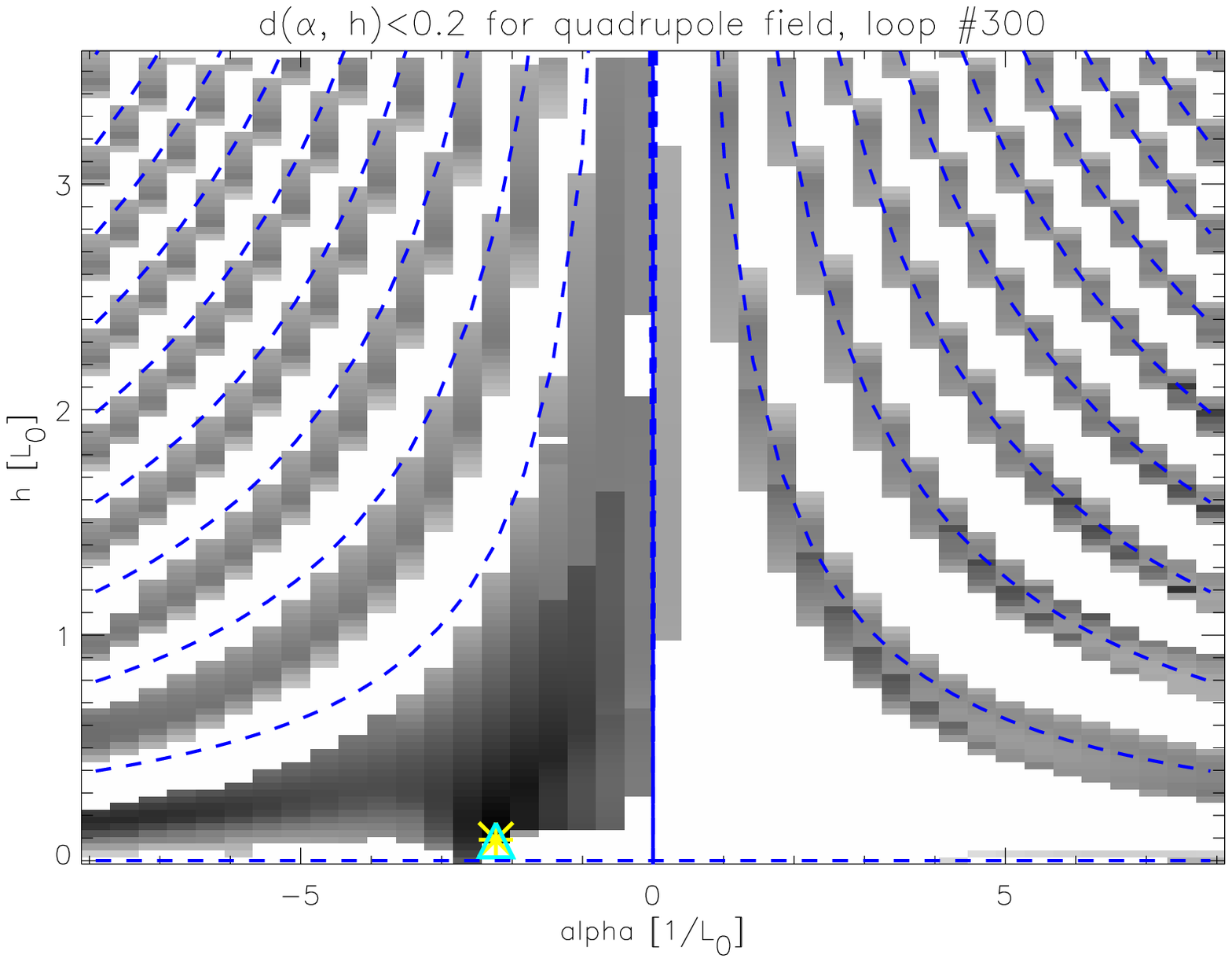} 
  \includegraphics[width=8cm]{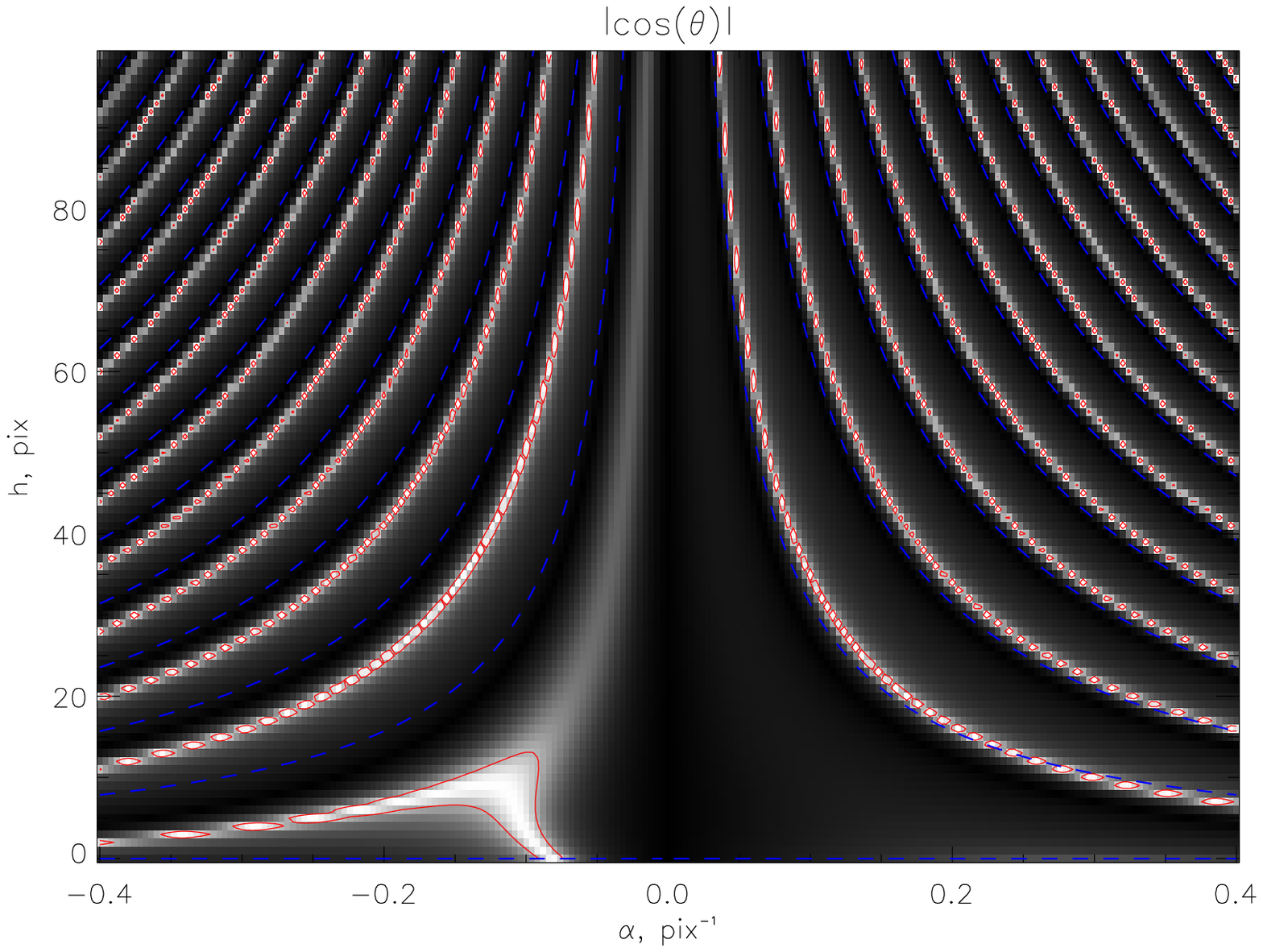} 
 \end{tabular}
 \end{center} 
 \caption{\small{\textit{(left)} -- The parameter space of $d(\alpha, h)$. The yellow asterisk shows the location of the ``real'' $(\alpha, h)$ of the field line and the cyan triangle shows the ``best-fit'' $(\alpha, h)$. The blue dotted lines are hyperbolae $h=n\pi/\alpha$, $n=0, \pm1, \pm2, ...$ For this field line, as for nearly all of them, the ``real'' minimum lies in the ``branch'' of local minima, that is within $-\pi/\alpha\leq h\leq\pi/\alpha$. We sped up the computation significantly by computing $d$ only for $(\alpha, h)$, for which at the initial point $(x_0, y_0)$ the magnetic field makes a relatively small angle with the normal of the loop: $|\cos(\theta)|=\left|\bvec\cdot\lhat\right|/|\bvec||\lhat|\leq\sqrt{2}/2$. Our study shows that for most of the ``loops'' the local minima lie within this range of $|\cos(\theta)|$. \textit{(right)} -- The parameter space of the Green's function for constant \als field in half-space. The function being plotted is $|\cos(\theta)|=\left|G_y(\rvec-\rvec_0, \alpha)/\sqrt{G_x^2(\rvec-\rvec_0, \alpha)+G_y^2(\rvec-\rvec_0, \alpha)}\right|$}, where $\rvec_0=(0, 0, 0)$ and $\rvec_1=(10\cos(320^{\circ}), 10\sin(320^{\circ}), h)$. The red contours are $|\cos(\theta)|=\sqrt{2}/2$, the blue dashes contours are $h=n\pi/\alpha$, $n=0, \pm1, \pm2, ...$}
 \label{fig_parspace0} 
 \end{figure}

\section{Shape of $d(\alpha, h)$ in the parameter space}\label{sec_shape}

Fig.~\ref{fig_parspace0} shows the function $d(\alpha, h)$ for one of the loops from the experiment described above. This function has valleys (dark) in the shapes of hyperbolae, located at or between the hyperbolae $h=n\pi/\alpha$, $n=\pm 1, \pm 2, ...$ After examining parameter spaces of many field lines we have concluded that there is one and only one ``branch'' of local minima in each $n\pi/\alpha\leq h\leq (n+1)\pi/\alpha$, except for $n=\pm1$; there are usually two or more ``branches'' in $-\pi/\alpha\leq h\leq \pi/\alpha$. 

The foregoing behavior can be explained by the Green's function in the far field. Far from the photospheric flux concentration at $\rvec_0$ the field is dominated by the monopole moment, $\bvec(\rvec)\propto\bf{G}(\rvec-\rvec_0)$. For a field restricted to half space, Green's function $\bf{G}$ is given in \citet{Chiu1977} and later in \citet{Lothian1995}. It depends on $\sin(\omega h+\phi_0)$ and $\cos(\omega h+\phi_0)$, where $\omega\propto\alpha$ and $\phi_0\propto\alpha\sqrt{(x-x_0)^2+(y-y_0)^2}$. If one changes both \als and $h$ in such a manner that $\alpha h=const$, that is, remaining on the same hyperbola in $(\alpha, h)$ parameter space, the $\sin(\omega h+\phi_0)$ and $\cos(\omega h+\phi_0)$ terms remain constant. 

This observation about the parameter space suggests a useful heuristic restriction to the search. 
Tracing field lines from $(x_0, y_0, h)$ for increasing values of $h$ (provided $\alpha\neq 0$) causes the angle between the field line and $\lvec$ at $(x_0, y_0, h)$ to increase or decrease monotonically. The cosine of this angle,
\[
\cos(\theta)={\bvec(x_0, y_0, h, \alpha)\cdot\lvec(x_0, y_0)
\over|\bvec||\lvec|}
\] 
will be a local maximum near the local minima of $d(\alpha, h)$, but not exactly at the same place. Evaluating the magnetic field at one point is, of course, much faster, than tracing a whole field line. We found that without loss of any information about local minima of $d$, we may restrict the search to only those $(\alpha, h)$, for which $|\cos(\theta)|\leq\sqrt{2}/2$.

It is clear why the local minima of $d(\alpha, h)$ are co-aligned with maxima in $|\cos(\theta)|$. Since $\lvec$ is a smooth curve, there will be a vicinity of $(x_0, y_0)$, where $\lvec$ is, to first order, a straight line. The same is true for field lines of a force-free field: the field lines are smooth curves, so the field line traced from $(x_0, y_0, h)$ in a close vicinity of this point is also to the first order a straight line. Suppose $|\cos(\theta)|=1$, i.e., $\bvec(x_0, y_0, h)$ is parallel to $\lvec(x_0, y_0)$. Then in the neighborhood of $(x_0, y_0)$ the two curves would be exactly the same, and $d$ would be close to zero, if averaged only in that vicinity. Farther from $(x_0, y_0)$ the two may differ significantly resulting in non-zero $d$ over the whole length of the loop. 

If the line-of-sight angle is such that the loop's projection is nearly a straight line, then there will be \textit{many} field lines, with different \al, that are high enough and long enough to appear nearly straight along all the length of the projection. In this case the fit may give poor results. The upshot is that even though one does not need the whole loop to perform the \al-h fit, the projection of the visible portion should not be ``too straight''. We develop a more quantitative measure for this criterion below. 
 
%\clearpage
%\section{\al-h-fit: motivation for use on non-constant-\als fields}\label{sec_motiv}
\section{\al-h fit: applied to Low \& Lou field}\label{sec_llf}

We next test the \al-h fit on a set of non-linear force-free fields from Low \& Lou (1990).~\nocite{LowLou1990}  Each of these can be viewed as the field of a singular point source placed below the photosphere and inclined. The field is specified by parameters $a$ (related to characteristic range of field's \al), $l$ (depth of the source under the photosphere, we used $l=0.3$ for all fields), $\Phi$ (orientation of the source, we used $\Phi=\pi/2$ for all fields) and $n$ (for explicit derivation and definitions, see the Appendix).

The experiments were conducted as follows. We generated several \llfs fields with different parameters. For each field, we traced a few hundred field lines, projected them on the $x$--$y$ plane and used them as synthetic loops. Then for each such loop we conducted an \al-h fit (by gridded search), with values of \als being within the range of Low \& Lou's field photospheric \al.

\begin{figure}[!hc]
 \begin{center}
 \begin{tabular}{cc}
  \includegraphics[width=10cm]{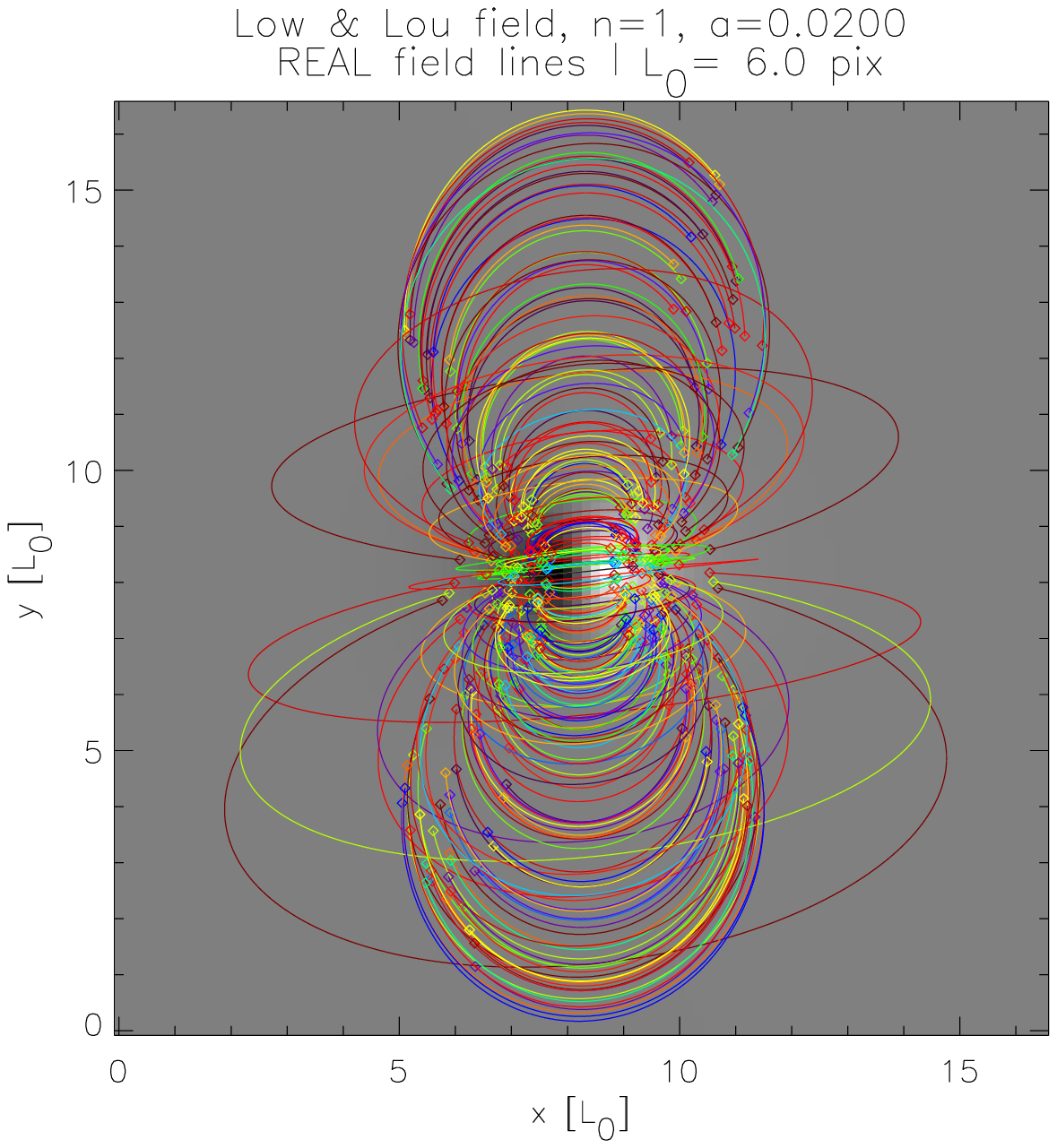} & 
  \includegraphics[width=10cm]{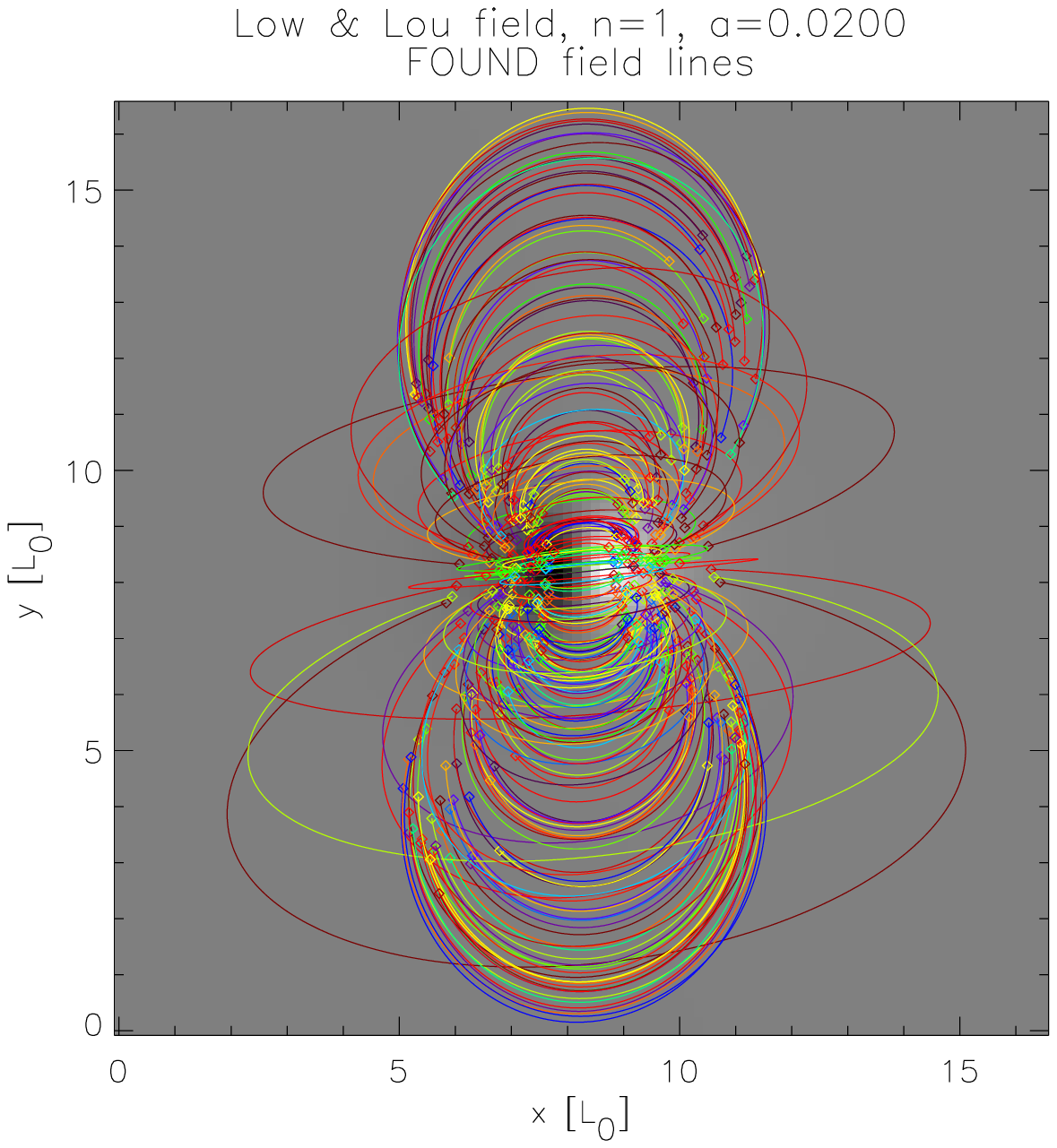} \\ 
 \end{tabular}
 \end{center} 
 \caption{\small{\textit{(left)} -- A synthetic magnetogram of a non-linear nearly-dipolar field at the photosphere, with synthetic ``loops'' -- field lines -- projected onto $x$--$y$ plane. For each of those lines we tried to approximate it with a line of a constant \als field, recording $\alpha_{found}$ and comparing it with the real \als of the original field line. \textit{(right)} -- Best-fit field lines of constant \als fields, each line belonging to a different constant \als field. Hereafter all lengths are given in the units of $L_0$ -- a characteristic separation distance between two polarities, we calculated it as the distance between the pixels with maximal and minimal magnetic fields. In this case, $L_0=6$ pix.}}
 \label{fig_low_lou_dipole_mg} 
 \end{figure}

\begin{figure}[!hc]
 \begin{center}
 \begin{tabular}{cc}
  \includegraphics[height=6cm]{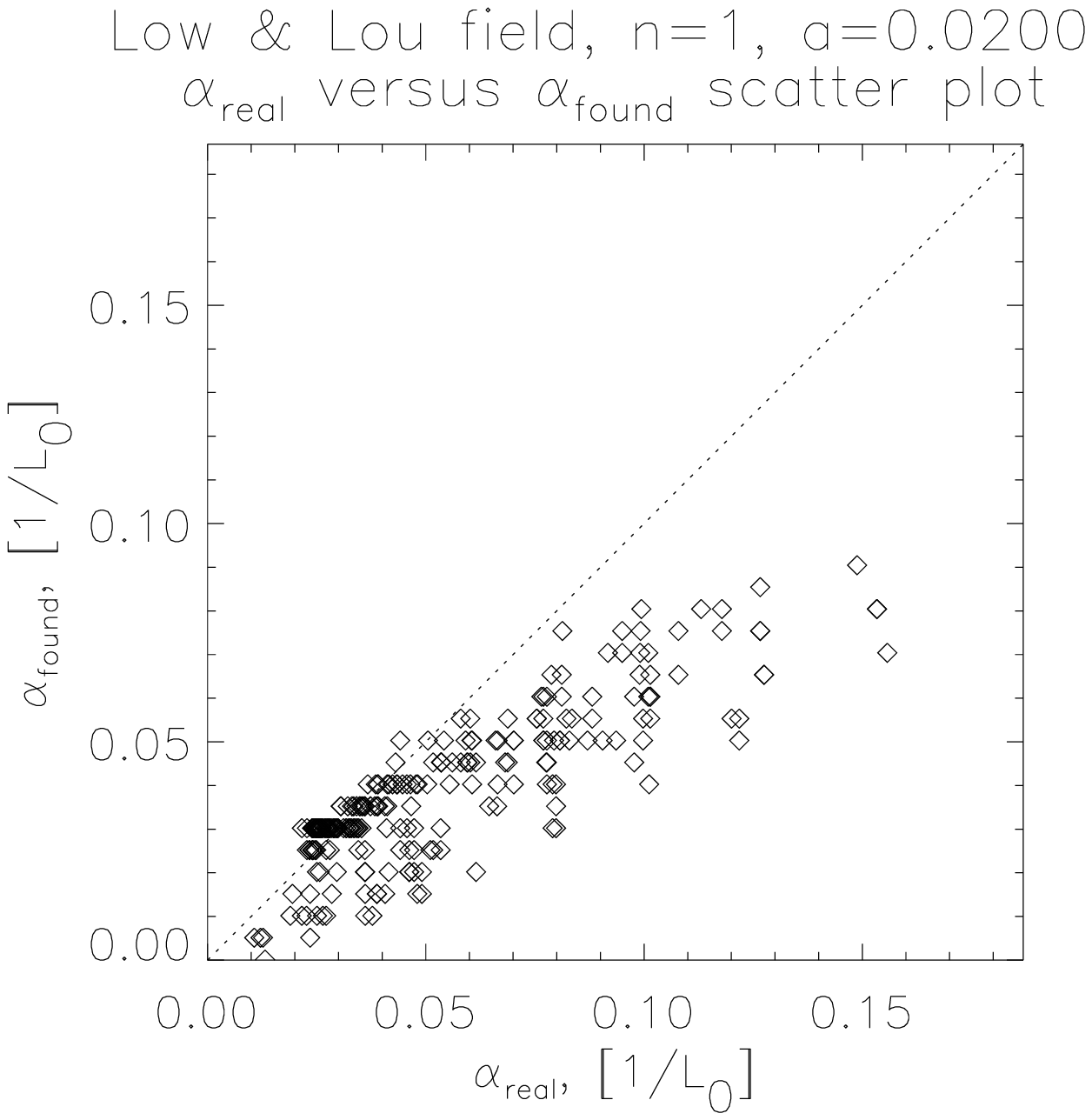} &
  \includegraphics[height=6cm]{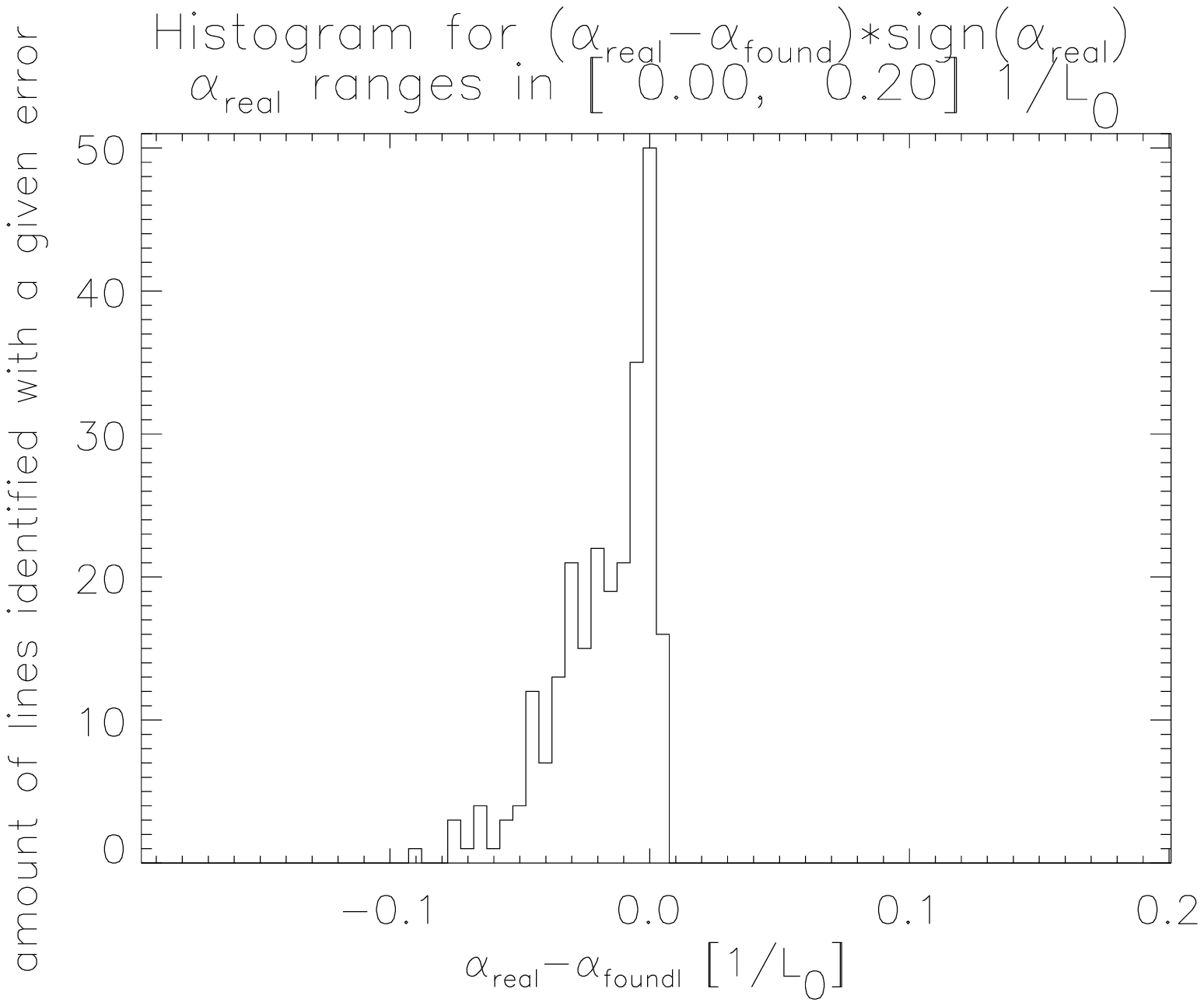} \\
 \end{tabular}
 \end{center} 
 \caption{\small{Results for the data, shown in Fig.~\ref{fig_low_lou_dipole_mg}. On the left is a scatter plot of $\alpha_{real}$ vs. $\alpha_{found}$. The correlation is evident. On the right is the histogram of the error $\alpha_{real}-\alpha_{found}$.}}
 \label{fig_low_lou_dipole_res} 
 \end{figure}

We found that for a ``dipolar'' field (in the sense of it having two distinct polarities, see Fig.~\ref{fig_low_lou_dipole_mg}) with $n=1$, $a=0.02$, the values resulting from constant \als fits do indeed correlate with the real values of the field lines, as shown in Fig.~\ref{fig_low_lou_dipole_res} (left plot).

\begin{figure}[!hc]
 \begin{center}
 \begin{tabular}{cc}
  \includegraphics[height=7cm]{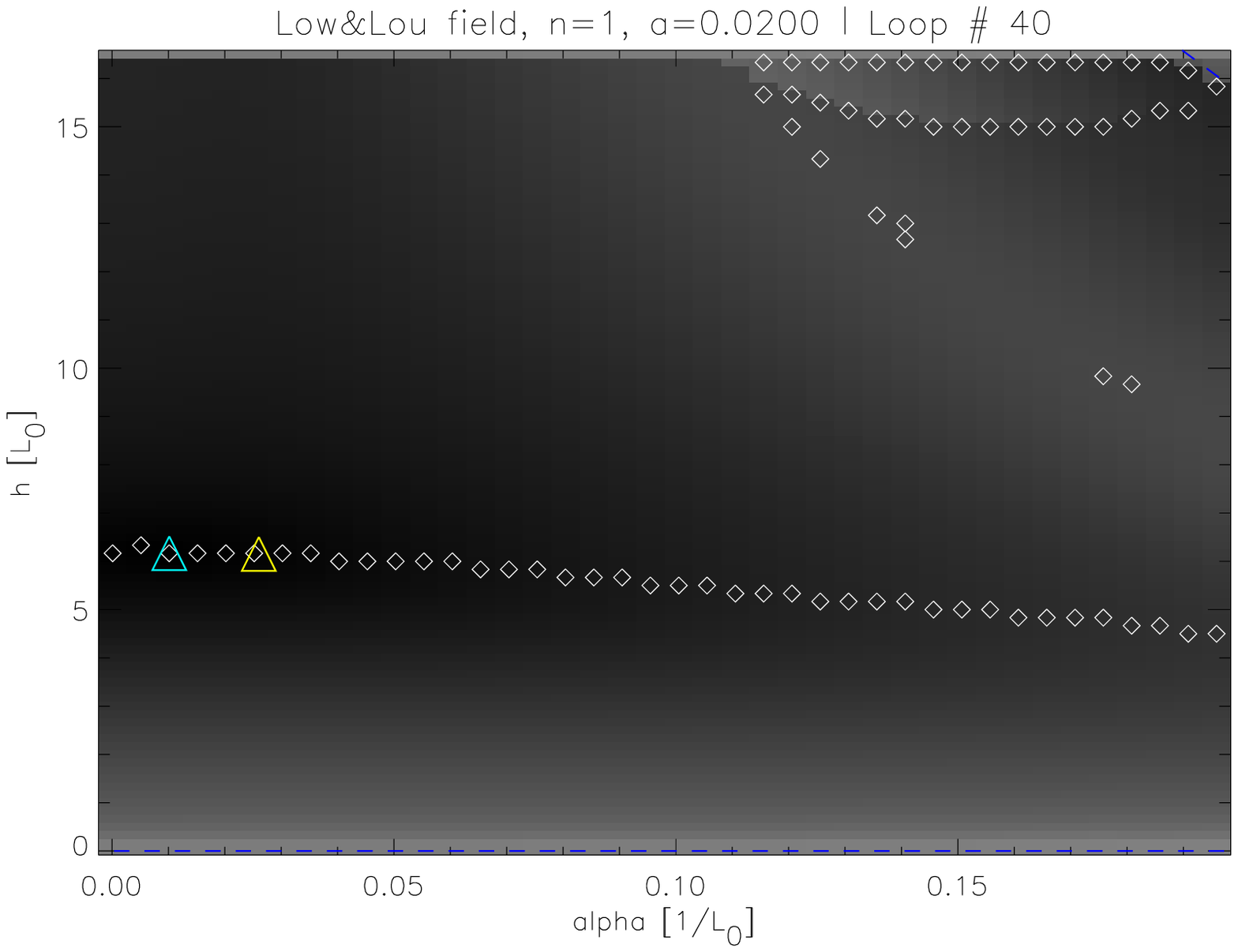} & 
  \includegraphics[height=7cm]{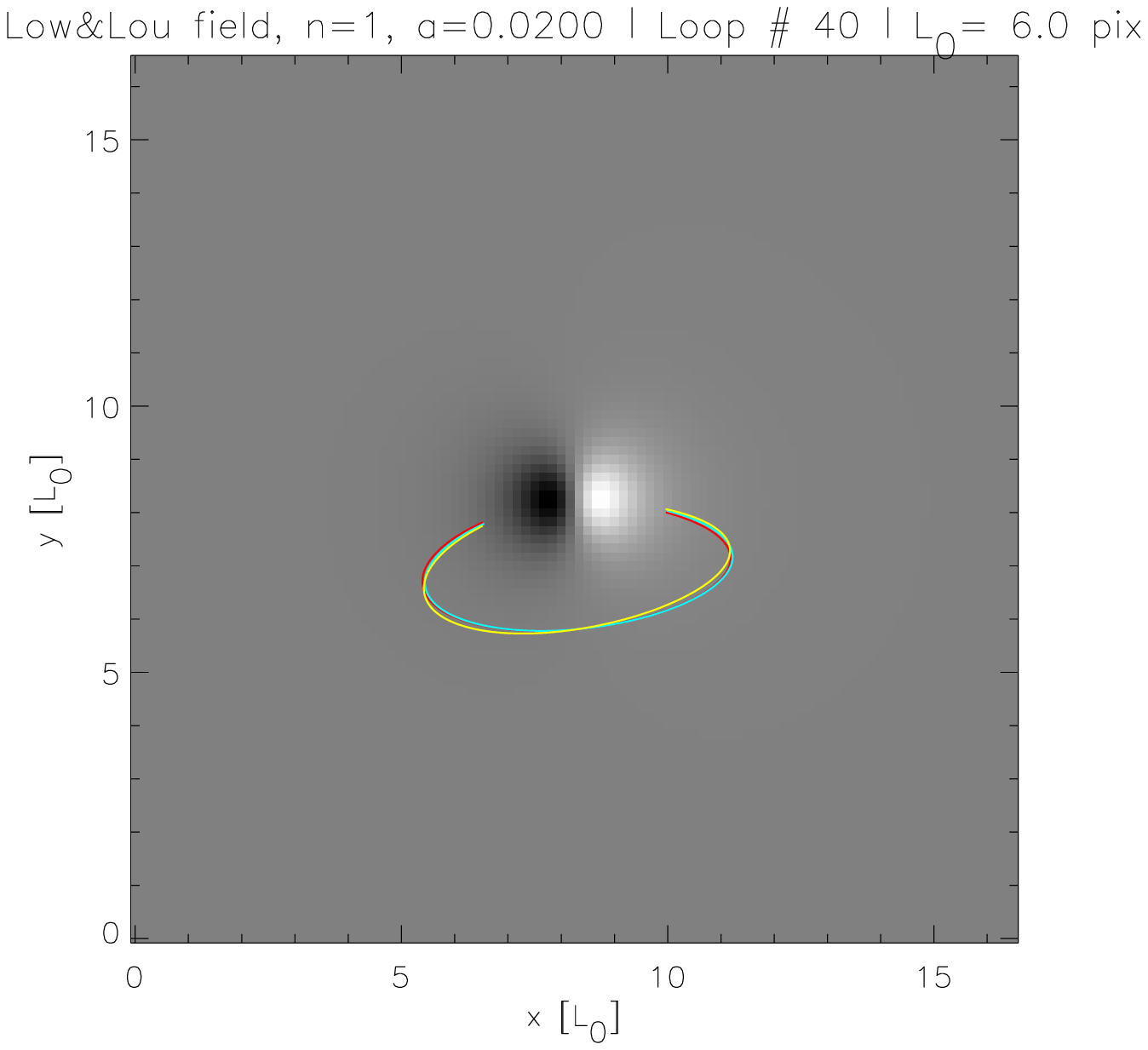} \\ 
 \end{tabular}
 \end{center} 
 \caption{\small{\textit{(left)} -- A typical $d(\alpha, h)$ parameter space for one of the field lines shown in Fig.~\ref{fig_low_lou_dipole_mg}. There is one nearly horizontal valley of local minima and a clue to other possible valleys for larger $(\alpha, h)$. White diamonds show the local minima in individual columns. Cyan triangle shows the location of the global minimum. A yellow triangle shows ``real'' $(\alpha, h)$ of the Low \& Lou field line. Blue dashed lines (barely visible on this plot, but more evident on the other plots of this kind) are hyperbolae $\alpha h=n\pi$, $n=0,\pm 1,\pm 2, ...$ 
 \textit{(right)} -- Best-fit for the same field line (cyan), field of the constant \als field, traced from the ``real'' values (yellow), and the the Low \& Lou field line (red). The difference is barely visible, however, the cyan line seems to match red line better than the yellow line.}}
 \label{fig_low_lou_dipole_mg_alpha} 
 \end{figure}

For this dipolar field the $d(\alpha, h)$ plots for most loops had one distinct valley of local minima (horizontal or nearly horizontal), and hints of other valleys at larger $\alpha$ or $h$. Another notable feature of the low valley is that it tends to cross the $\alpha=0$ line rather than approach it asymptotically. The parameter space for one of these loops is shown in Fig.~\ref{fig_low_lou_dipole_mg_alpha} (left plot). Notable in that plot is \emph{the global minimum was always in the lowest nearly-horizontal valley}. The ``true location'' (from the original Low \& Lou field line) was also within that valley; however, it is sometimes offset with respect to the global minimum. In general, \emph{\als of the global minima are correlated with \als of the original Low \& Lou field lines}, as shown in Fig.~\ref{fig_low_lou_dipole_res} (left plot). Finally, Fig.~\ref{fig_low_lou_dipole_mg_alpha} (right plot) shows that the constant \als field line of the global minimum seems to approximate the original Low \& Lou field line quite well, although a tendency to under-estimate \als is evident, and still clearer in the histogram of Fig.~\ref{fig_low_lou_dipole_res} (right plot).

For notational convenience, we hereafter refer to lines of Low \& Lou fields as \textit{real field lines} and to their best-fits of constant \als field as \textit{found field lines}. We will also use $(\alpha_{real}, h_{real})$ to denote the parameters of the real field line (recall: $h$ is a height in the midpoint of the line's projection into the photosphere). Similarly, we will use $(\alpha_{found}, h_{found})$ notation to refer to the parameters of the found field line. 

\begin{figure}[!hc]
 \begin{center}
 \begin{tabular}{cc}
  \includegraphics[height=7cm]{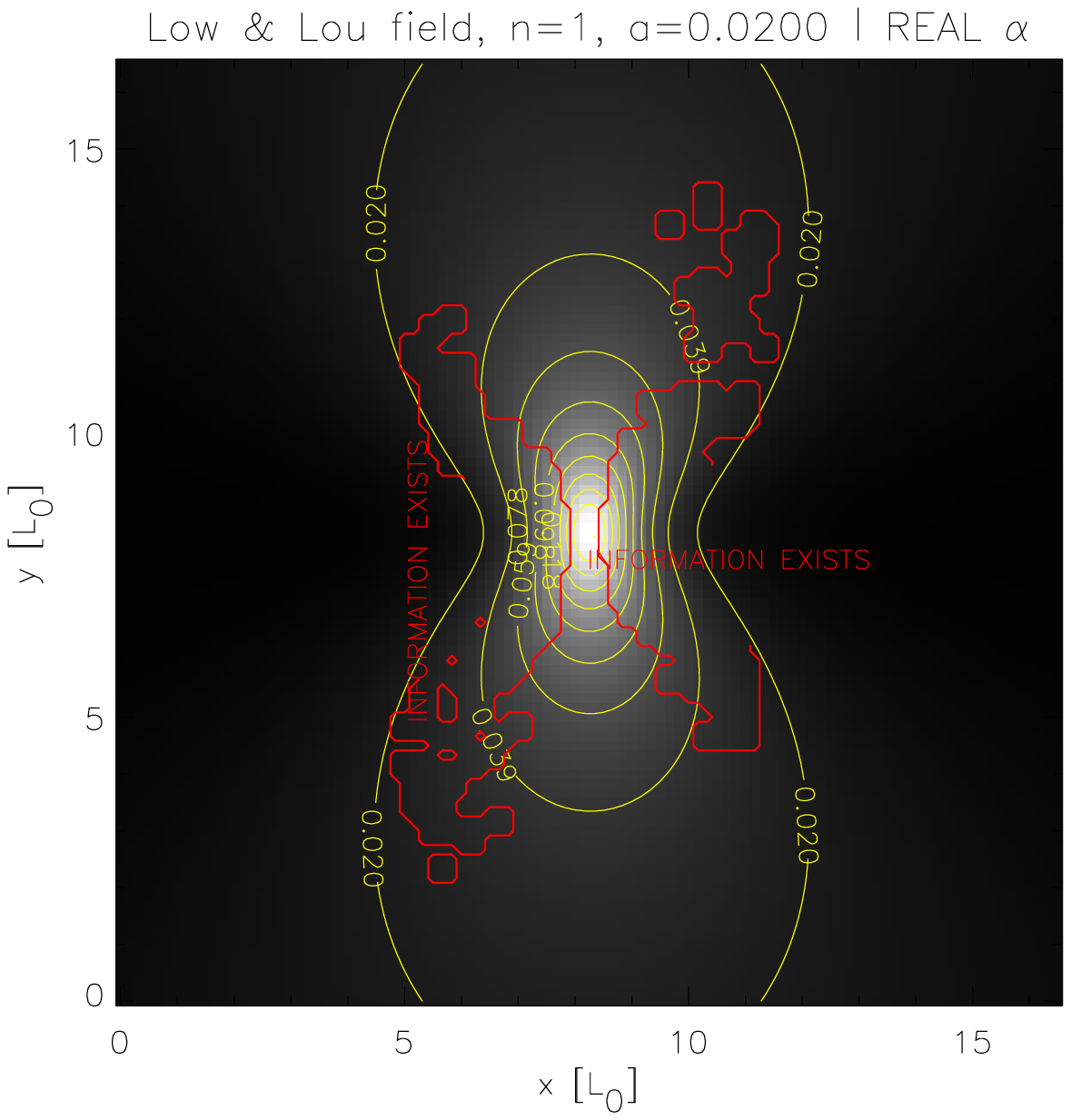} & 
  \includegraphics[height=7cm]{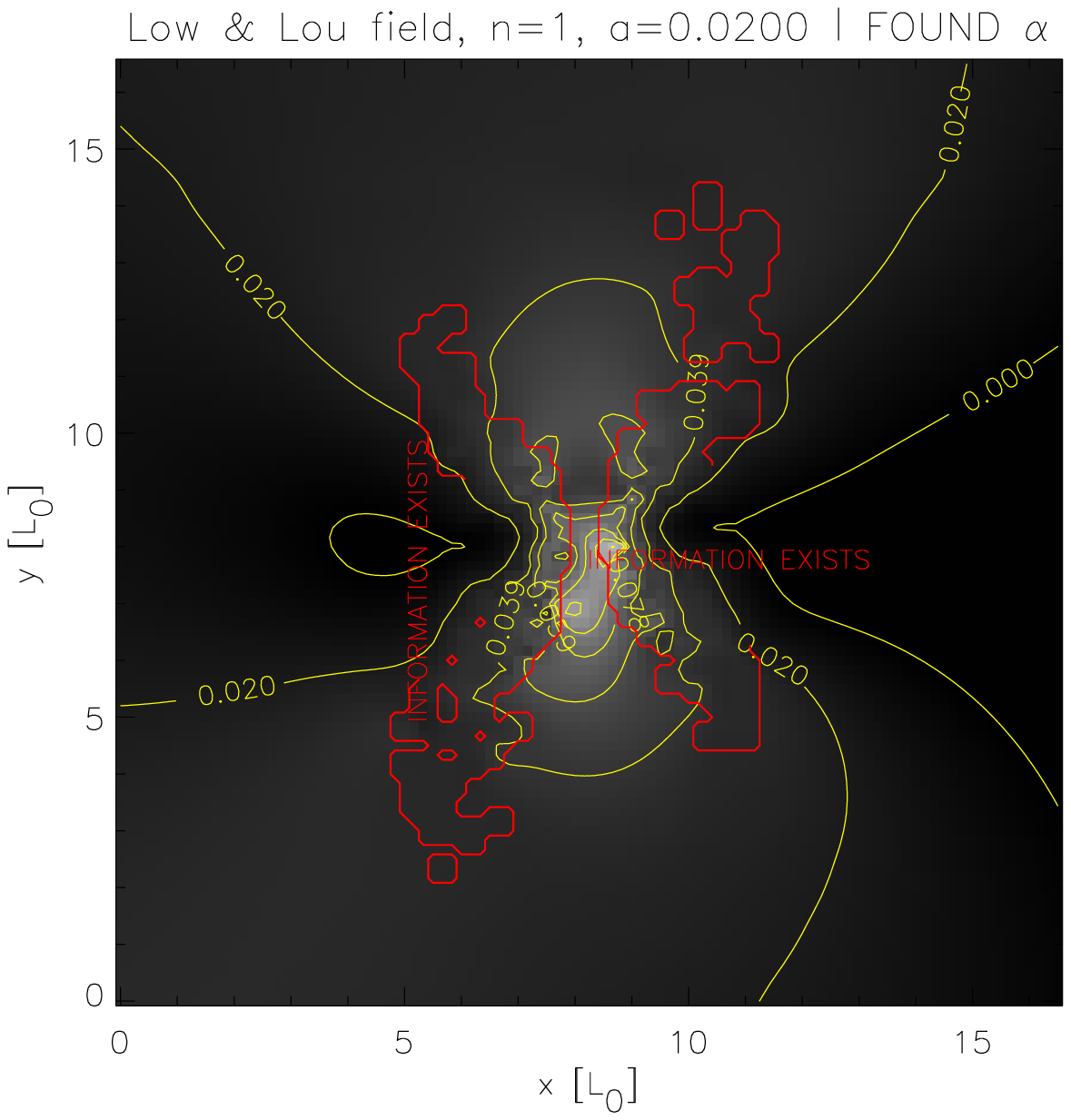} \\ 
 \end{tabular}
 \end{center} 
 \caption{\small{\textit{(left)} -- Photospheric distribution of \als for the field from Fig.~\ref{fig_low_lou_dipole_mg}. The yellow contours are contours of \al.
 \textit{(right)} -- the result of the reconstruction using \al-h fit. The grayscale and contours on this reconstruction are identical to those on the real distribution. The red contours show the location of the footpoints of the found lines, that is, \emph{there is no information outside of these contours}, and whatever is outside is shown solely for easy viewing. The only meaningful part is inside of the red contours. The result was extrapolated using thin plate splines fit into the set of footpoints with found \al. This robust fit is sensitive to individual noisy points, and it is intended only to illustrate of the potential possibility of such reconstruction. Yet, with all these remarks, such robust fit is capable of reconstructing the principal shape of the distribution.}}
 \label{fig_low_lou_dipole_parspace} 
 \end{figure}

Best-fits are potentially useful in reconstructing \emph{the photospheric distribution of \al}. We constructed a photospheric map of \als by assigning the coronal value to the footpoints of the reconstructed field lines. A full map requires a smoothing, averaging or interpolation, to assign \als to photopsheric points around the footpoints of observed loops. To illustrate this possibility we did a robust reconstruction with bicubic spline interpolation \citep[see, for example,][]{Press1986}, shown in Fig.~\ref{fig_low_lou_dipole_parspace}. The fit that we did is simple and robust, nevertheless it is able to reconstruct the general shape of the actual distribution of \als in the Low and Lou field.

\begin{figure}[!hc]
 \begin{center}
 \begin{tabular}{cc}
  \includegraphics[height=6cm]{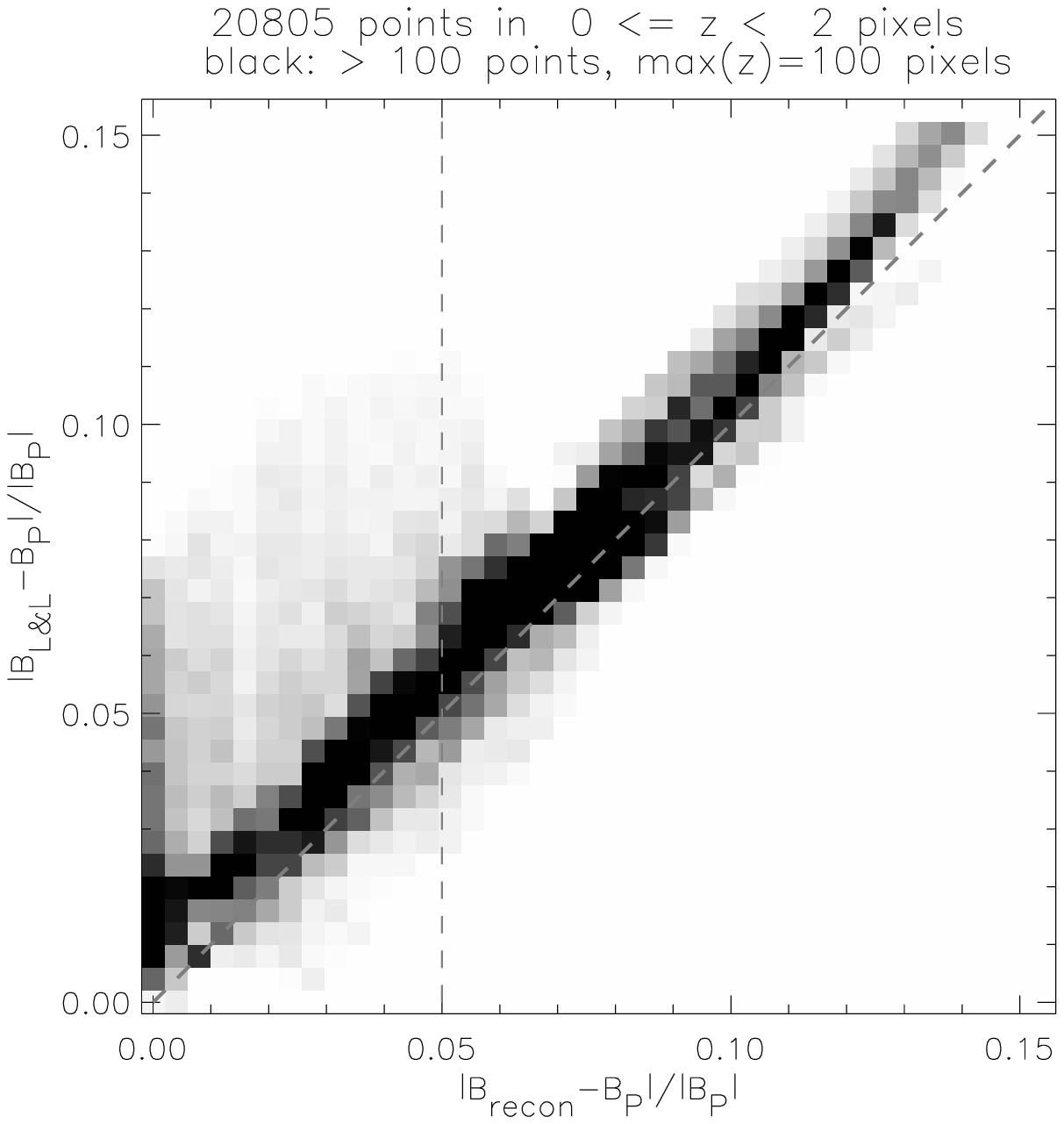} &
  \includegraphics[height=6cm]{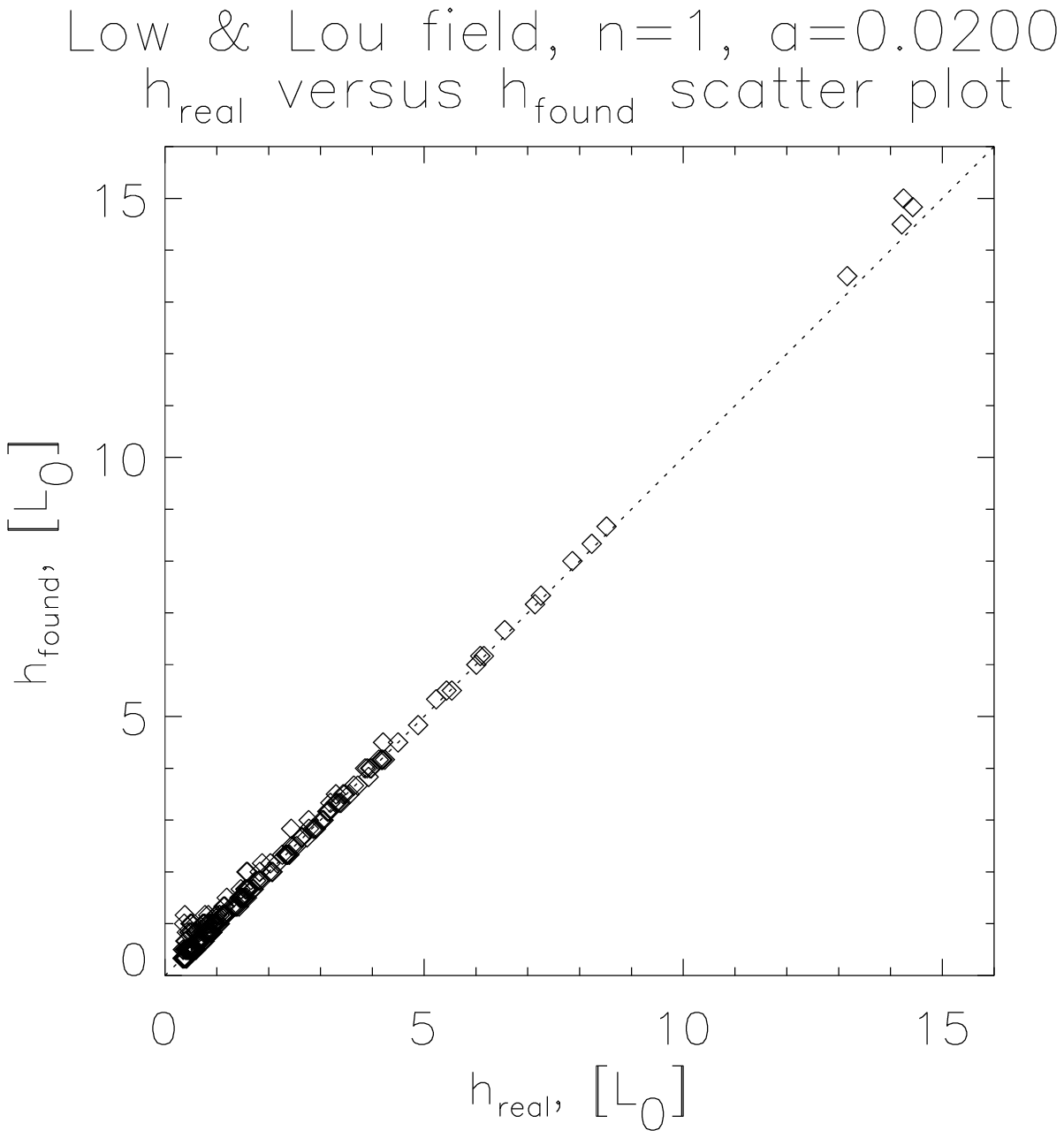} \\
 \end{tabular}
 \end{center} 
 \caption{\small{\textit{(left)} -- the comparison of reconstructed magnetic field, $\bvec_{recon}$ to the \llfs field, $\bvec_{L\&L}$, at the photospheric level ($0<h<2$ pix). We follow the ``found'' (best-fit) field lines and evaluate $\bvec_{lfff}$ (with different \als for each field line), $\bvec_{L\&L}$ and $\bvec_{pot}$ along them. By $\bvec_{recon}$ we mean the composition of $\bvec_{lfff}$'s for all loops (strictly speaking, the way it's constructed makes it in general not force-free and probably not even source-free, so it could hardly be called a magnetic field at all, rather, an approximation of the reconstructed field, evaluated along different field lines). As discussed in the text, most of the field is potential, so we compare the ``non-potential'' contributions only, normalized by the potential field. Here $\bvec_{pot}$ is a potential magnetic field, that is restricted to the upper half-space and has the same Dirichlet boundary conditions as $\bvec_{L\&L}$ and the same as $\bvec_{recon}$. It seems that the non-potential part of the magnetic field is reconstructed well. \textit{(right)} -- scatter plot of $h_{real}$ vs. $h_{found}$. It seems, that $h$ is found with much better confidence than \al.}}
 \label{fig_low_lou_dipole_b} 
 \end{figure}

Another measure of the quality of the fit is its reconstruction of magnetic field $\bvec$, for example, at $z=0$. We utilize the form $\bvec_{ff}=\bvec_{pot}+\bvec_{np}$, where $\bvec_{ff}$ is the full force-free field, $\bvec_{pot}$ is the potential field with the same normal component at the boundary, and $\bvec_{np}$ is a ``current contribution'' --- a non-potential force-free field with $\bvec_{np}\cdot\nhat|_{\partial\vol}=0$. Note that $\bvec_{pot}$ is the same for $\bvec_{L \& L}$ and $\bvec_{recon}$ (reconstructed), since it is uniquely defined by the volume and by the Dirichlet boundary conditions. For ``weakly non-potential'' field $|\bvec_{pot}|\gg|\bvec_{np}|$; this is true of some of our cases. Rather than comparing $\bvec_{L \& L}$ to $\bvec_{recon}$, we compare their ``current contribution'' terms, normalized by the potential field: $|\bvec_{L \& L}-\bvec_{pot}|/|\bvec_{pot}|$ to $|\bvec_{recon}-\bvec_{pot}|/|\bvec_{pot}|$. The histogram for $z\in [0, 2]$~pix is shown in Fig.~\ref{fig_low_lou_dipole_b} (left). To make it, we evaluated $\bvec_{L \& L}$, $\bvec_{recon}$ and $\bvec_{pot}$ \textit{along the found field lines}. It seems that for most of points the two fields were nearly identical, suggesting the accuracy of the reconstruction.

From both Fig.~\ref{fig_low_lou_dipole_res} and Fig.~\ref{fig_low_lou_dipole_b} it seems that the reconstruction does a better job for smaller \als and for weaker $\bvec_{np}$, than for larger \als and for stronger $\bvec_{np}$. This and the reasoning from the previous section suggest that \emph{\al-h fit might not work for strongly twisted, or maybe strongly non-linear, fields}. We tried to determine the range of \als for which the fit would yield reliable results. For that, we generated several more Low \& Lou fields, this time quadrupolar (in the sense of it having three polarities, like the field of a point quadrupole): we kept $n=2$ and gradually increased $a$ over the following values: $a\in[0.05, 0.1885, 0.3, 0.6, 1.0, 1.5, 2.0]$; in addition we computed a field for $n=3$, $a=0.4$. We generated both signed and unsigned Low \& Lou fields; both have identical photospheric $B_z$, but the first one has $\alpha>0$ and $\alpha<0$, while the second one has only $\alpha>0$ (see Appendix). This was done in order to relate the errors of the fit with the ``non-linearity'', that is, with how much \als changes over a fixed length. %The results are shown on Figs.~\ref{low_lou_res_a_0_05_n_2_s}-\ref{low_lou_res_a_0_4_n_3_u}.

\begin{figure}[!hc]
\begin{center}
\begin{tabular}{cc}
\includegraphics[height=5cm]{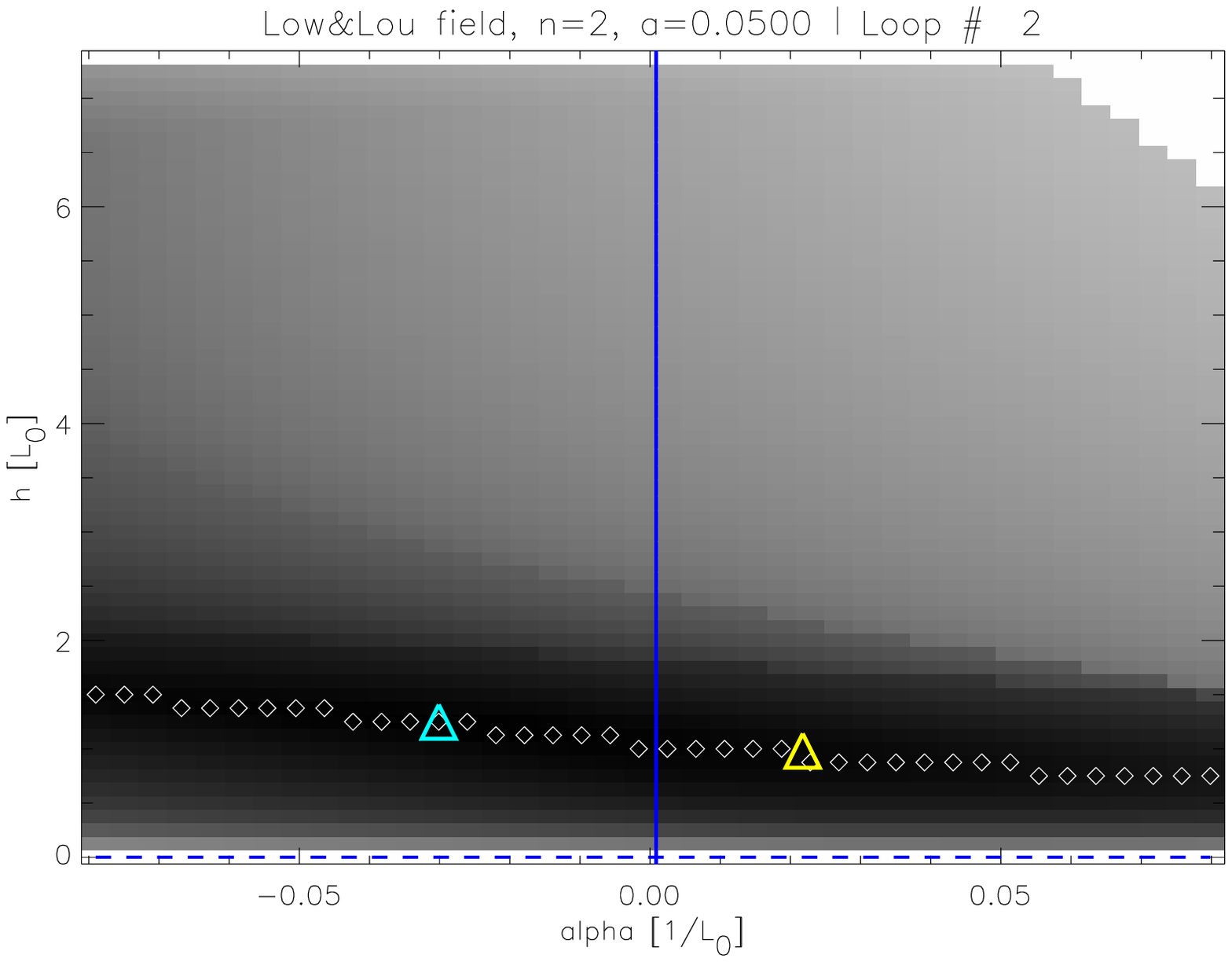} &
\includegraphics[height=5cm]{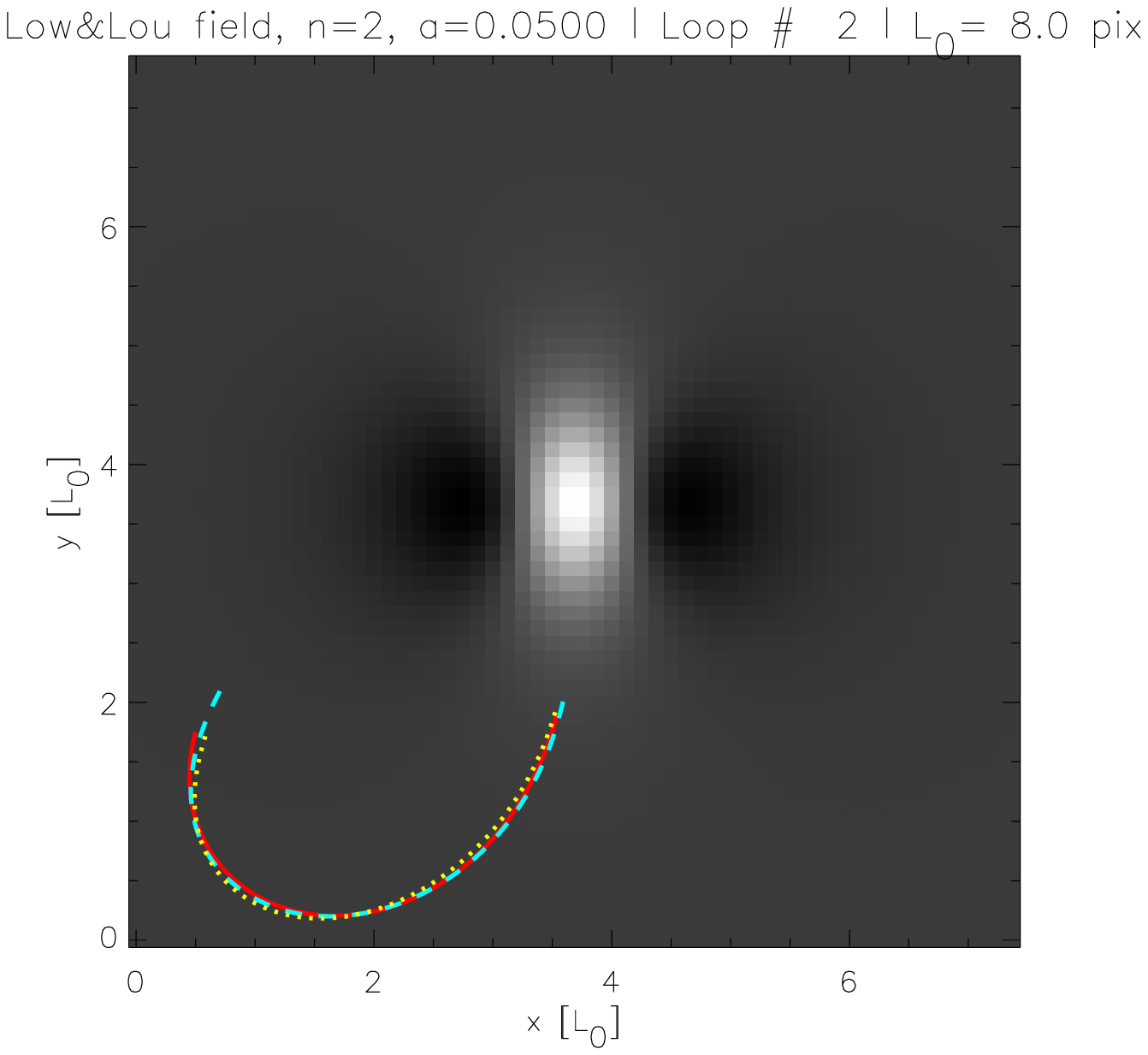} \\
\includegraphics[height=5cm]{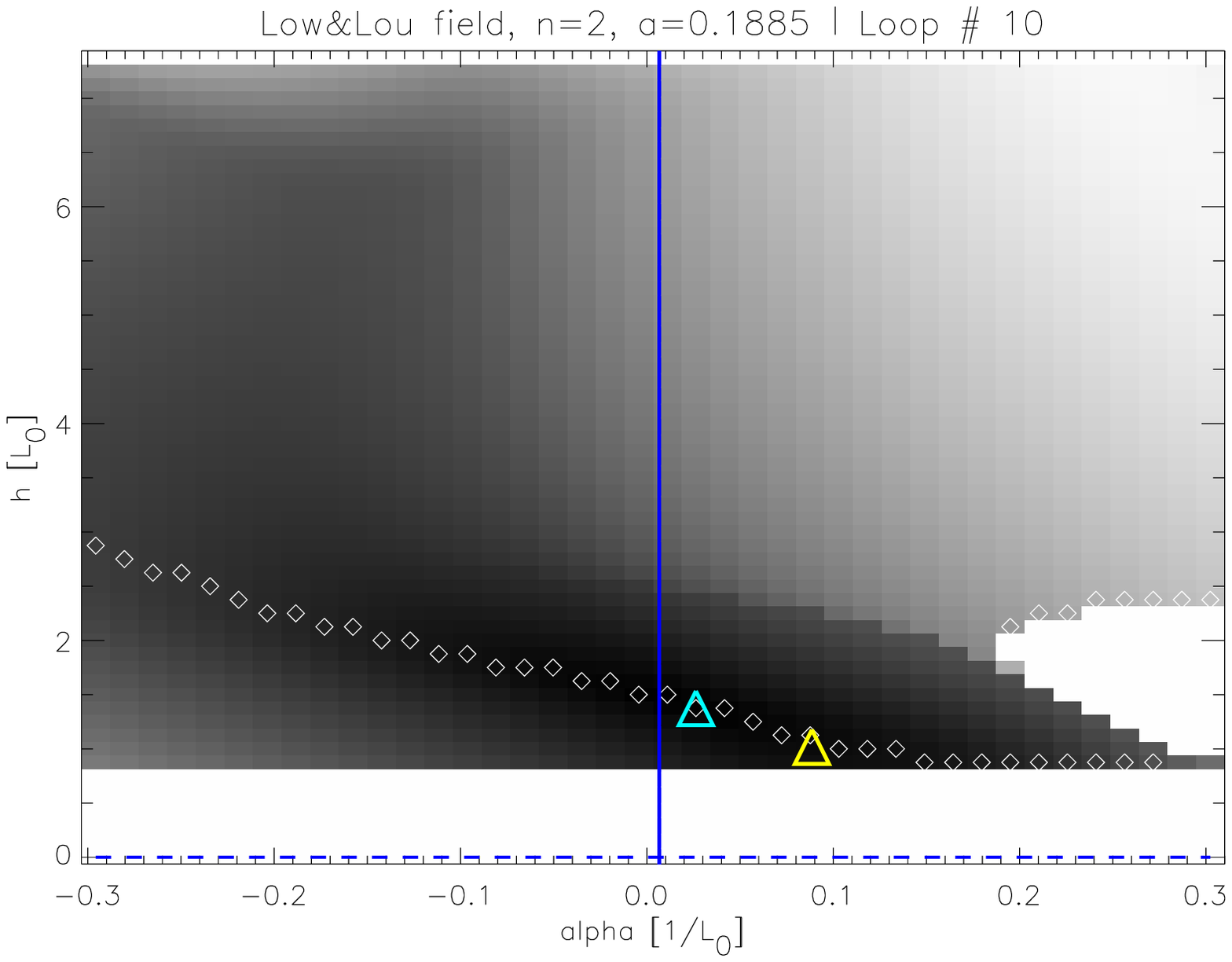} &
\includegraphics[height=5cm]{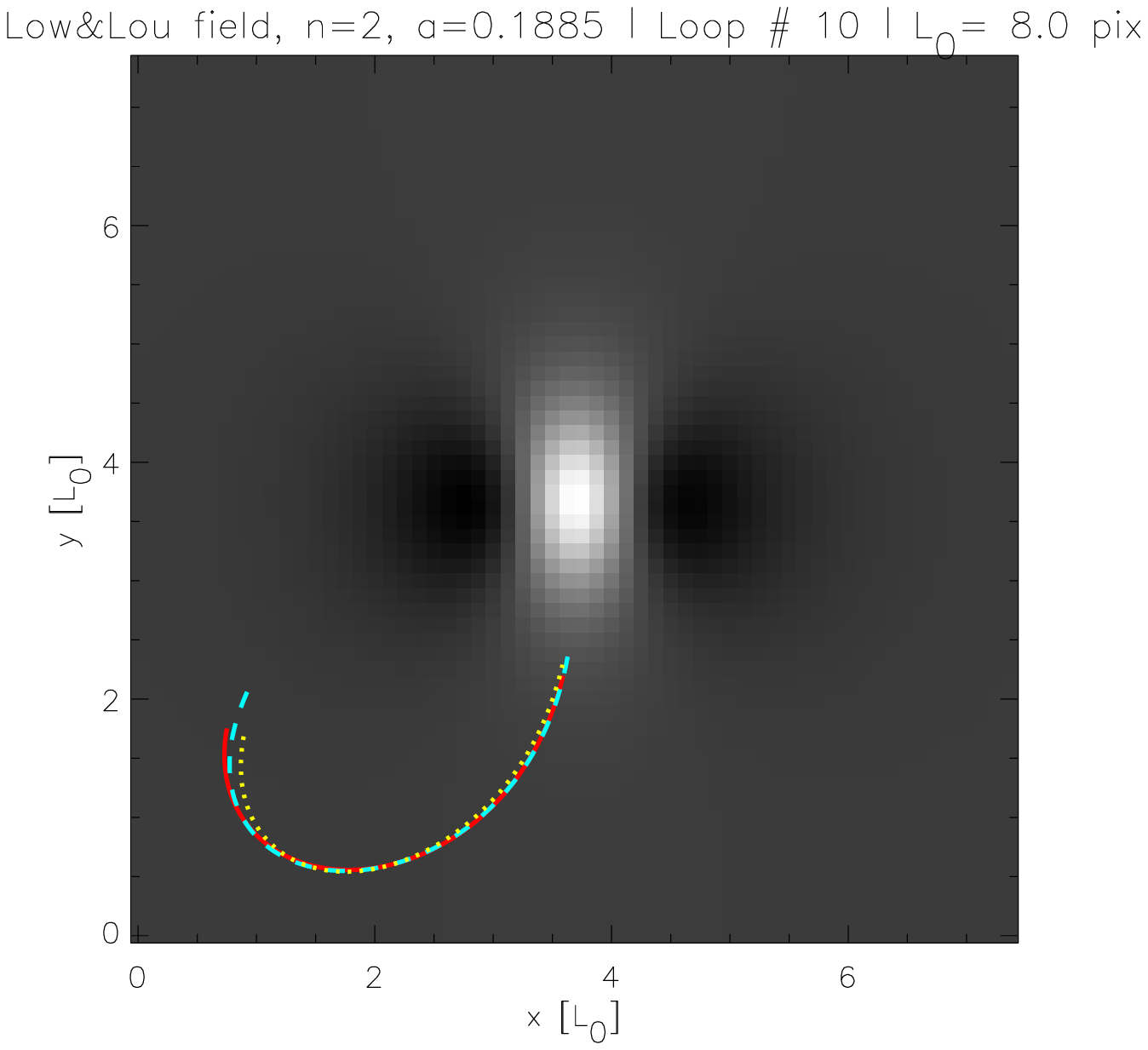} \\
\includegraphics[height=5cm]{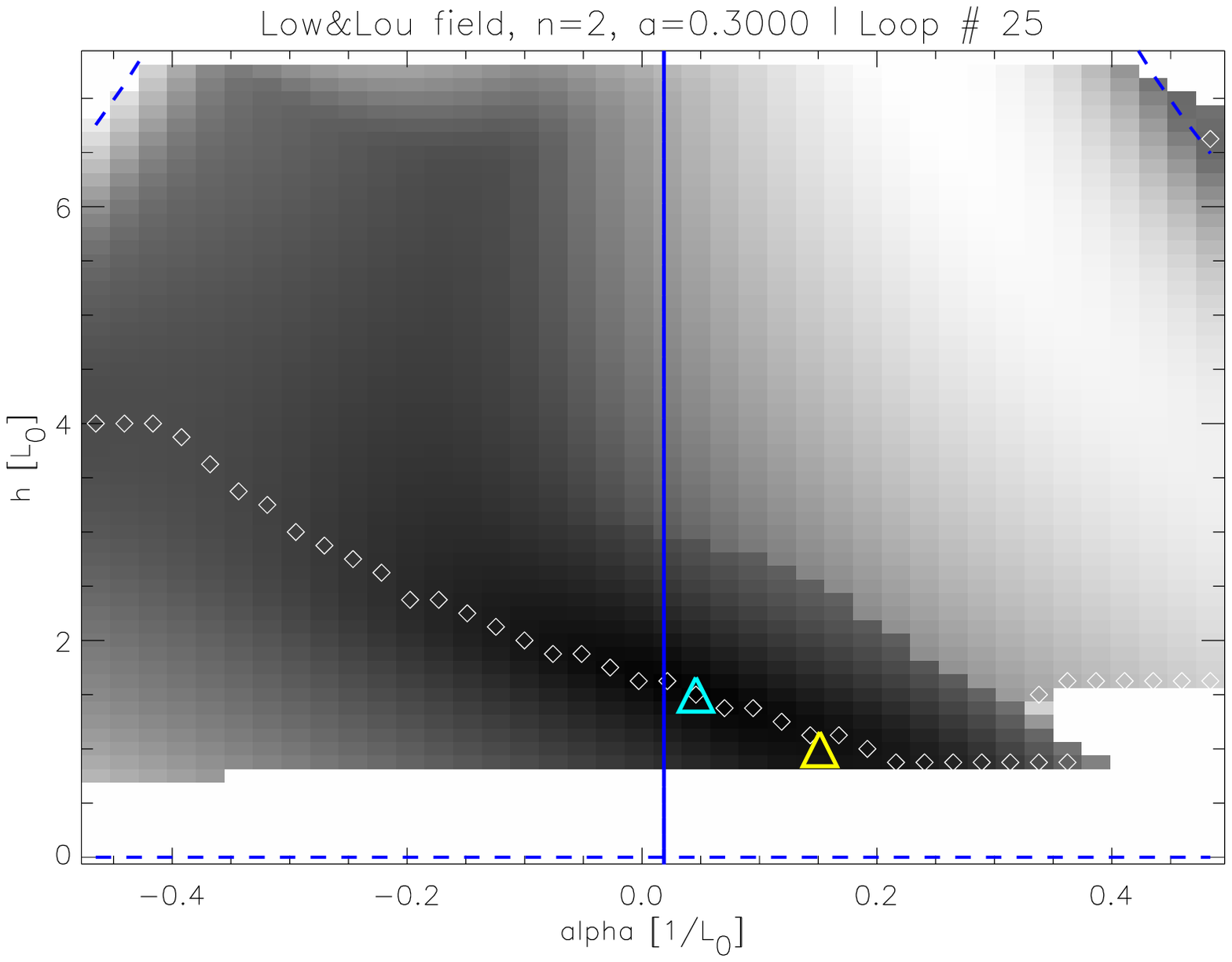} &
\includegraphics[height=5cm]{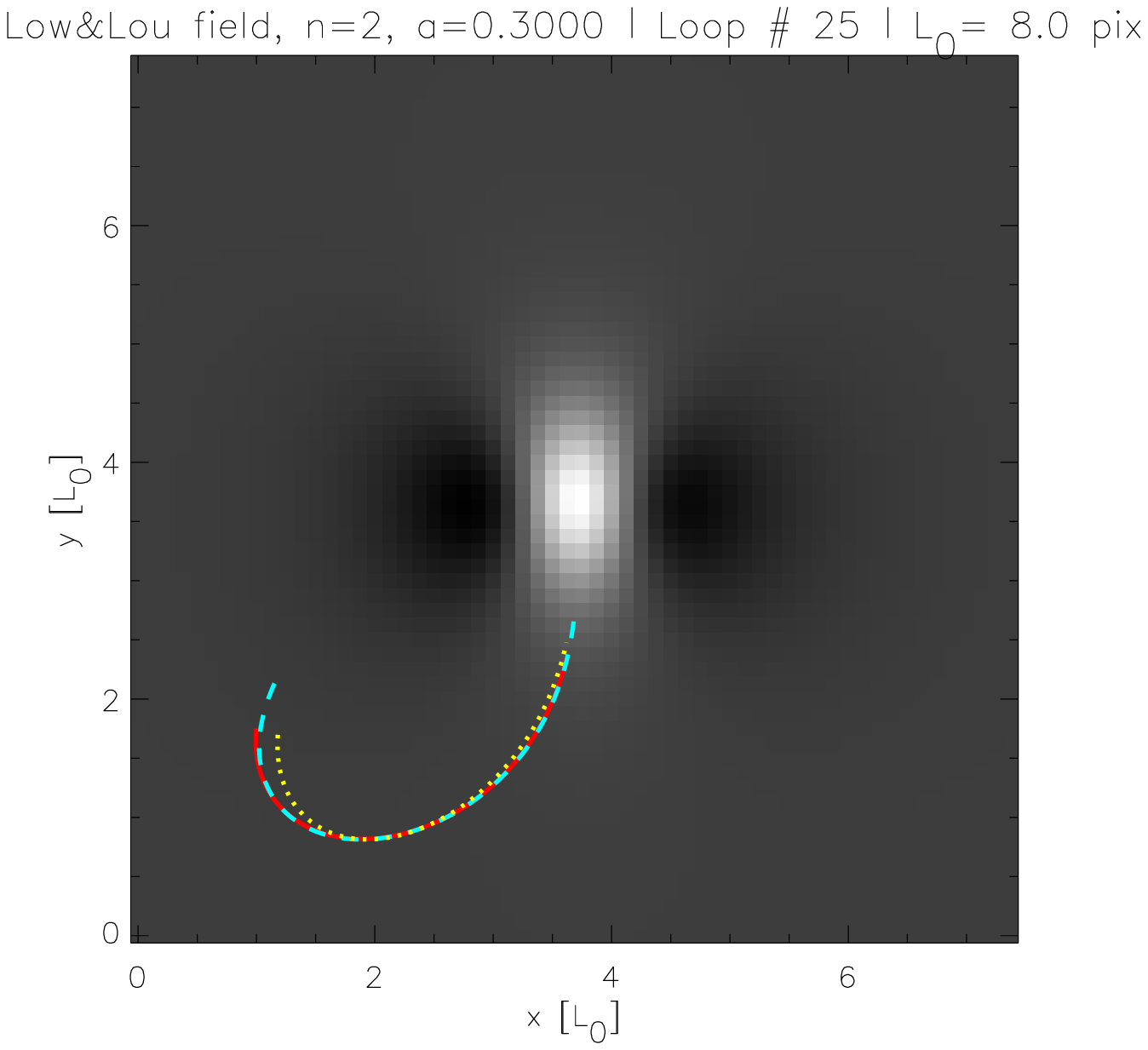} \\
\includegraphics[height=5cm]{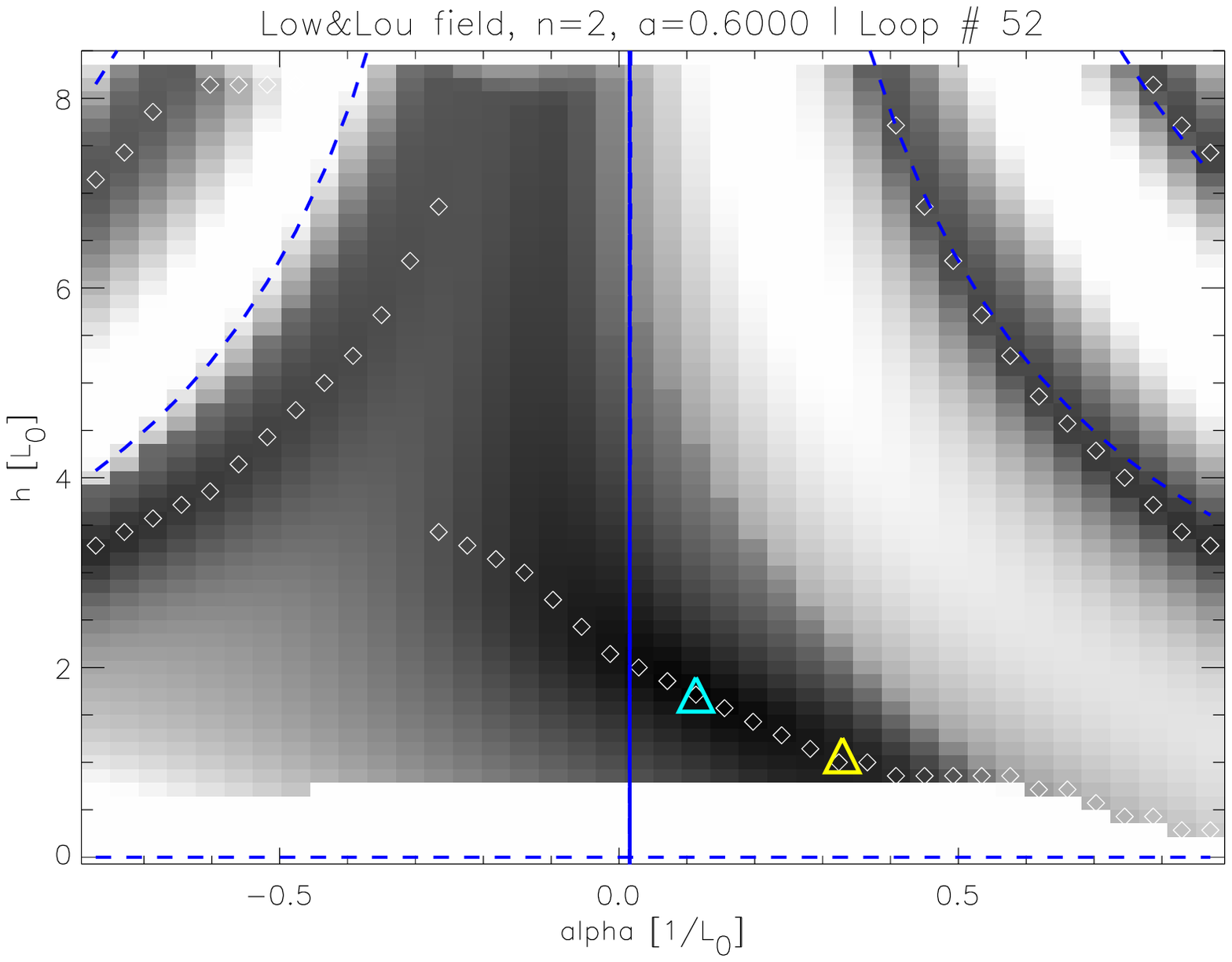} &
\includegraphics[height=5cm]{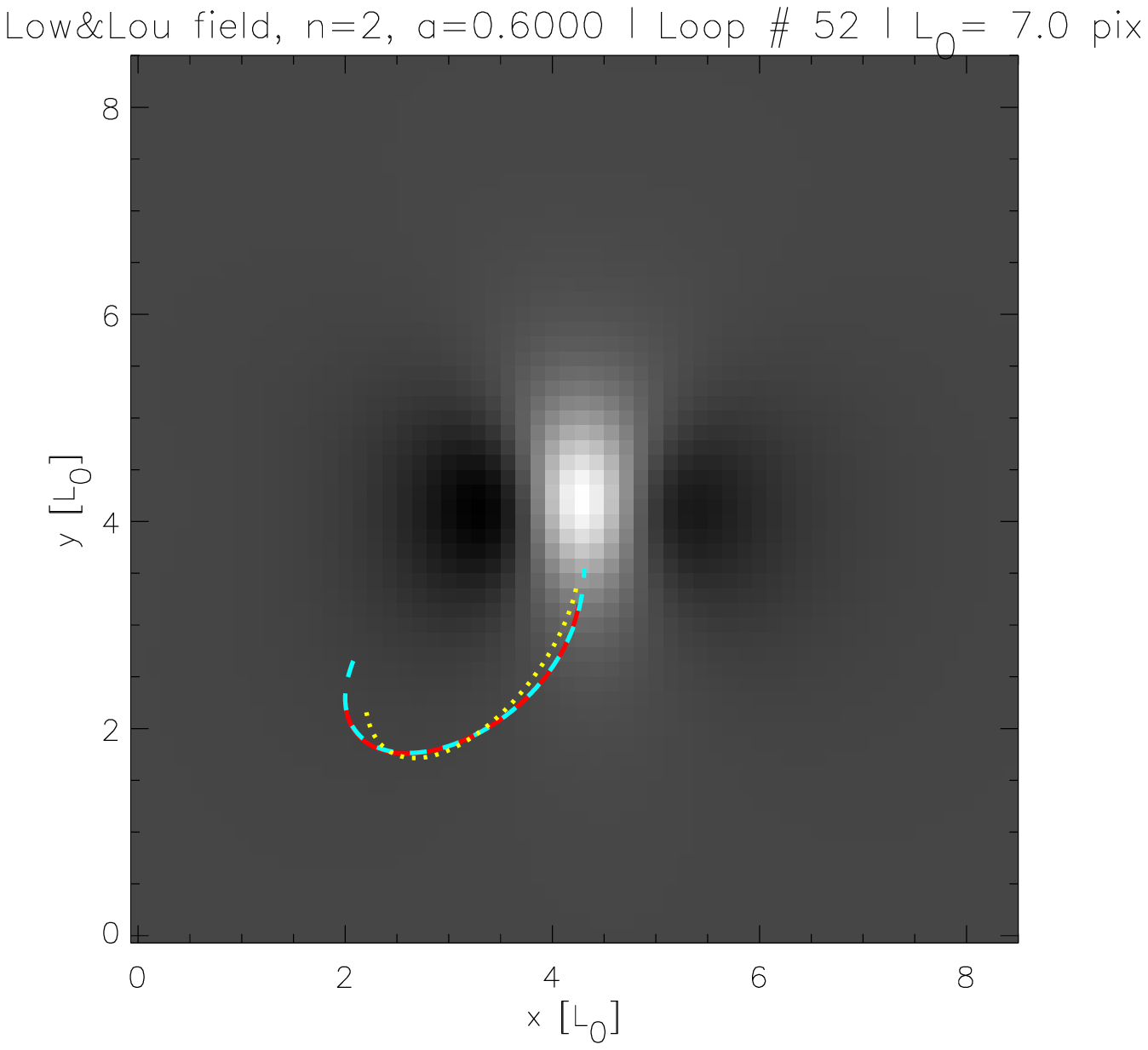} \\
 \end{tabular}
 \end{center}
 \caption{\small{\textit{(left column)} -- typical parameter spaces for \llfs fields with gradually increasing $a$. Here dashed blue lines are hyperbolae $\alpha h=0,\pm\pi,\pm 2\pi, ...$, white diamonds show local minima in each column $\alpha=const$, yellow triangle shows the location of $(\alpha_{real}, h_{real})$ and cyan triangle shows the location of the global minimum, that is, $(\alpha_{found}, h_{found})$. \textit{(right column)} -- the original ``loop'' of each of those parameter spaces (red), the ``global minimum'' field line (cyan) that has $(\alpha_{found}, h_{found})$ and a constant \als field line, that has $(\alpha_{real}, h_{real})$ (yellow).}}
 \label{parspaces_diff_alpha_p1}
 \end{figure}

\begin{figure}[!hc]
\begin{center}
\begin{tabular}{cc}
\includegraphics[height=5cm]{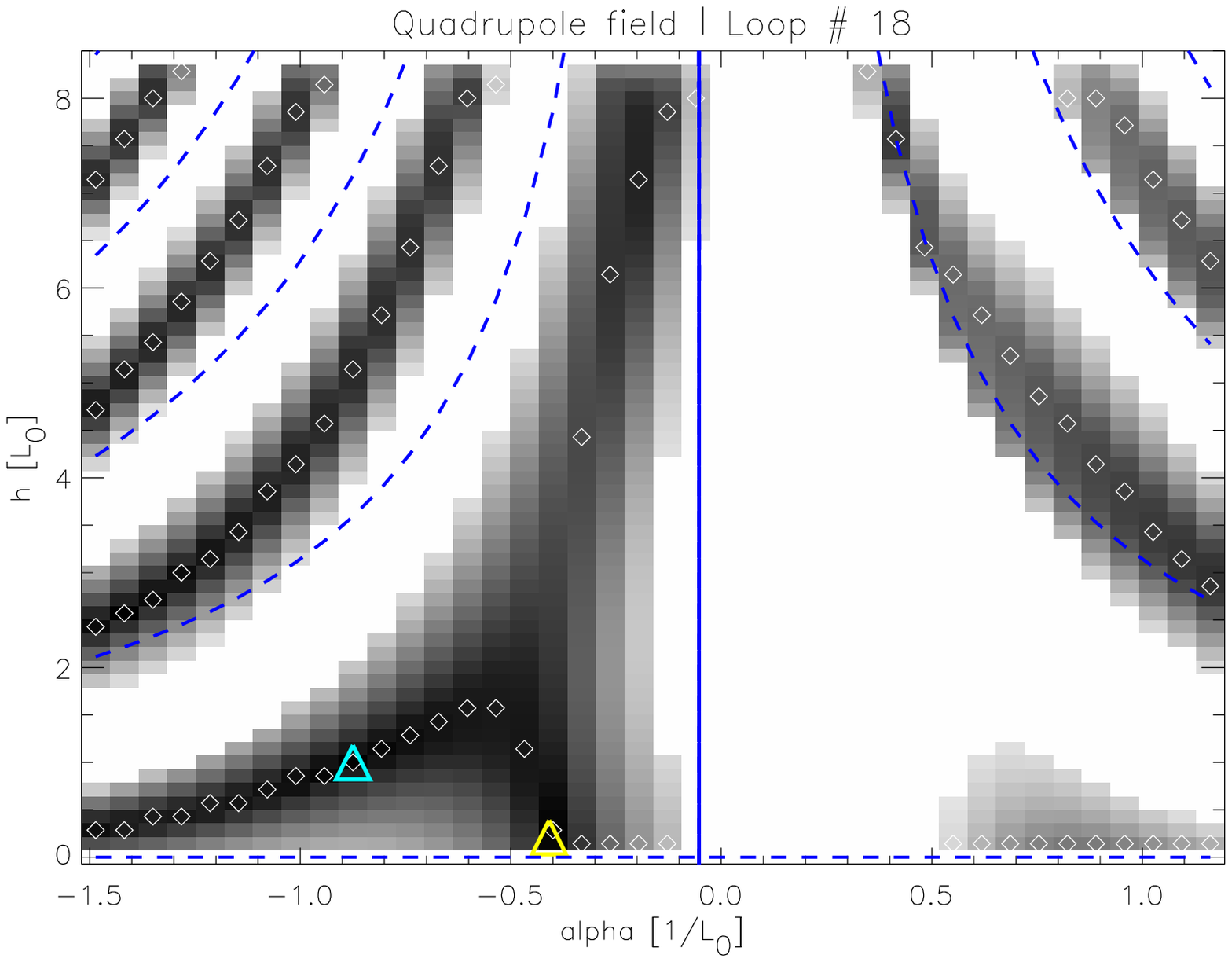} &
\includegraphics[height=5cm]{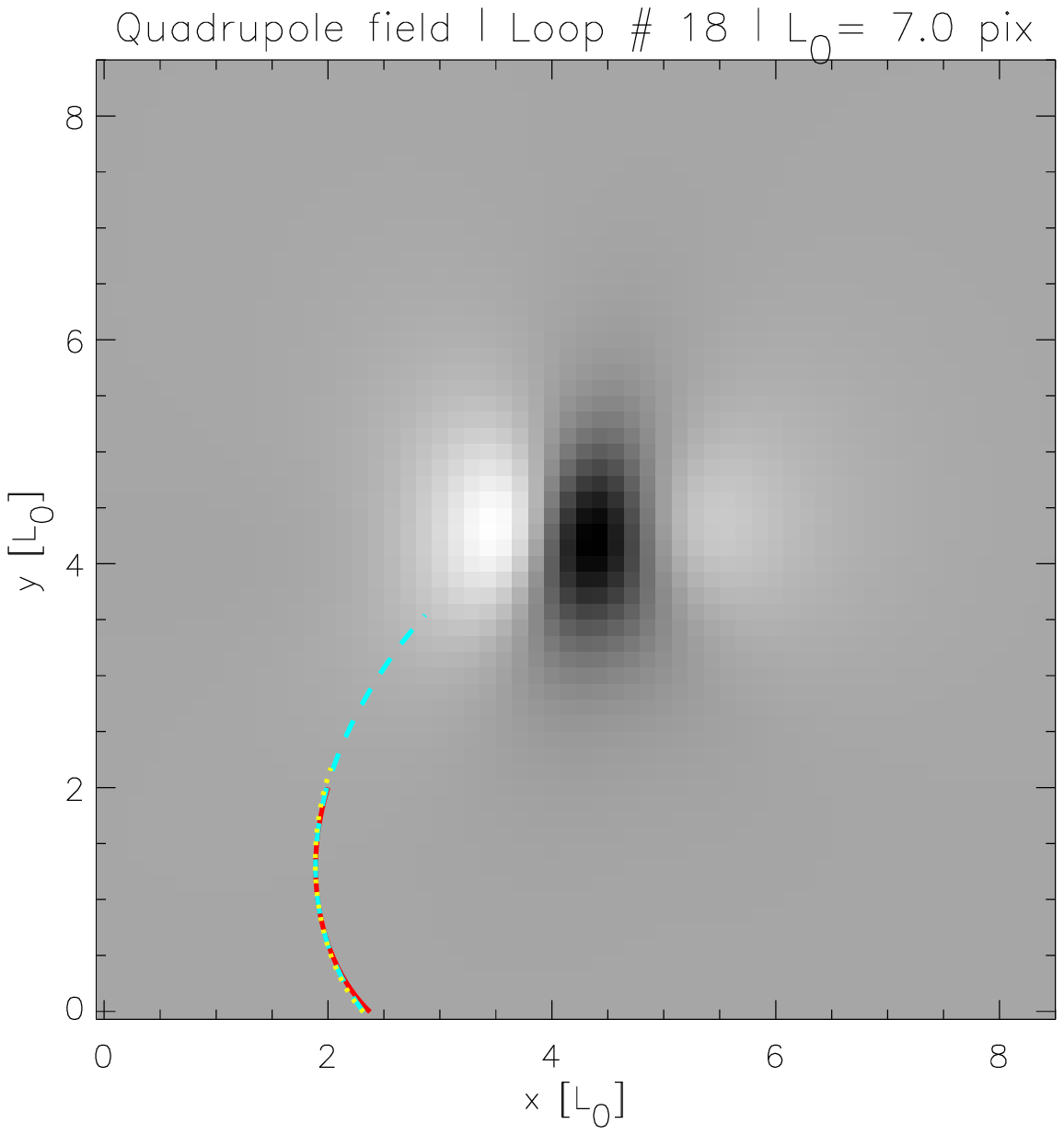} \\
\includegraphics[height=5cm]{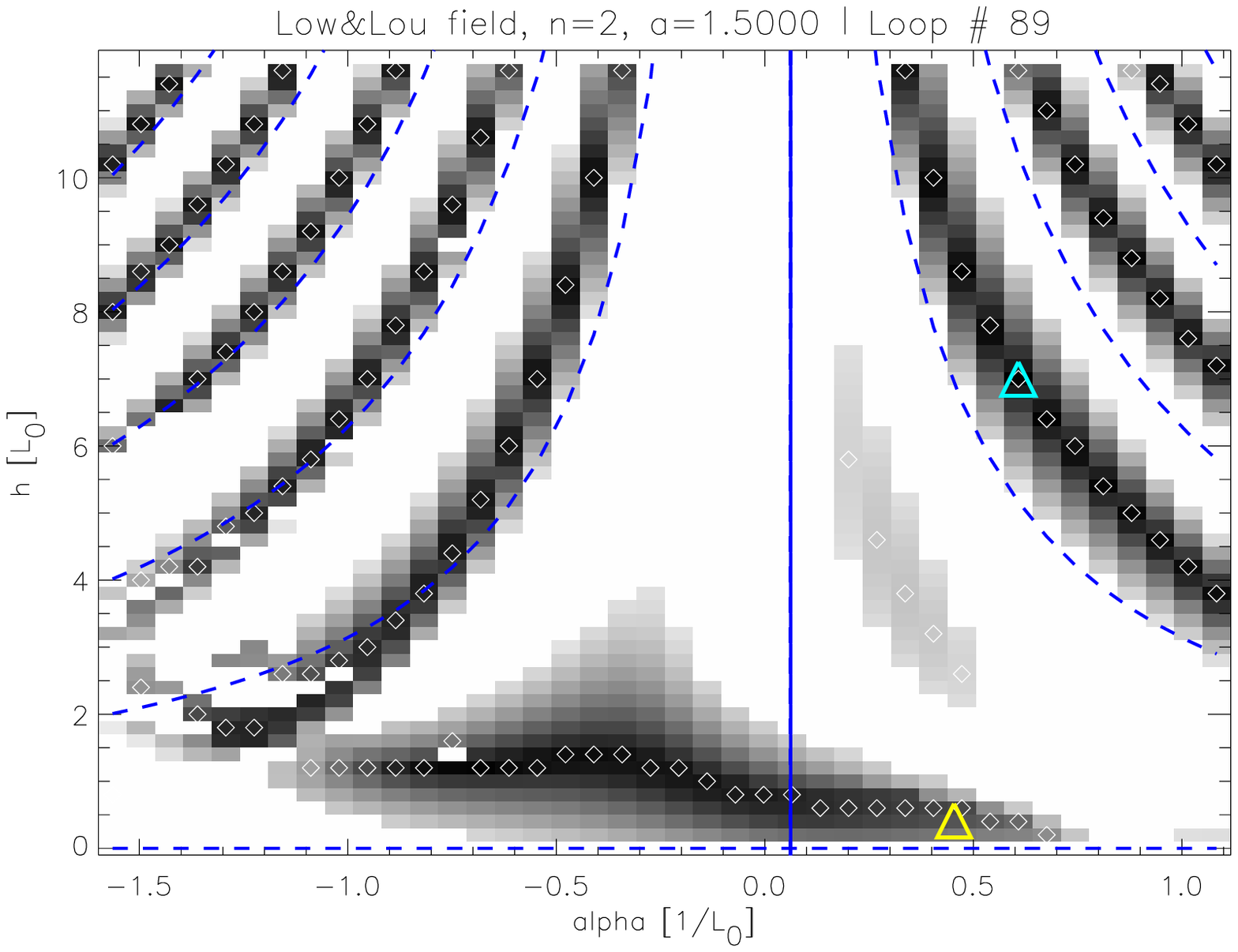} &
\includegraphics[height=5cm]{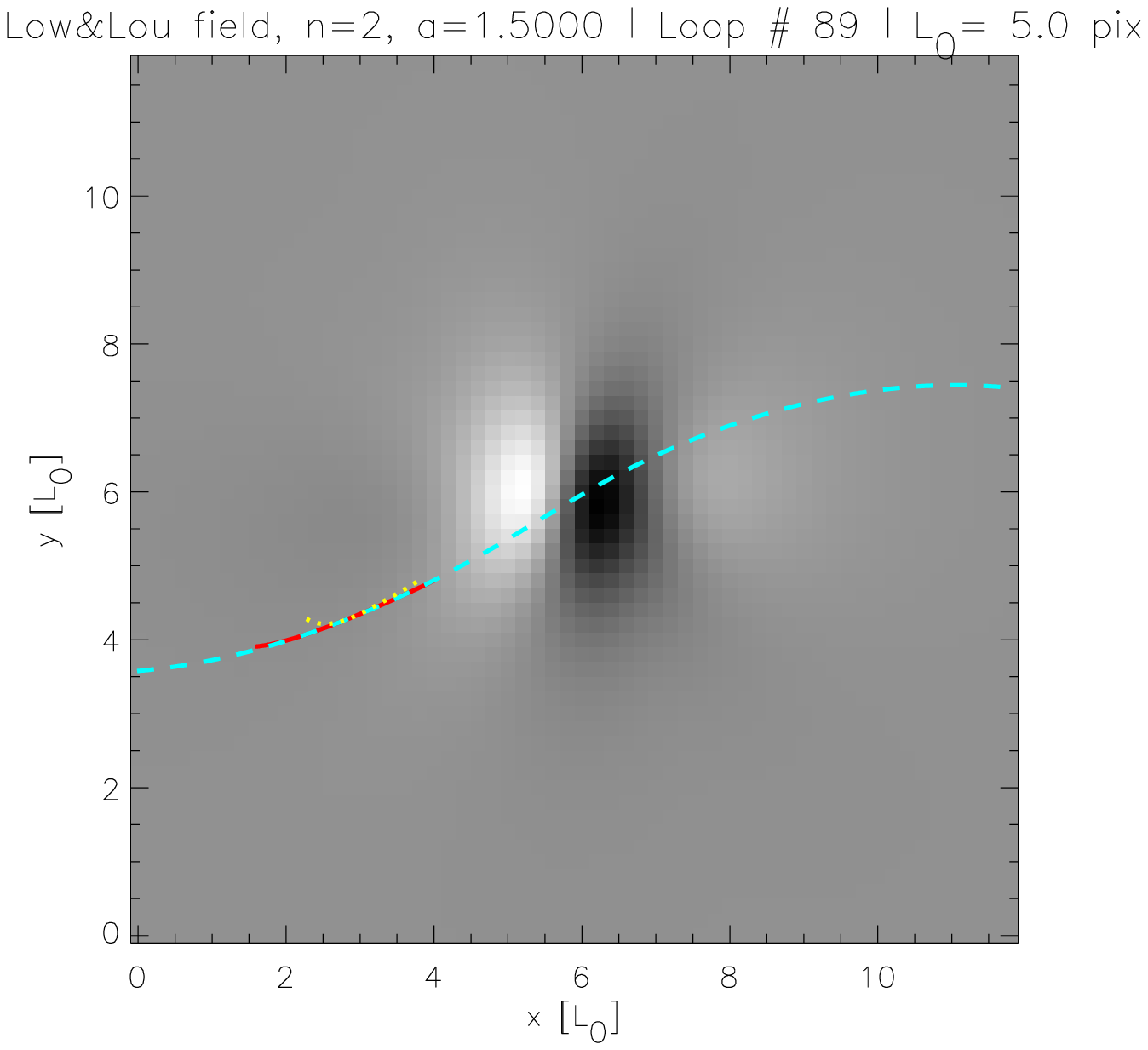} \\
\includegraphics[height=5cm]{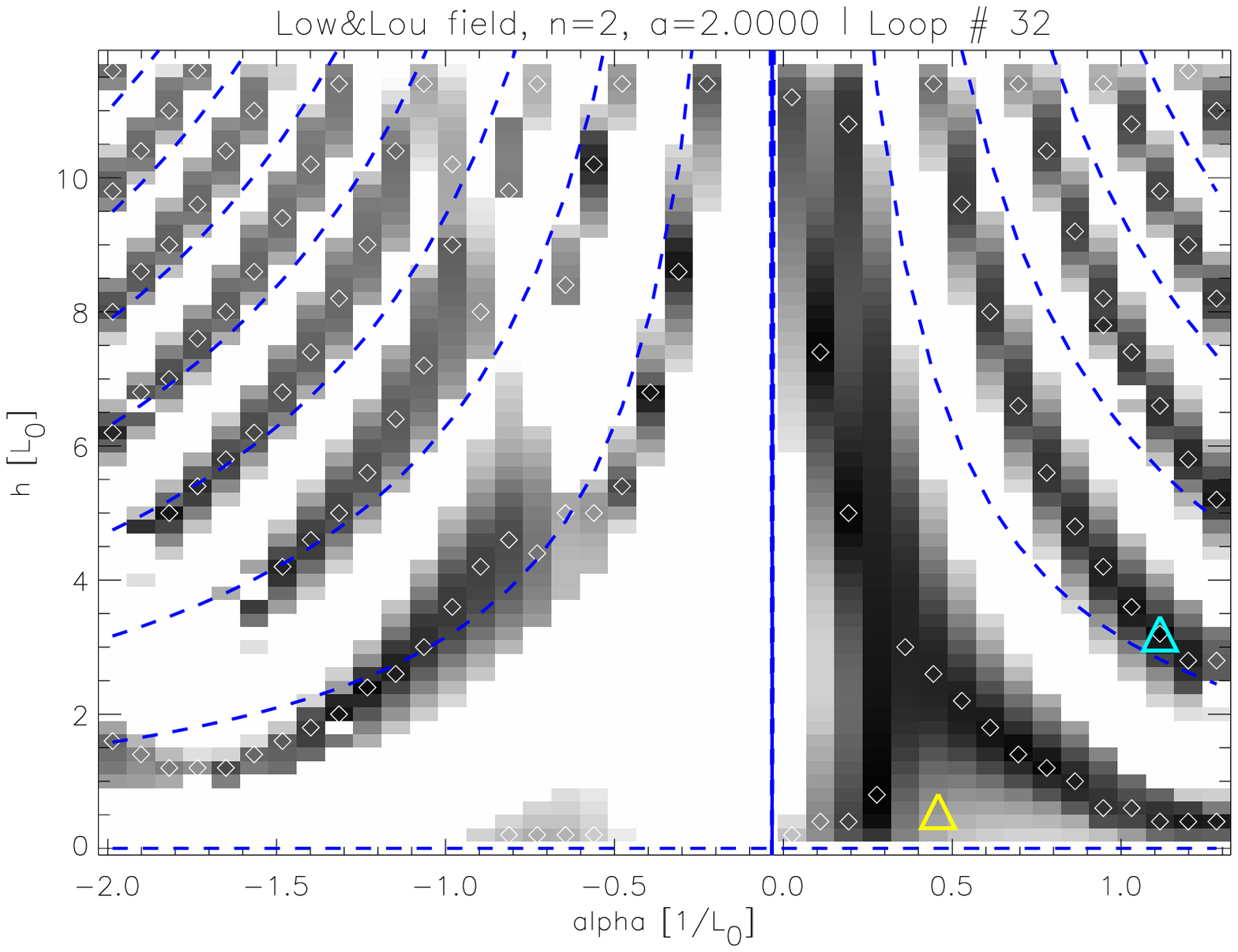} &
\includegraphics[height=5cm]{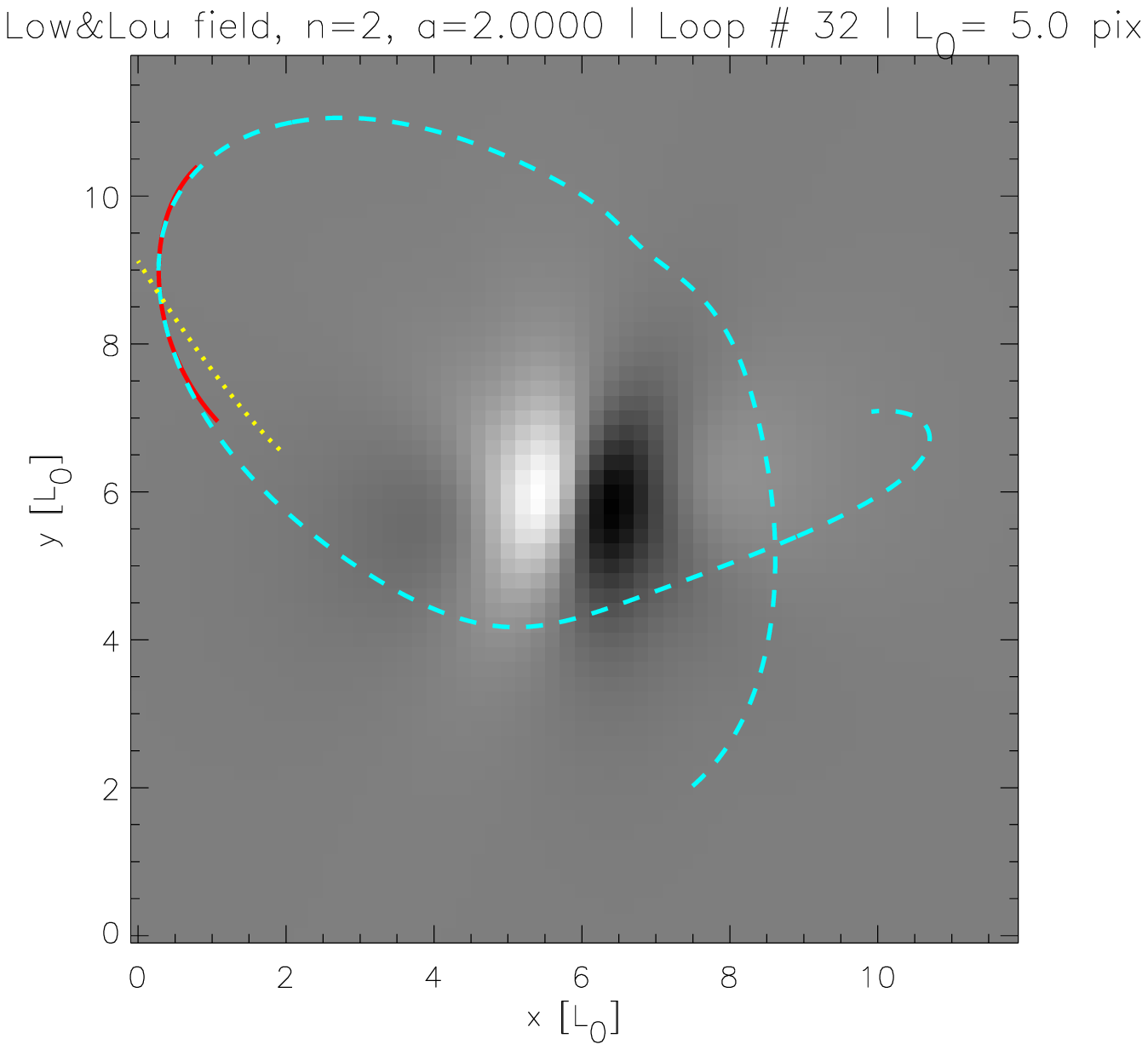} \\
 \end{tabular}
 \end{center}
 \caption{\small{The continuation of Fig.~\ref{parspaces_diff_alpha_p1} for larger $a$, the notation is the same. Note that $(\alpha_{real}, h_{real})$ is still at or near the ``non-hyperbolic'' valley, while the global minimum could be at one of the other valleys. The field line, corresponding to the global minimum, is much longer than the loop and part of this field line happened to match the loop.}}
 \label{parspaces_diff_alpha_p2}
 \end{figure}

Typical parameter spaces for those fields are shown in Fig.~\ref{parspaces_diff_alpha_p1} and Fig.~\ref{parspaces_diff_alpha_p2}. We found that for weakly twisted fields (that is, $a=0.05$, $a=0.1885$ and marginally $a=0.3$, for which $|\bvec_{recon}-\bvec_{pot}|/|\bvec_{pot}|$ is at most $0.15$, $0.3$ and $0.4$ respectively within the computational box close to the photosphere) a typical $d(\alpha, h)$ parameter space has one valley of local minima, with the same characteristics as the parameter space for $n=1.0$, described earlier. It does not seem to approach $\alpha=0$ or $h=0$ asymptotically like a hyperbola would, but rather it crosses $\alpha=0$ axis. The global minimum always lies on this valley. The point $(\alpha_{real}, h_{real})$ also lies on this valley. It seems that the more horizontal this valley is, the more offset could be the ``real'' and ``found'' points along the valley, so the more different could be $\alpha_{real}$ and $\alpha_{found}$; $h_{real}$ and $h_{found}$ always appear to be very close. 

For more strongly twisted fields ($a>0.3$) the parameter space within the range of \als and $h$ of the real field reveals more valleys, although the larger-scale behaviour seems to follow the analytic Green's function behavior shown in Fig.~\ref{fig_parspace0}. That is, $d(\alpha, h)$ seems to have valleys of local minima that look like hyperbolae and are located at or in between the hyperbolae $\alpha h=\pm\pi,\pm 2\pi, ...$. Except for $|\alpha h|<\pi$, there is one and only one valley in between every two hyperbolae $n\pi<\alpha h<(n+1)\pi$, $n=\pm 1,\pm 2, ...$. Within $|\alpha h|<\pi$ there are usually two or more valleys, and one of them is usually ``non-hyperbolic'' in the sense described above. In these more strongly twisted cases we observed that the global minimum could be in one of the ``higher'' valleys. It seems that the field line corresponding to $(\alpha, h)$ of the global minimum is much longer than the ``loop'' (line of \llf$\mbox{ }$field) and morphologically is quite different. Its smaller $d$ results from a small portion of the long line coinciding with the ``loop''. This happens especially often for loops that are ``too straight'' in some sense. This is qualitatively described in Section~\ref{sec_shape} and quantitatively described further in the text.

\begin{figure}[!hc]
 \begin{center}
 \begin{tabular}{p{5.0cm}p{5.0cm}p{5.0cm}}
  \includegraphics[height=5cm]{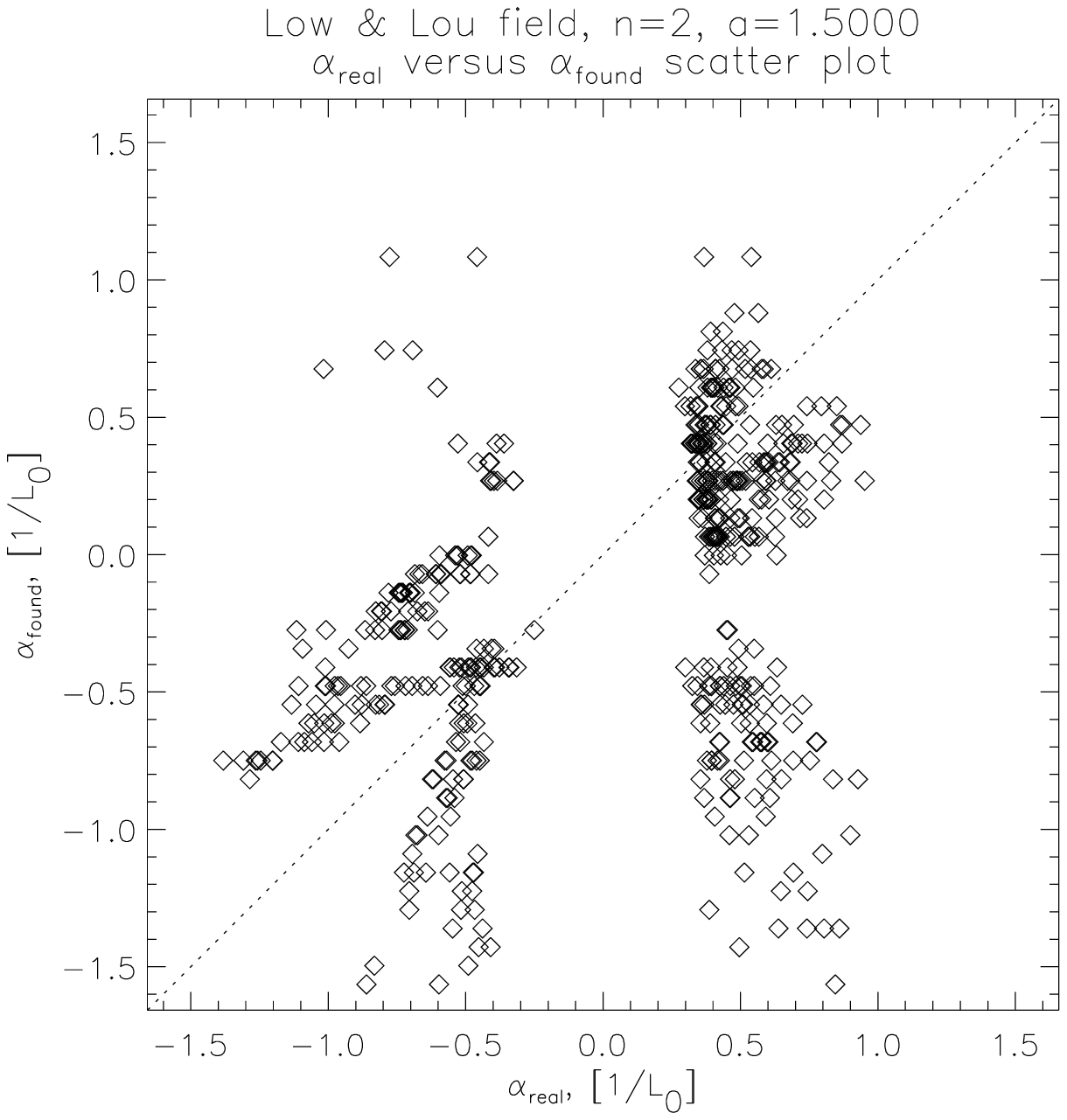} &
  \includegraphics[height=5cm]{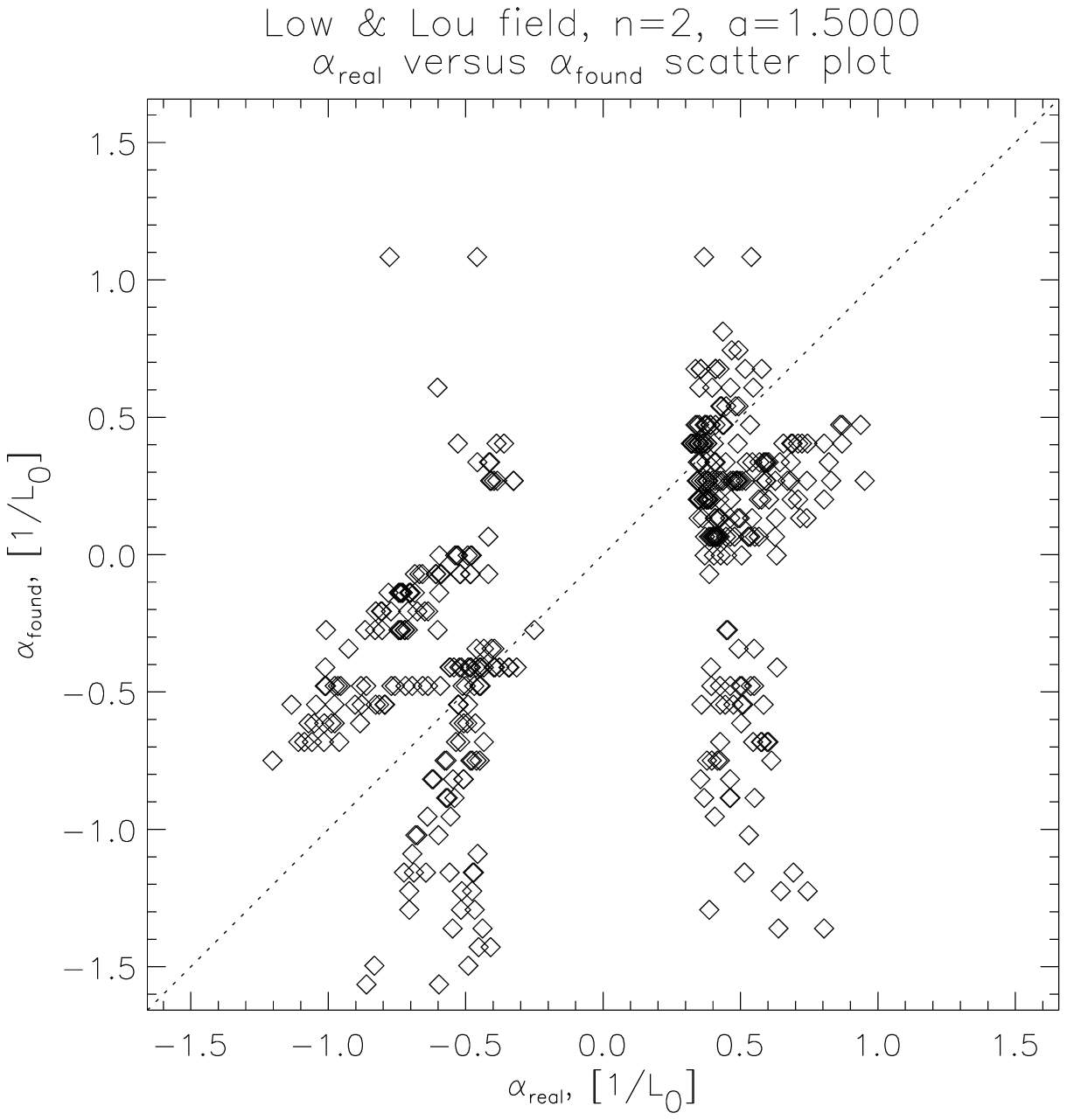} &
  \includegraphics[height=5cm]{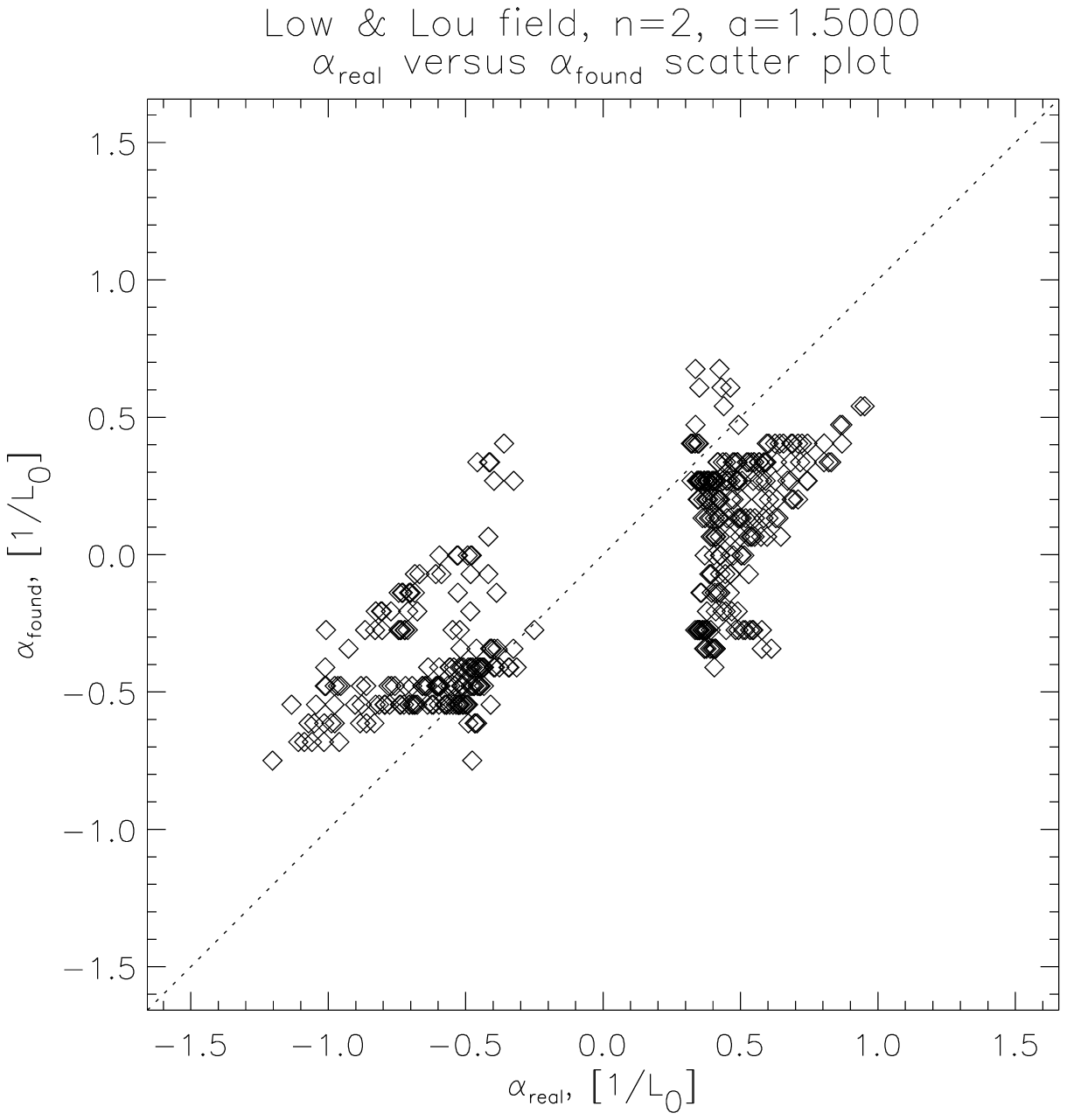} \\
 \end{tabular}
 \end{center}
 \caption{\small{\textit{(left)} -- a scatter plot of $\alpha_{real}$ vs. $\alpha_{found}$ for ``signed'' $a=1.5$, $n=2$ \llfs field. \textit{(middle)} -- same, but with ``too straight'' loops removed. \textit{(right)} -- same, but with ``too straight'' loops removed and with the minima selected only along ``non-hyperbolic'' valleys rather than the global minimum.}}
 \label{parspaces_cleaned}
 \end{figure}

After inspecting a great number of these parameter space plots we have noticed that $(\alpha_{real}, h_{real})$ still tends to correlate with the location of the ``non-hyperbolic'' valley. To prove this point, we conducted the following experiment. First we excluded loops that were ``too straight''. Second, we chose as the best fit for each of the loops a local minimum on the non-hyperbolic valley, rather than the global miminum. The results of this two-step procedure are shown in Fig.~\ref{parspaces_cleaned}. The explicit description of the procedure is below.

As a definition of ``too straight'' we adopted the ratio of sides of a box circumscribing the loop. The box is aligned with the least-square line fit to the loop, its length being the length of the loop along this line and its width being twice the maximal deviation. Based on visual examination we chose the minimum width-to-length ratio to be $0.05$ for the ``loop'' to be eligible for the analysis.

As for selection of the ``non-hyperbolic'' valley, we developed and followed an algorithm based on the shape of the parameter space. We have found that for \llfs fields this algorithm yields good results. First of all, for a given parameter space plot we identified several one-dimensional local minima for each column $(\alpha_i, h)$. Then we manually select some of those local minima that belong to only one of the valleys and find a local minimum of $d(\alpha, h)$ within this valley. 

\begin{figure}[!hc]
 \begin{center}
 \begin{tabular}{cc}
  \includegraphics[height=5cm]{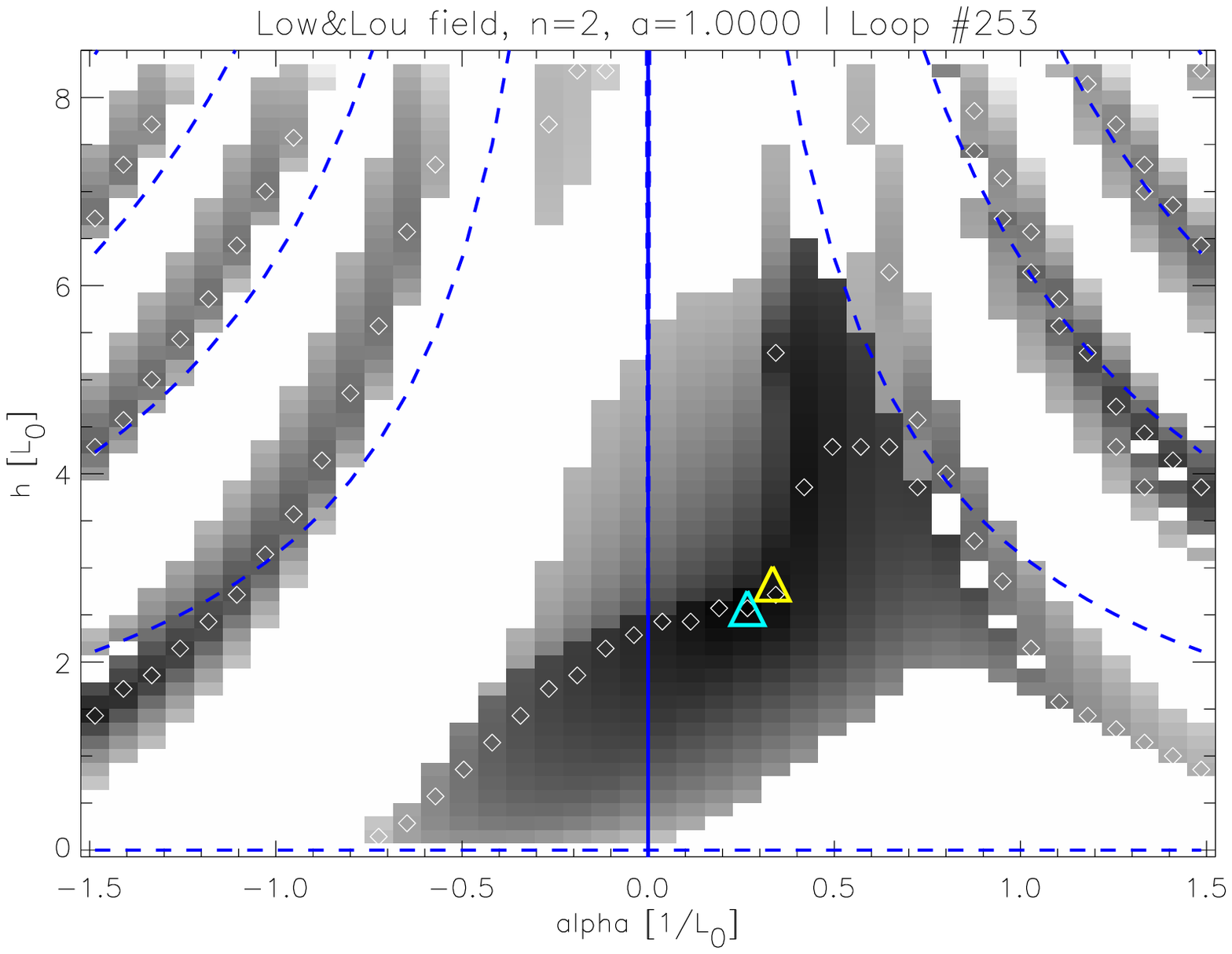} &
  \includegraphics[height=5cm]{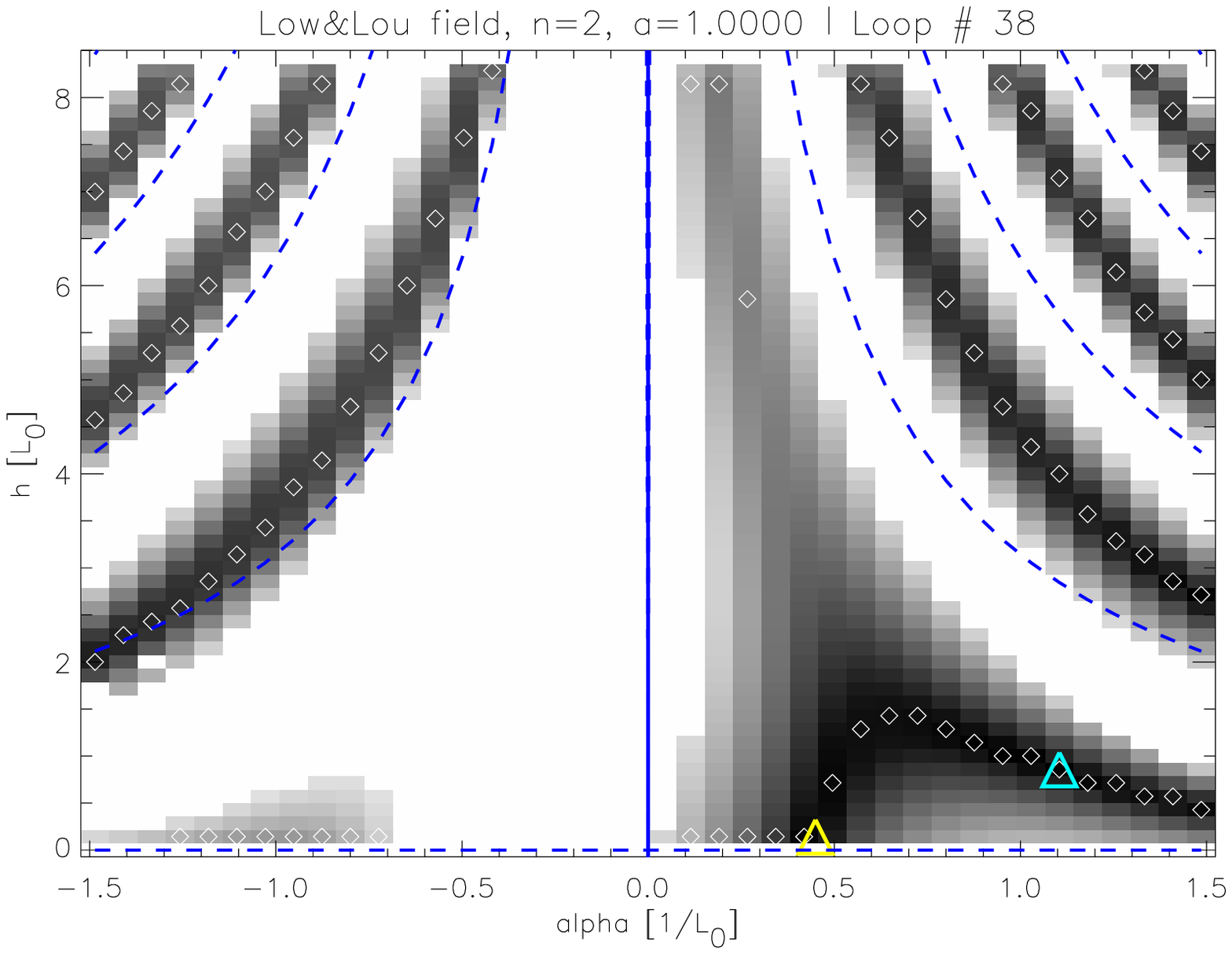} \\
  \includegraphics[height=5cm]{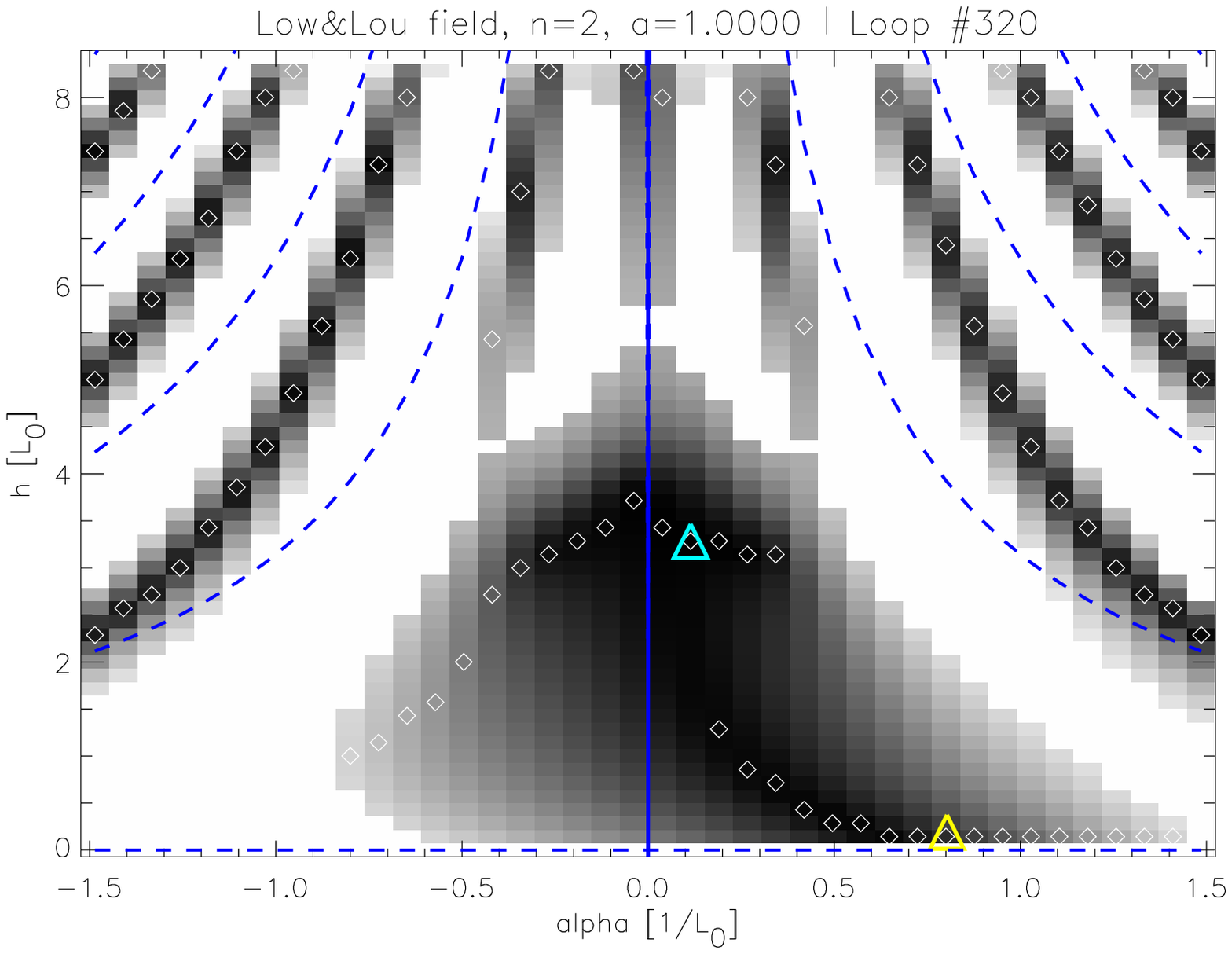} &
  \includegraphics[height=5cm]{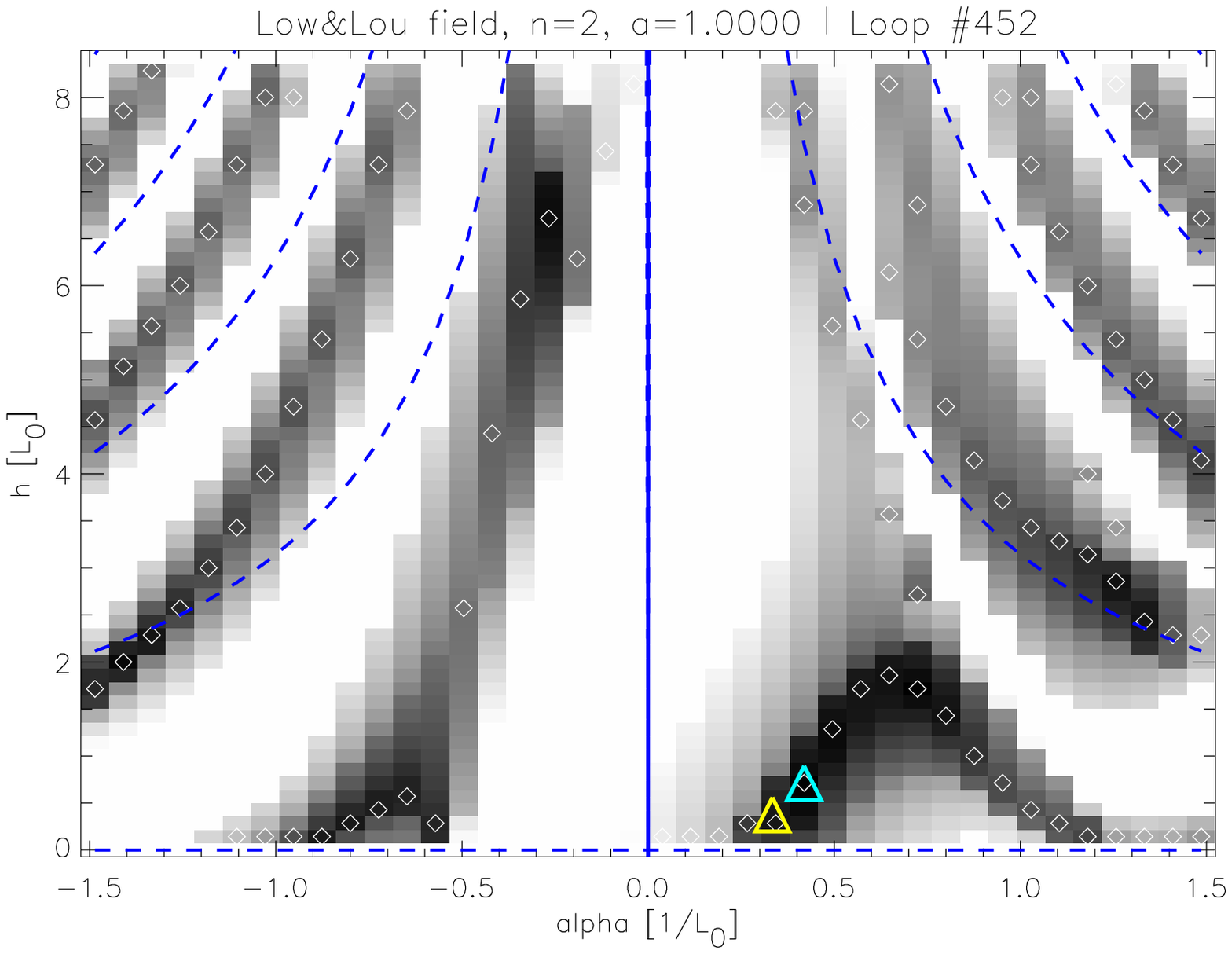} \\
  \includegraphics[height=5cm]{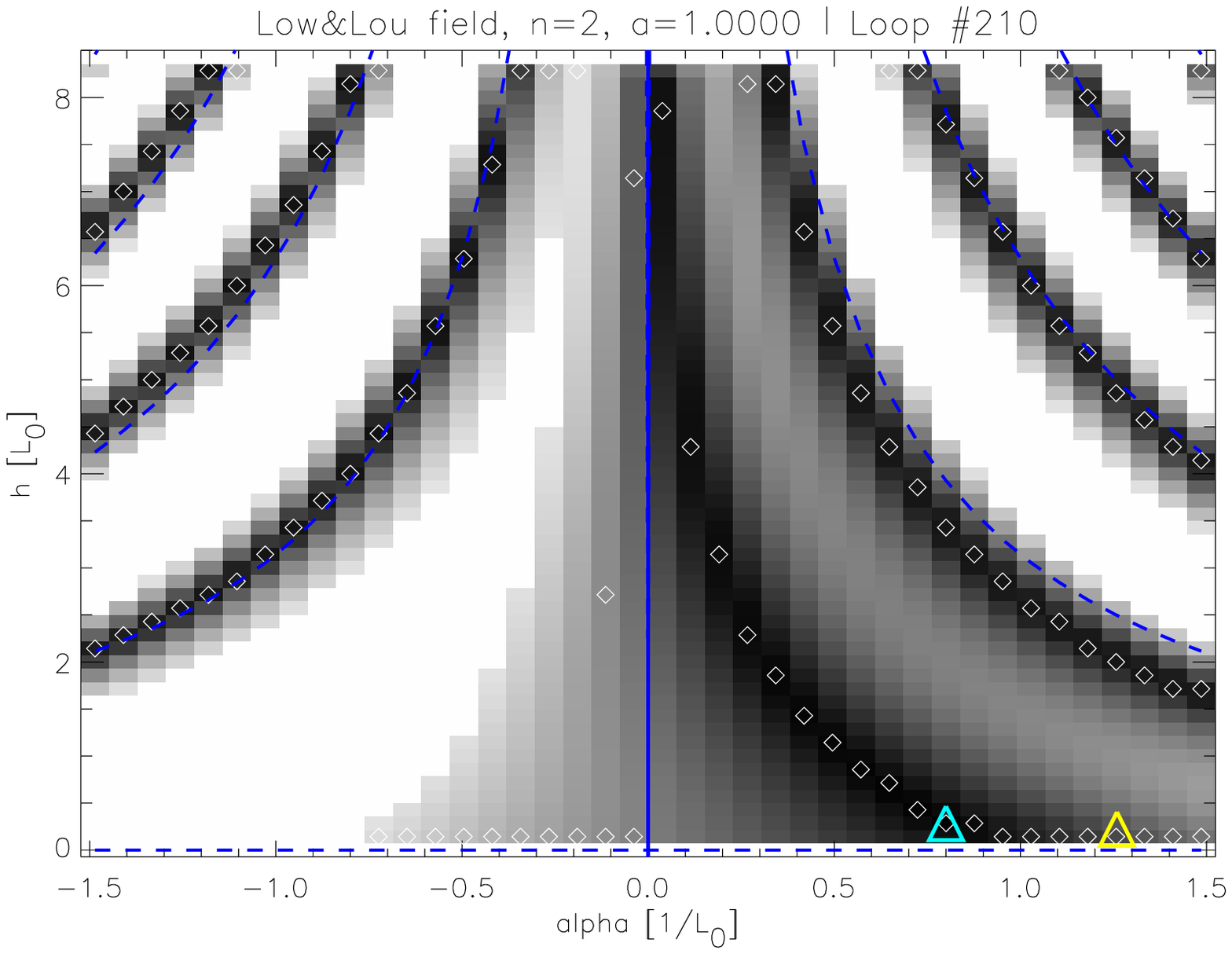} &
  \includegraphics[height=5cm]{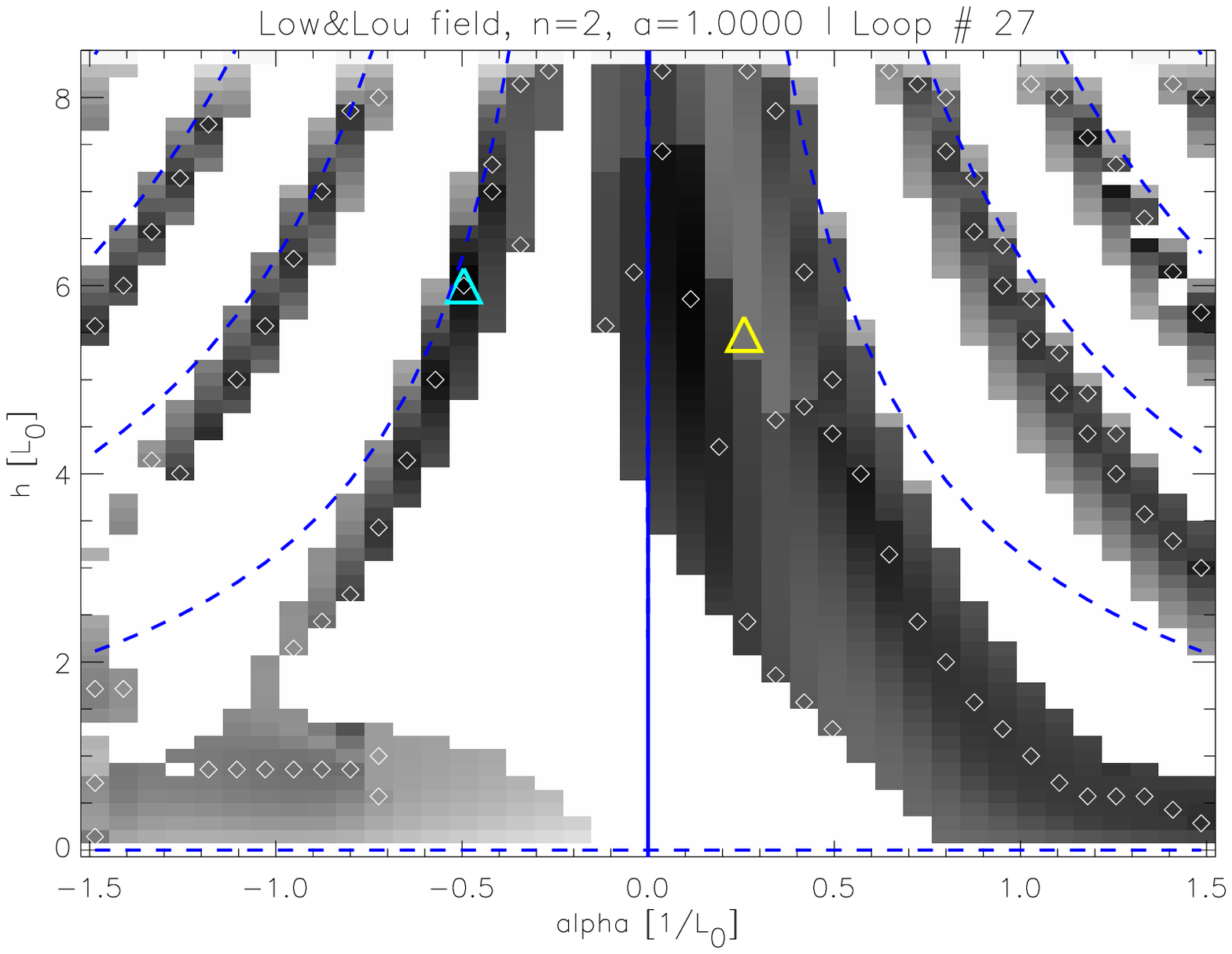} \\
 \end{tabular}
 \end{center}
 \caption{\small{The illustrations of the algorithm that helps to select local minimum on ``non-hyperbolic'' valley. The notation is the same as in Fig.~\ref{parspaces_diff_alpha_p1}. For description, please refer to the different options in the algorithm.}}
 \label{parspaces_cases}
 \end{figure}

For the selection of the valley, we followed these steps: 
\begin{enumerate}
	\item[0.]{Consider only the valleys for which $|\alpha h|<\pi$.}
	\item{Is there one ``non-hyperbolic'' valley in this region? If yes, select the local minima within it. If definitely no, proceed to the next step. If not sure, discard this loop from consideration. If there are several local minima along this valley, select the one that has the lowest $h$. Example in Fig.~\ref{parspaces_cases}, top left.}
	\item[1a.]{If the ``non-hyperbolic valley'' merges with a ``hyperbolic-like'' loop, select the local minima in the ``non-hyperbolic'' part. If unclear, discard this loop from the consideration. Example in Fig.~\ref{parspaces_cases}, top right.}
	\item{Does this ``non-hyperbolic'' valley seem to change directions, possibly crossing $\alpha=0$ more than once? If yes, select local minima on \textit{the lowest (smallest $h$) section of it}. If definitely no, proceed to the next step. If not sure, discard this loop from consideration. Example in Fig.~\ref{parspaces_cases}, middle left.}
	\item{Are there two ``non-hyperbolic'' valleys on either side of $\alpha=0$, and neither of them crosses $\alpha=0$ line? If yes, select local minima on the one that extends to a bigger range of $h$. If definitely no, proceed to the next step. If not sure, discard this loop from consideration. (We found that such parameter space plots often happens for a ``too straight'' loop, and threshold of $2\sigma/L=0.05$ seems to eliminate the majority of them. For the latter ones, $\alpha_{real}$ seems to be on the higher-extending valley.) Example in Fig.~\ref{parspaces_cases}, middle right.}
	\item{If there is no such special valley, among the ``hyperbolic'' valleys in $|\alpha h|<\pi$ there is a ``lowest-order'' one, that is, the one that has smallest $|\alpha|$ for $h\rightarrow\infty$. Is there enough of this loop presented? (I.e., that did not fall below the threshold on $|cos\theta|\leq\pi/2$, as described in Section~\ref{sec_shape}, or that did not fall below any other threshold that was used, such as difference in length being too big, or the length of the field line being significantly smaller than the length of the loop, or the amount of self-crossovers of a field line being two large -- we use the second and third thresholds, but not the first one.) If yes, select the \textit{lowest in h} local minima on this valley. If definitely no or not sure, discard this loop from consideration. Example in Fig.~\ref{parspaces_cases}, bottom left.}
	\item{Hard to classify cases: discard from consideration. Example in Fig.~\ref{parspaces_cases}, bottom right.}
\end{enumerate}

As shown in Fig.~\ref{parspaces_cleaned}, for signed field with $a=1.5$, the global minima selection does not work very well. The above-mentioned algorithm of selection of only ``non-hyperbolic'' minima works much better; it significantly improves the correlation of $\alpha_{real}$ and $\alpha_{found}$ for large $a$ (and big ranges of $\alpha$). We also tested this algorithm for when the loops belong to linear force-free fields and verified that it yields the correct results at least within the range $|\alpha|L\leq 5$, which is far beyond the range of all \llfs fields studied in this paper.

\begin{figure}[!hc]
 \begin{center}
  \includegraphics{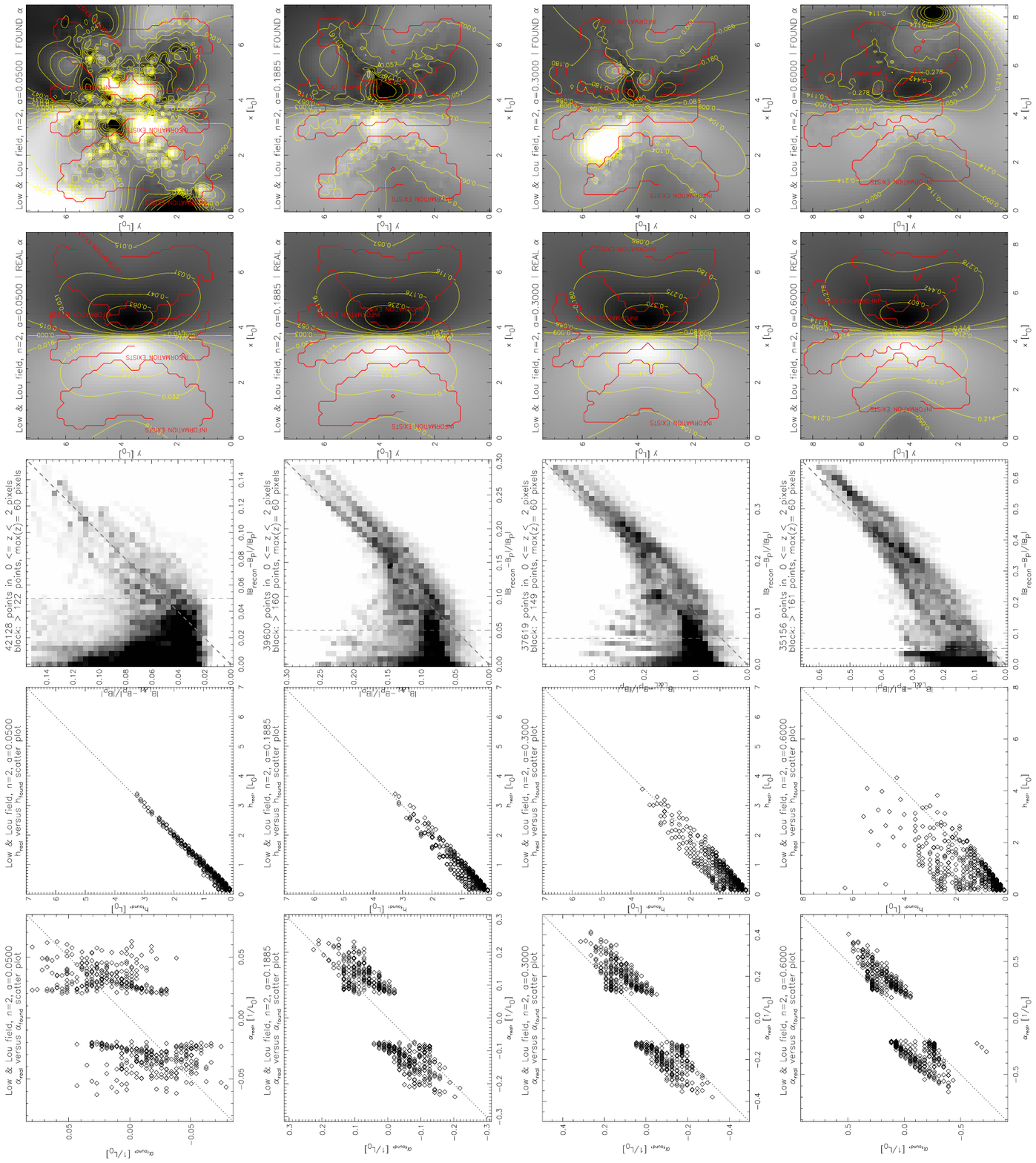}
 \end{center}
 \caption{\small{The results of the fit for ``signed'' \llfs fields: scatter plots of $\alpha_{real}$ vs. $\alpha_{found}$ (first column), $h_{real}$ vs. $h_{found}$ (second column), histogram of $|\bvec_{ff}-\bvec_{pot}|/|\bvec_{pot}|$ (third column), real and reconstructed photospheric distributions of \als (fourth and fifth columns respectively, the notation is same as in Fig.~\ref{fig_low_lou_dipole_parspace}). Note that as the range of \als increases, the reconstructed photospheric distribution of \als gets better, but the correlation of $h_{real}$ and $h_{found}$ gets worse.}}
 \label{strip_all_p1}
 \end{figure}

\begin{figure}[!hc]
 \begin{center}
  \includegraphics{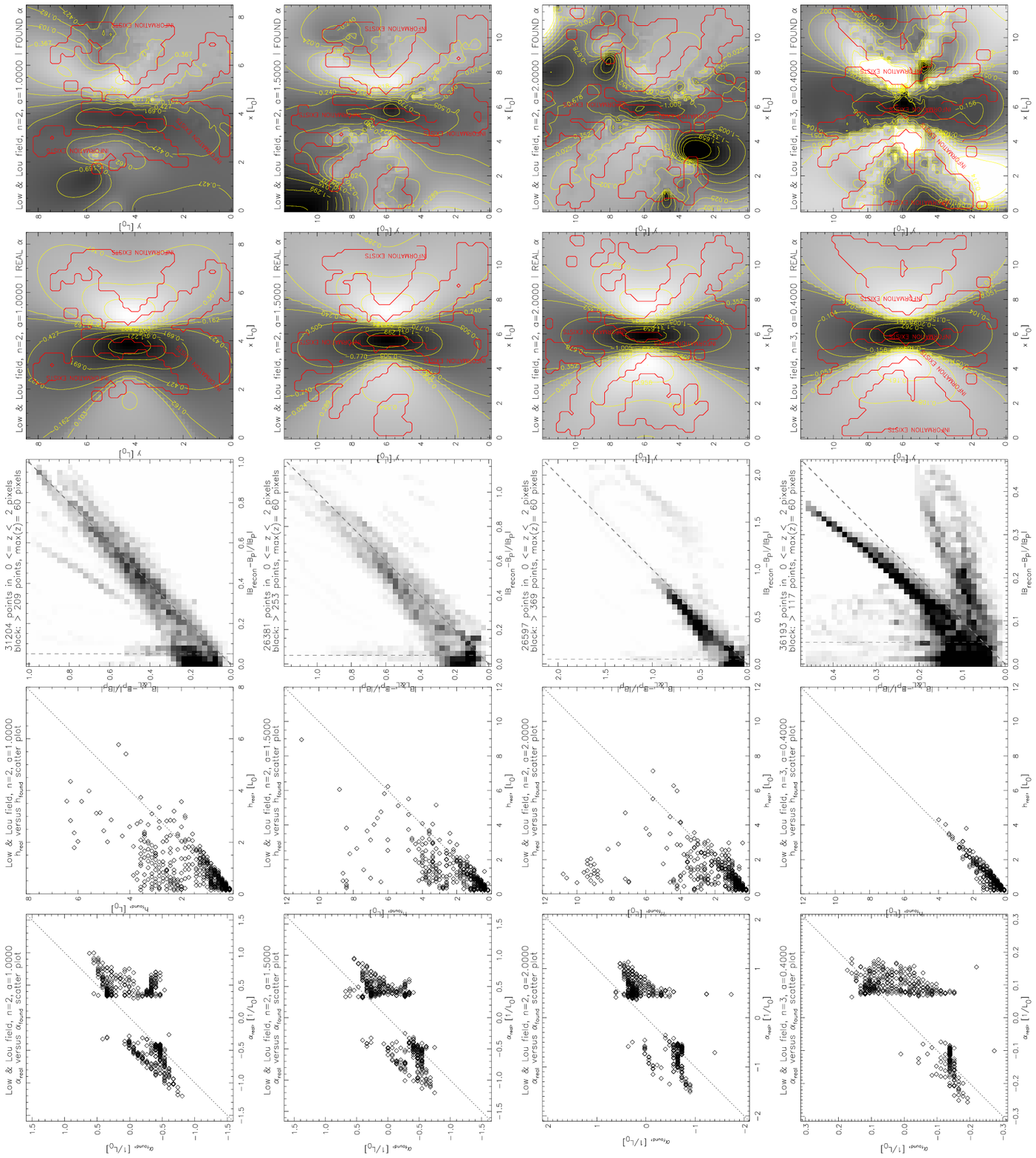}
 \end{center}
 \caption{\small{The results of the fit for ``signed'' \llfs fields. The notation is the same as in Fig.~\ref{strip_all_p1}. Note that the last row corresponds to an $n=3$ field.}}
 \label{strip_all_p2}
 \end{figure}

\begin{figure}[!hc]
 \begin{center}
  \includegraphics{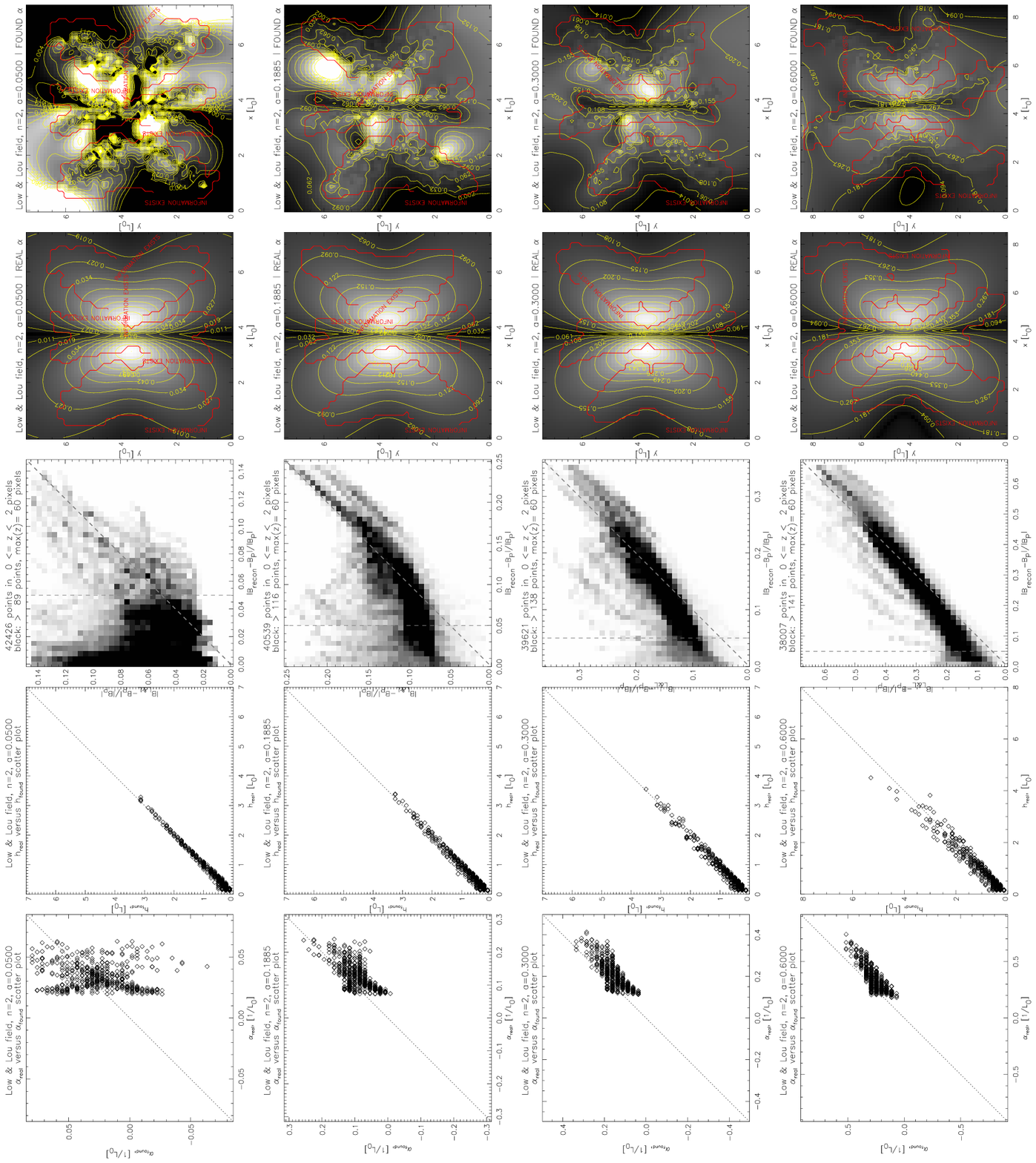}
 \end{center}
 \caption{\small{The results of the fit for ``unsigned'' \llfs fields. The notation is the same as in Fig.~\ref{strip_all_p1}.}}
 \label{strip_all_p3}
 \end{figure}

\begin{figure}[!hc]
 \begin{center}
  \includegraphics{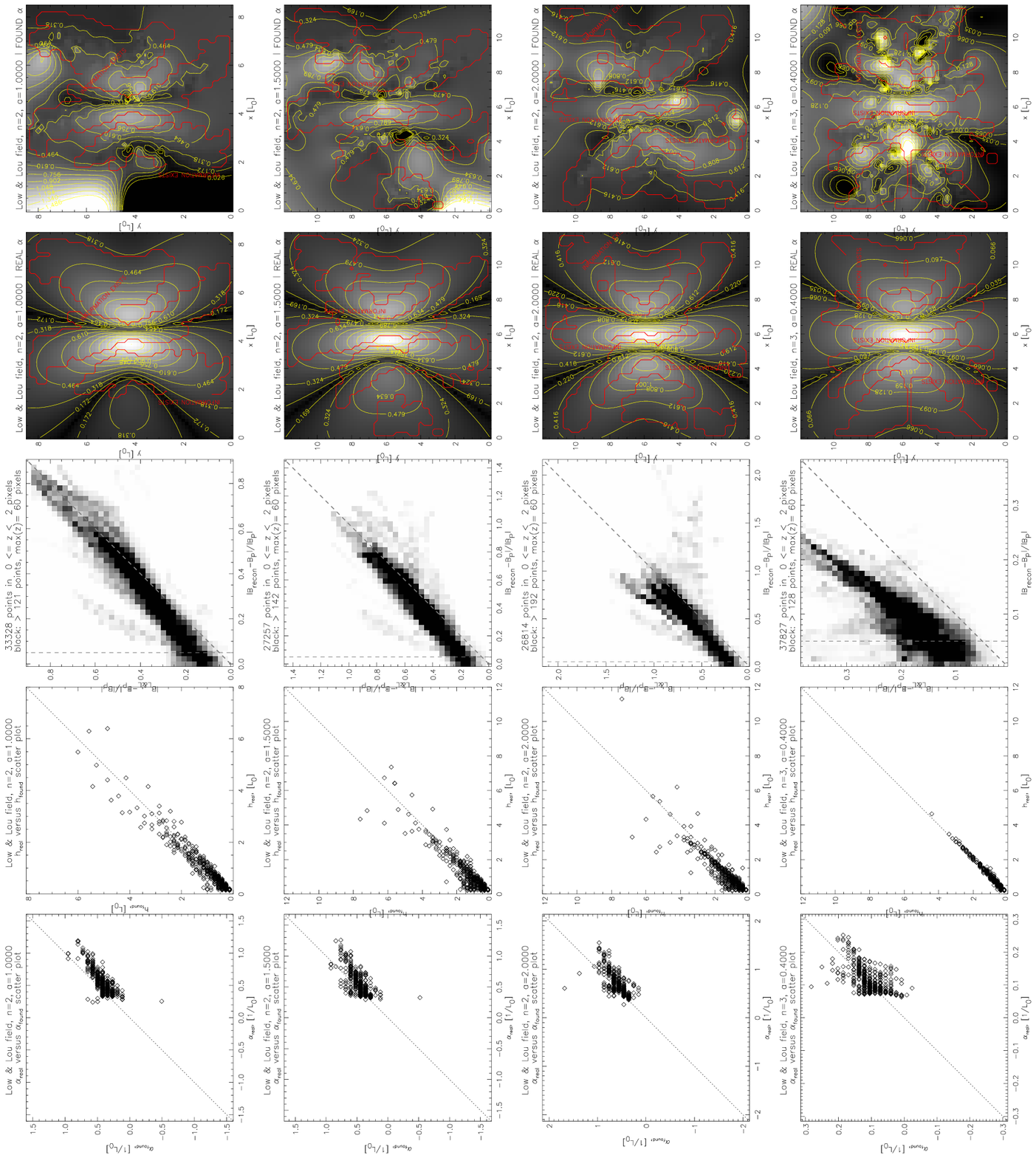}
 \end{center}
 \caption{\small{The results of the fit for ``unsigned'' \llfs fields. The notation is the same as in Fig.~\ref{strip_all_p1}. Note that the last row corresponds to an $n=3$ field.}}
 \label{strip_all_p4}
 \end{figure}

The results for all \llfs fields are summarized in Table~\ref{table_results}. The individual results are shown in Figs.~\ref{strip_all_p1}~-~\ref{strip_all_p4}. This includes scatter plots of $\alpha_{real}$ versus $\alpha_{found}$, $h_{real}$ versus $h_{found}$, the comparison of $\bvec_{ff}$ and $\bvec_{recon}$ and photospheric distributions of $\alpha_{real}$ and $\alpha_{found}$. The magnetic fields are compared in the same manner as described in Fig.~\ref{fig_low_lou_dipole_b}: $\bvec_{ff}$ and $\bvec_{recon}$ are evaluated at the photospheric level for each of the reconstructed field lines, and a two-dimensional histogram is computed. The photospheric distributions of $\alpha_{found}$ are plotted in the same color table and with the same contours as $\alpha_{real}$ and are obtained in the same manner as described in Fig.~\ref{fig_low_lou_dipole_parspace}: $\alpha_{found}(x, y, z=0)$ is collected from all reconstructed field lines; the resulting set of points is used for two-dimensional spline interpolation. 

We draw several conclusions based on the results of this analysis, First, at least for some range of \al, field lines of \llfs fields could indeed be approximated with the field lines of constant \als fields of similar \als and $h$. The reconstructed photospheric distribution of \als seems to recover the general shape of the original field. Amazingly, it is also able to recover the area of the strongest gradients of \al. Second, the height of the loops is reconstructed very well for the fields with a small range of \als and less well for the fields with a greater range of \als (see correlation coefficients and errors in Table~\ref{table_results}; note that for unsigned fields the range of \als is about half the range in signed fields). Third, this method is also capable of reconstructing the magnetic field, at least near the photosphere. 

\begin{figure}[!hc]
 \begin{center}
 \begin{tabular}{p{7.0cm}p{7.0cm}}
  \includegraphics[height=8cm]{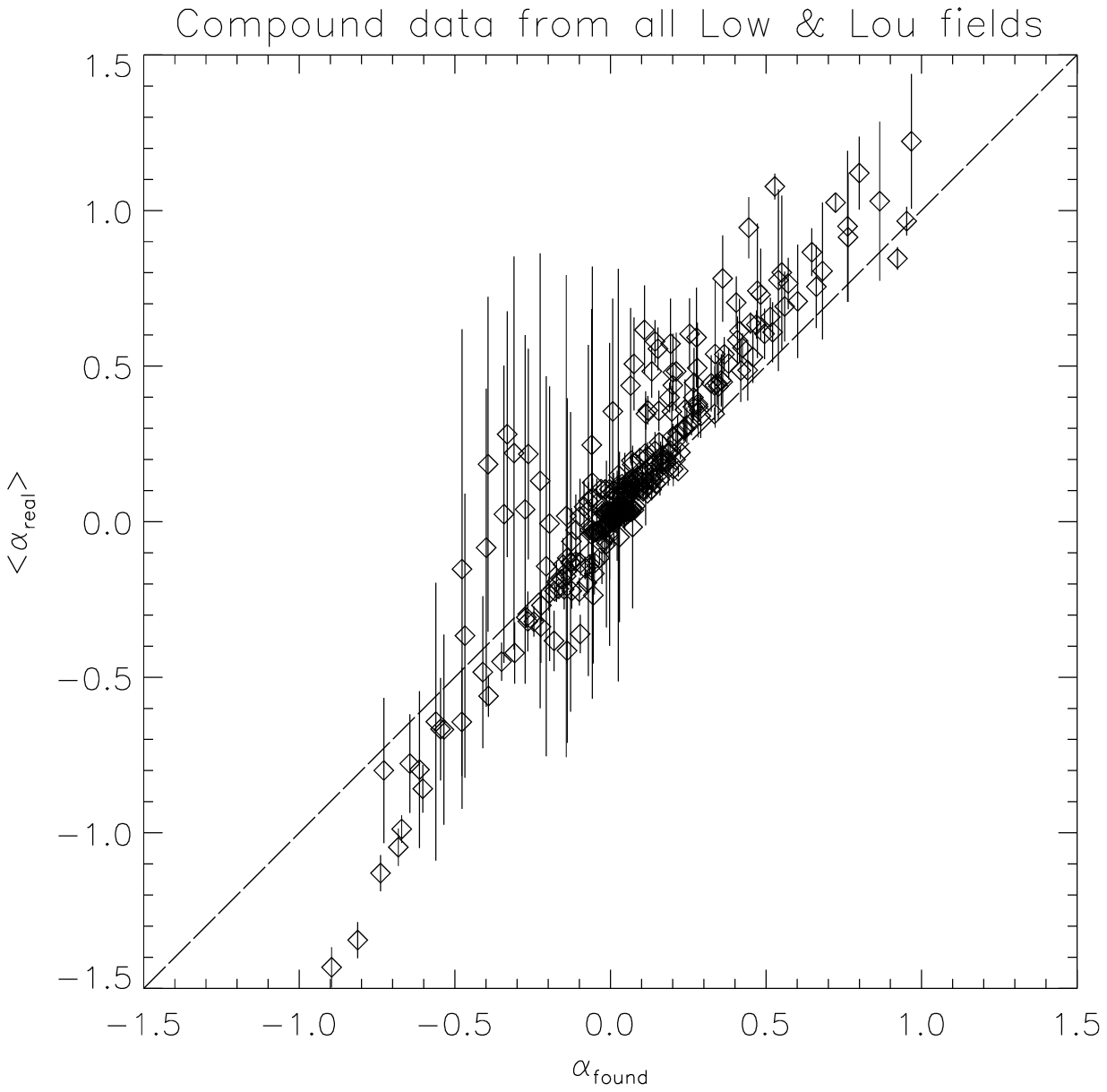} &
  \includegraphics[height=8cm]{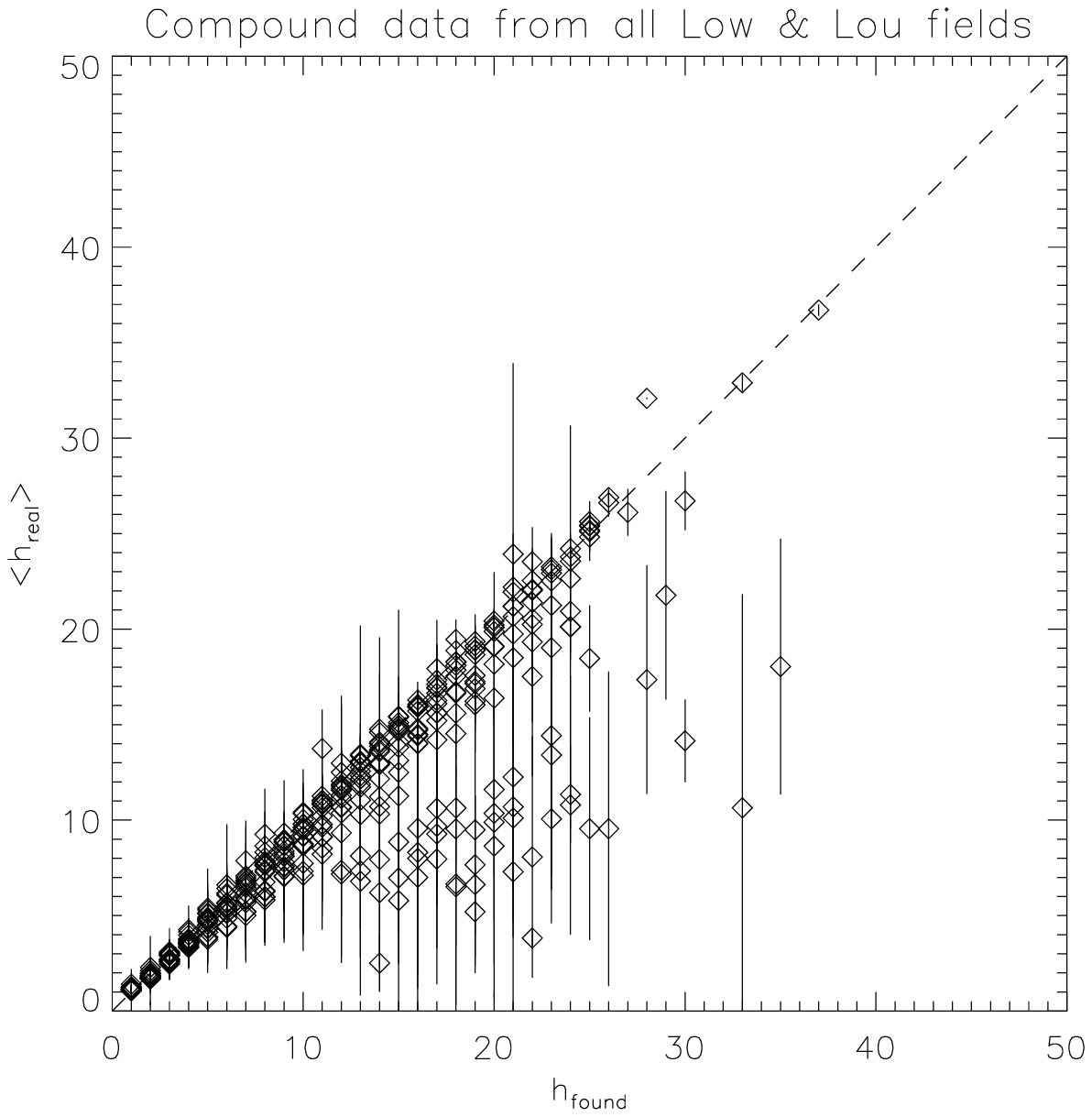} \\
 \end{tabular}
 \end{center}
 \caption{\small{\textit{(left)} -- Compound plot based on all \llfs fields measurements. For each field we went through all $\alpha_{found}$ and measured mean and standard deviation of $\alpha_{real}$. Each such measurement corresponds to a point and error bar on this plot. \textit{(right)} -- same for $h$. The units for \als are $1/L_0$ and for $h$ are $L_0$, where $L_0$ is a separation distance between two polarities.}}
 \label{all_on_one_plot}
 \end{figure}

We summarize all the results on two plots in Fig.~\ref{all_on_one_plot}. For each \llfs field, we looped through $\alpha_{found}$ and measured the mean and standard deviation of $\alpha_{real}$, and plotted $\langle\alpha_{real}\rangle\pm\sigma$ versus $\alpha_{found}$. We did the same for $h$. It seems that the method systematically underestimates \als by a small amount and it sometimes overestimates $h$ by a small amount. The least-squares line fit of the mean values, including standard deviation, gives an estimate $\langle\alpha_{real}\rangle \propto 1.23\alpha_{found}$ and $\langle h_{real}\rangle\propto 0.79h_{found}$.

% \input ../low_lou_a_0.05_n_2.0_sign/strip.tex
% \input ../low_lou_a_0.05_n_2.0_unsign/strip.tex

% \input ../low_lou_a_0.1885_l_0.3_sc_1_n_2.0_sign/strip.tex
% \input ../low_lou_a_0.1885_l_0.3_sc_1_n_2.0_unsign/strip.tex

% \input ../low_lou_a_0.3_n_2.0_sign/strip.tex
% \input ../low_lou_a_0.3_n_2.0_unsign/strip.tex

%\clearpage
% \input ../low_lou_a_0.6_n_2.0_sign/strip.tex
% \input ../low_lou_a_0.6_n_2.0_unsign/strip.tex

% \input ../low_lou_a_1.0_n_2.0_sign/strip.tex
% \input ../low_lou_a_1.0_n_2.0_unsign/strip.tex

% \input ../low_lou_a_1.5_n_2.0_sign/strip.tex
% \input ../low_lou_a_1.5_n_2.0_unsign/strip.tex

% \input ../low_lou_a_2.0_n_2.0_sign/strip.tex
% \input ../low_lou_a_2.0_n_2.0_unsign/strip.tex

% \input ../low_lou_a_0.4_n_3.0_sign/strip.tex
% \input ../low_lou_a_0.4_n_3.0_unsign/strip.tex

%\input strip_res_2.0_man_rej_lscape_table

\begin{landscape}
{\tiny
\begin{table}
	\centering
%\begin{tabular}{ccp{1.0cm}|p{0.7cm}p{1.3cm}|p{1.0cm}p{1.0cm}cp{1.7cm}|p{1.0cm}p{1.0cm}cp{1.7cm}|p{1.0cm}p{1.0cm}cp{1.7cm}}
\begin{tabular}{p{0.1cm}cp{1.0cm}|p{1.0cm}p{1.5cm}|p{0.8cm}p{0.7cm}c|p{0.8cm}p{0.7cm}c|p{0.8cm}p{0.7cm}c}
\hline
 $n$ & $a$ & sign of \als & $L_0$, pix & \als range, $1/L_0$ & 
 \multicolumn{3}{c}{\als fit results} & \multicolumn{3}{c}{$h$ fit results} & 
 \multicolumn{3}{c}{$|\bvec-\bvec_{pot}|/|\bvec|$ fit results} \\
&&&&& LAD slope & LAD mean abs. dev. & $r_s$ & LAD slope & LAD mean abs. dev. & $r_s$ & LAD slope & LAD mean abs. dev. & $r_s$ \\
\hline
1 & 0.02 & $+$ & 6.0 & 0.196 & 1.63 & 6.65 & 0.805 & 0.99 & 0.55 & 0.967 & 1.03 & 3.21 & 0.934 \\
\hline
2 & 0.05 & $+$ & 8.0 & 0.080 & 0.17 & 10.15 & 0.160 & 1.01 & 0.73 & 0.968 & 1.06 & 7.29 & 0.563 \\
2 & 0.1885 & $+$ & 8.0 & 0.302 & 0.60 & 7.55 & 0.585 & 1.01 & 0.92 & 0.966 & 0.59 & 6.99 & 0.647 \\
2 & 0.3 & $+$ & 8.0 & 0.484 & 0.86 & 6.49 & 0.743 & 0.99 & 1.13 & 0.954 & 0.64 & 5.79 & 0.812 \\
2 & 0.6 & $+$ & 7.0 & 0.871 & 1.03 & 5.63 & 0.805 & 0.93 & 1.59 & 0.944 & 0.75 & 4.01 & 0.937 \\
2 & 1.0 & $+$ & 7.0 & 1.486 & 1.19 & 5.22 & 0.799 & 0.87 & 1.86 & 0.926 & 0.82 & 3.21 & 0.943 \\
2 & 1.5 & $+$ & 5.0 & 1.564 & 1.09 & 6.36 & 0.696 & 0.90 & 2.22 & 0.825 & 0.78 & 6.74 & 0.896 \\
2 & 2.0 & $+$ & 5.0 & 1.985 & 1.19 & 5.45 & 0.789 & 0.87 & 2.11 & 0.870 & 0.79 & 9.02 & 0.809 \\
\hline
3 & 0.4 & $+$ & 5.0 & 0.315 & 0.45 & 7.50 & 0.567 & 1.06 & 0.60 & 0.965 & 1.32 & 3.37 & 0.775 \\
\hline
2 & 0.05 & $\pm$ & 8.0 & 0.159 & 0.72 & 15.00 & 0.601 & 1.00 & 0.85 & 0.964 & 0.94 & 6.60 & 0.593 \\
2 & 0.1885 & $\pm$ & 8.0 & 0.598 & 1.36 & 9.14 & 0.879 & 0.96 & 1.80 & 0.921 & 0.67 & 6.84 & 0.708 \\
2 & 0.3 & $\pm$ & 8.0 & 0.949 & 1.41 & 8.78 & 0.889 & 0.89 & 2.70 & 0.855 & 0.68 & 7.38 & 0.749 \\
2 & 0.6 & $\pm$ & 7.0 & 1.642 & 1.40 & 10.02 & 0.875 & 0.53 & 4.83 & 0.715 & 0.73 & 6.58 & 0.885 \\
2 & 1.0 & $\pm$ & 7.0 & 2.647 & 1.30 & 12.14 & 0.660 & 0.35 & 5.41 & 0.621 & 0.81 & 5.56 & 0.895 \\
2 & 1.5 & $\pm$ & 5.0 & 2.647 & 1.42 & 10.16 & 0.770 & 0.47 & 3.48 & 0.756 & 0.88 & 6.49 & 0.818 \\
2 & 2.0 & $\pm$ & 5.0 & 3.268 & 1.49 & 8.36 & 0.795 & 0.56 & 3.89 & 0.689 & 0.64 & 3.50 & 0.821 \\
\hline
3 & 0.4 & $\pm$ & 5.0 & 0.529 & 0.95 & 11.38 & 0.704 & 0.98 & 1.25 & 0.889 & 0.22 & 6.23 & 0.386 \\
\hline
\end{tabular}
\caption{A brief summary of the results, shown in Figs.~\ref{strip_all_p1}-\ref{strip_all_p4}. LAD fit and Spearman's rank order correlation $r_s$ were computed for $\alpha_{real}(\alpha_{found})$. LAD mean absolute deviation for \als is given in percent of the range of \als along the photosphere, for $h$ in percent of the maximal height, and for $|\bvec-\bvec_{pot}|/|\bvec_{pot}|$ in percent of the maximal value along all measured field lines in the same height range that is used for the histograms. All statistics for the latter one is measured for $|\bvec-\bvec_{pot}|/|\bvec_{pot}|>0.05$. The confidence of Spearman's rank-order correlation is bigger than $99.95\%$ of \als for signed $n=2$, $a=0.05$ and bigger than $99.999\%$ for all other entries.}
\label{table_results}
\end{table}}
\end{landscape}

\clearpage
\section{\al-h-Fit Applied To Solar Data}\label{sec_realdata}

%We applied the algorithm described above to real data. We chose the unnumbered active region close to the disk center on 2007-02-10T11:22 (SOL2007-02-10T11:22:L290C101 according to the Solar Object Locator\footnote{{\verb =http://www.iac.es/proyecto/iau\_divii/IAU-DivII/documents/target\_naming\_convention.html}}. We then used XRT data to obtain the two-dimensional shapes of 12 visible loops and MDI data to perform the constant-\als extrapolation.

As a further test of the applicability of the method, we applied the algorithm described above to real solar data. We chose the unnumbered active region close to disk center on 2007 February 10. This region, denoted SOL2007-02-10T11:22:L290C101 according to the Solar Object Locator\footnote{http://www.iac.es/proyecto/iau\_divii/IAU-DivII/documents/target\_naming\_convention.html}, was observed by the instruments on Hinode \citep{HinodeRef}, particularly the X-Ray Telescope (XRT, \citep{XRTRef1, XRTRef2}). As reported by \citet{McKenzie2008}, the region had a sigmoidal shape indicative of highly sheared coronal loops. Such regions are believed to comprise strongly twisted non-potential magnetic fields, and thus store significant amounts of energy, and commonly erupt in flares and/or coronal mass ejections. For this reason, coronal sigmoids present a very reasonable target structure for application of the proposed method of magnetic field modeling. The 2007 February sigmoid is particularly useful because the high-resolution observations of Hinode/XRT (1 arcsecond per pixel) and the sigmoid's location near disk center facilitate tracing of the individual loops forming the sigmoid. We utilized the XRT image from 2007 February 10, 11:22:06UT, to identify and trace 12 coronal loops. The magnetic models were generated from LOS magnetograms from the MDI instrument on SOHO \citep{MDIRef}, and then utilized for fittings to obtain the 3D shape and \als of the coronal loops.

To get the two-dimensional shape from XRT image we manually drew smooth curves (3-point cubic spline) over each of the loops. We then visually co-aligned XRT and MDI images and obtained $(x, y)$ of the loops in MDI coordinates. No de-rotation was needed since MDI and XRT data were within one minute of each other. Since the region of interest was small enough, and close to disk center, we worked in a tangent plane approximation where the photospheric plane, $z=0$, was taken to be the plane of the sky. 

To perform the fitting we extracted a region measuring $506''\times 506''$ from the full-disk MDI magnetogram ($257\times 257$ pixels). In order to save computation time, we downsized the magnetogram by a factor of two in each dimension. Then we generated constant \als fields confined to half space (using Green's function from \citet{Chiu1977}) in a $506''\times 506''\times 200''$ computational box. We generated 41 different fields with $-0.04\leq\alpha\leq0.04\mbox{ arcsec}^{-1}$.

\begin{figure}[!hc]
 \begin{center}
 \begin{tabular}{p{7cm}p{7cm}}
 \multicolumn{2}{c}{\includegraphics[width=20cm]{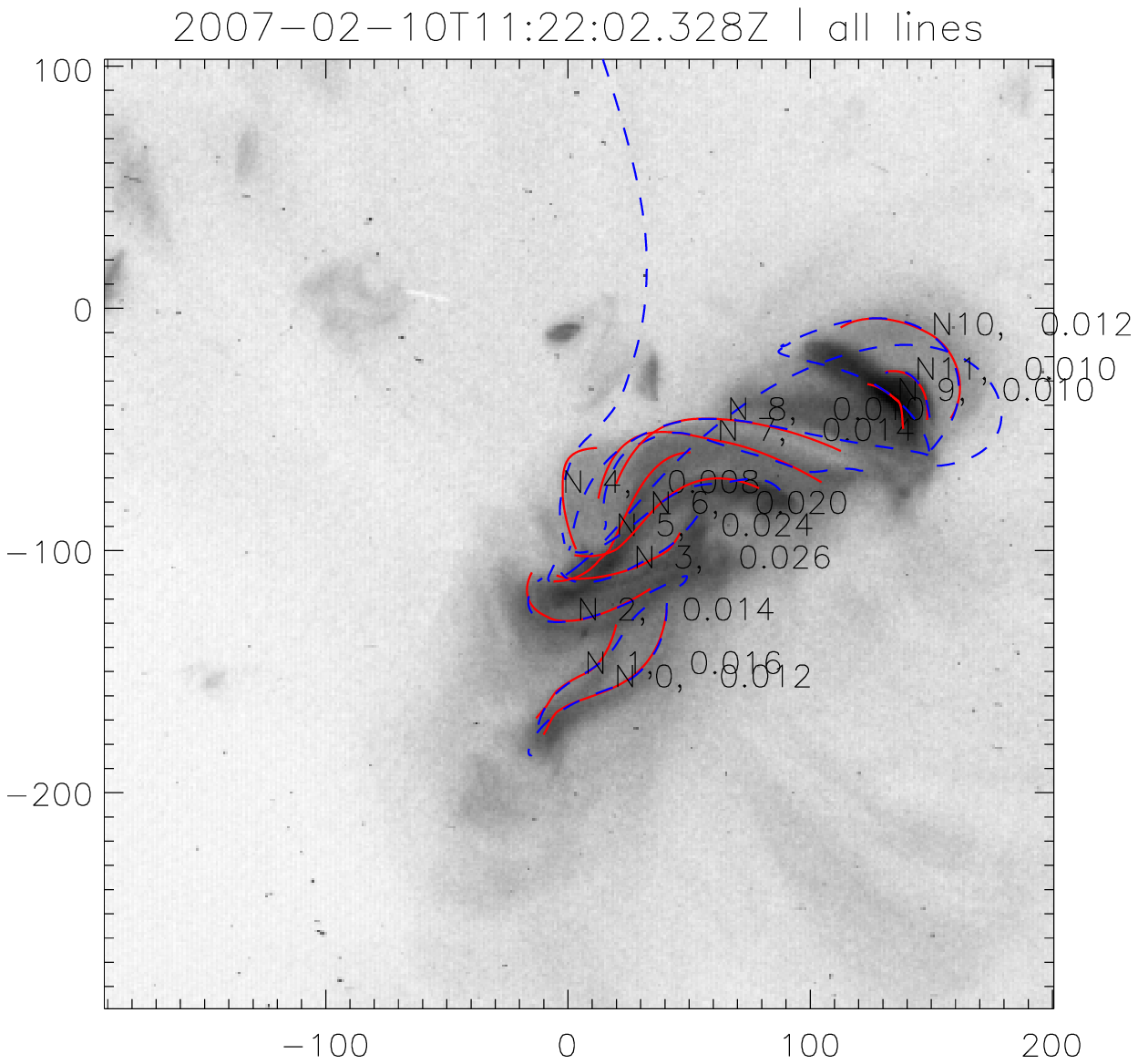}}\\
  \includegraphics[height=8cm]{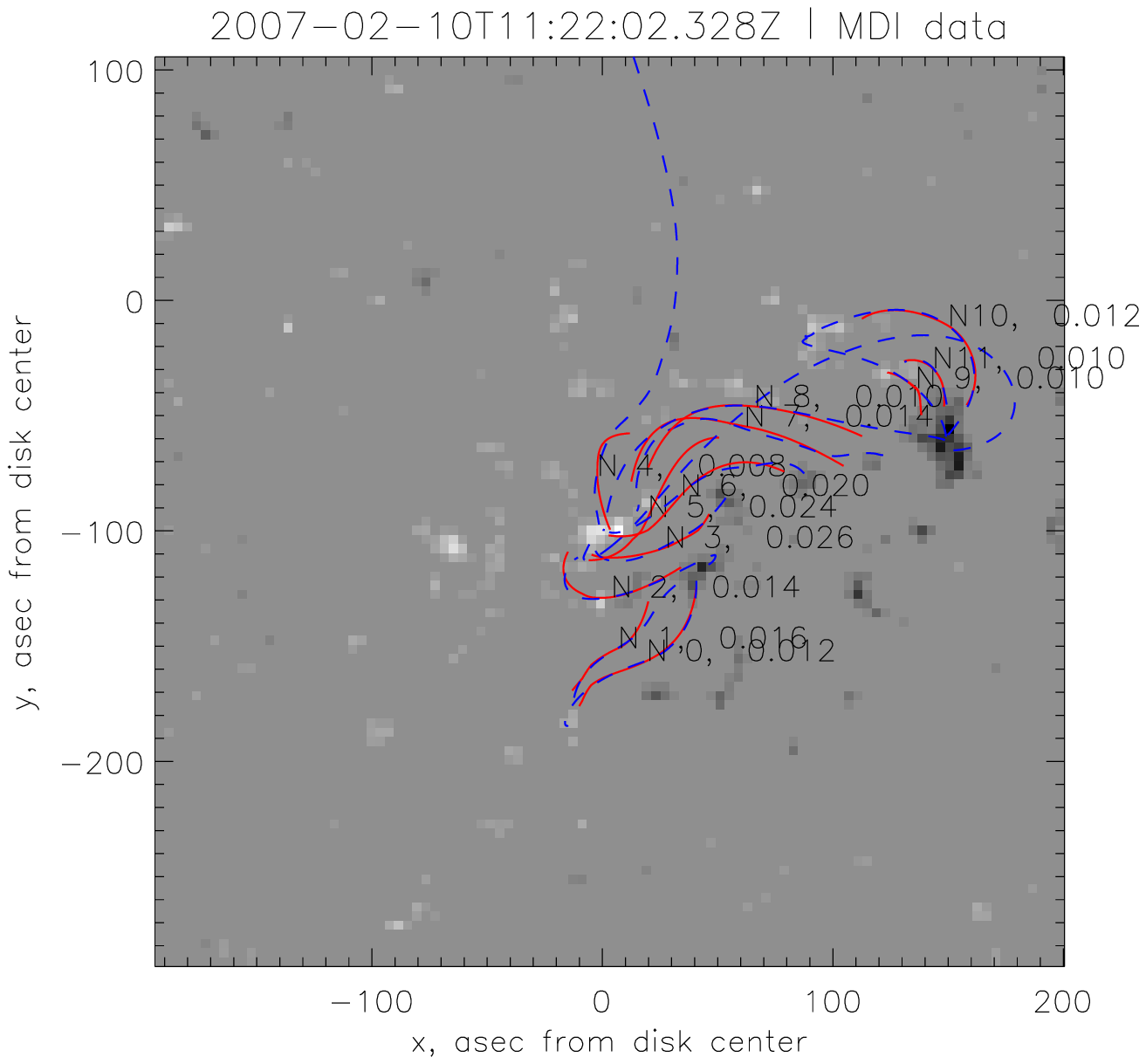} &
  \includegraphics[height=8cm]{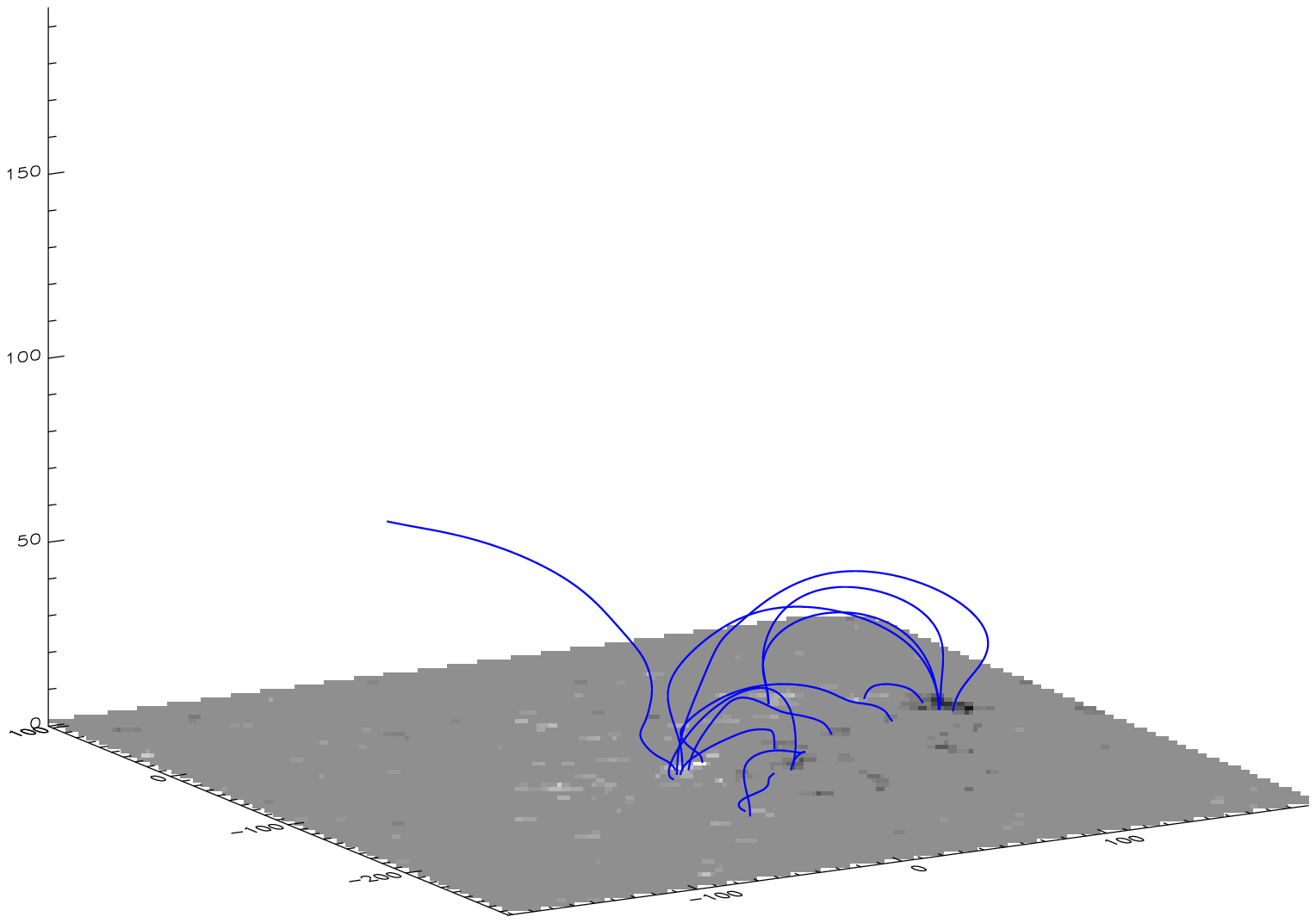} \\
 \end{tabular}
 \end{center} 
 \caption{\small{Out of 12 loops that we selected, three, namely, loops 4, 5, 9 don't seem to give a good fit to the data, and the rest seem to give fairly good fit.}}
 \label{real_data} 
 \end{figure}

\begin{figure}[!hc]
 \begin{center}
 \begin{tabular}{p{5.5cm}p{5.5cm}p{5.5cm}}
  \includegraphics[height=4.5cm]{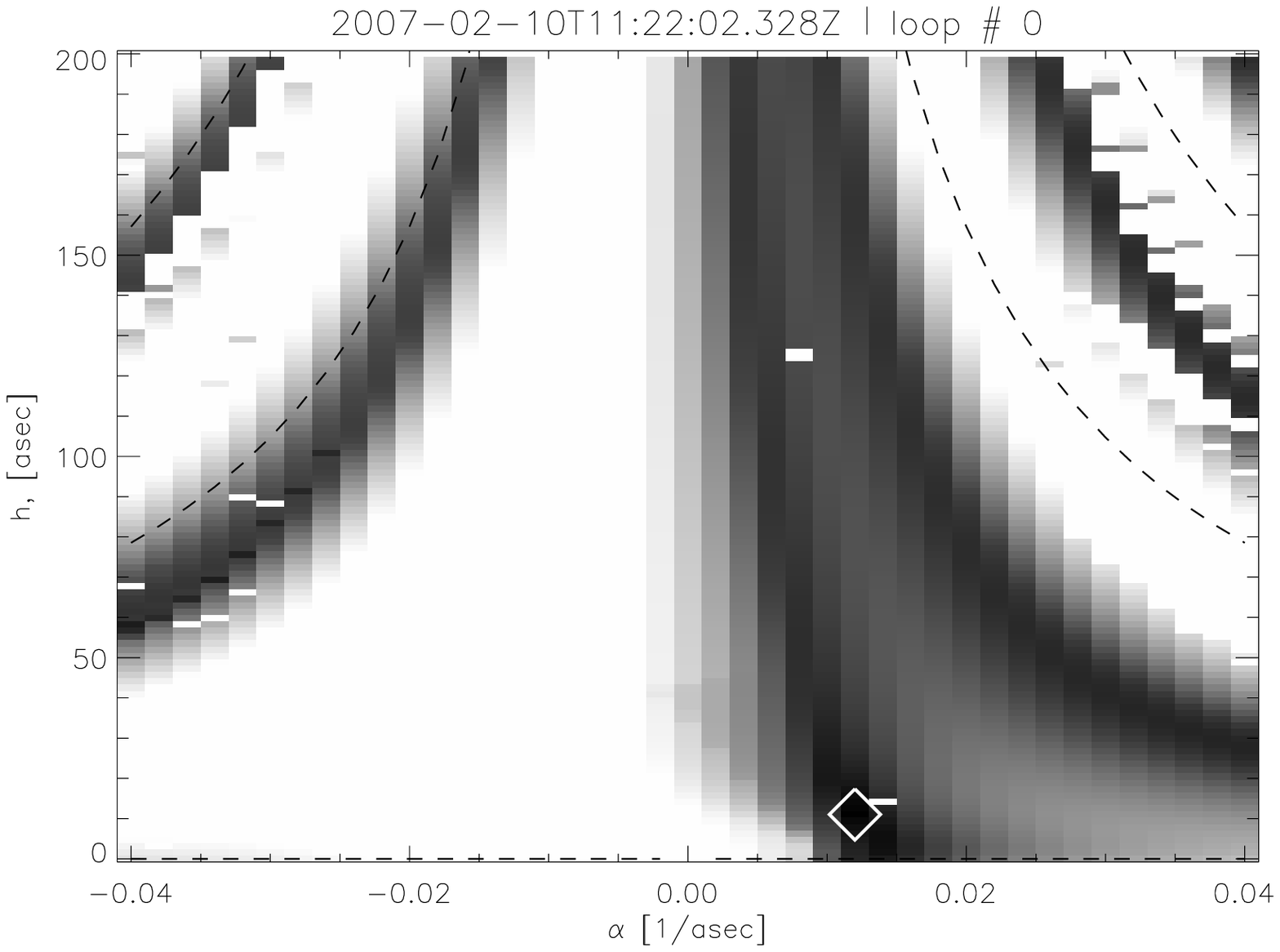} &
  \includegraphics[height=4.5cm]{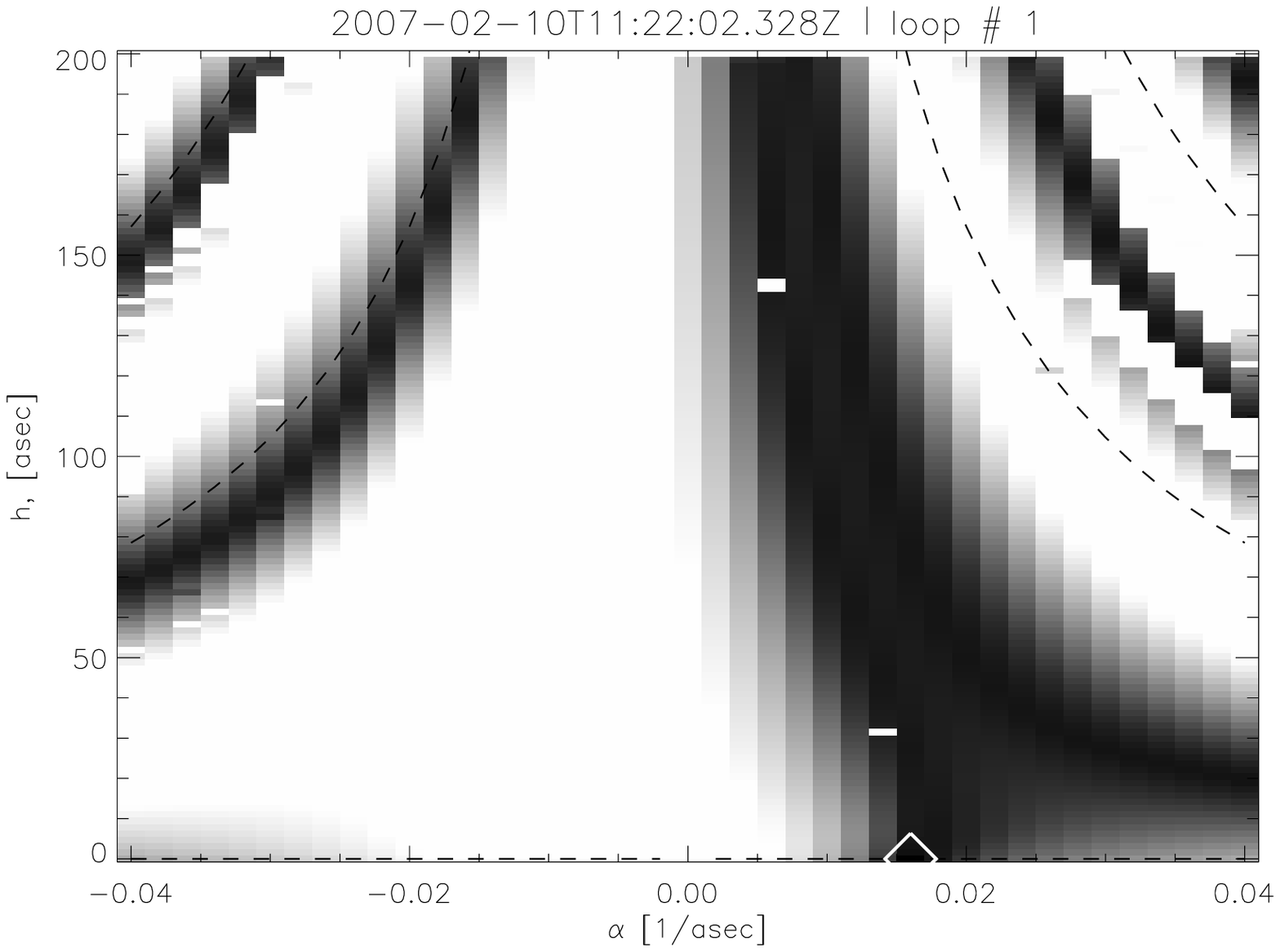} &
  \includegraphics[height=4.5cm]{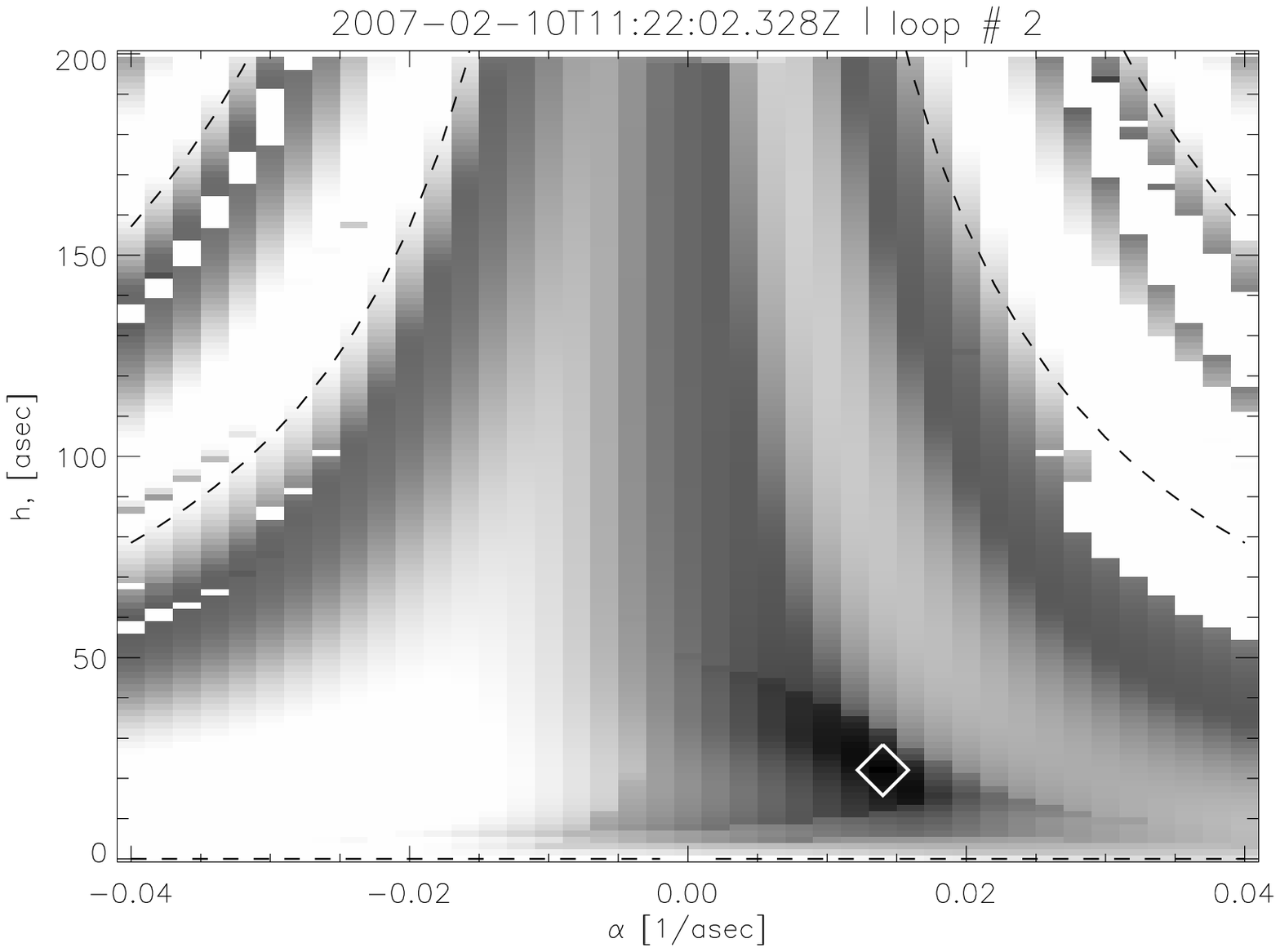} \\
  \includegraphics[height=4.5cm]{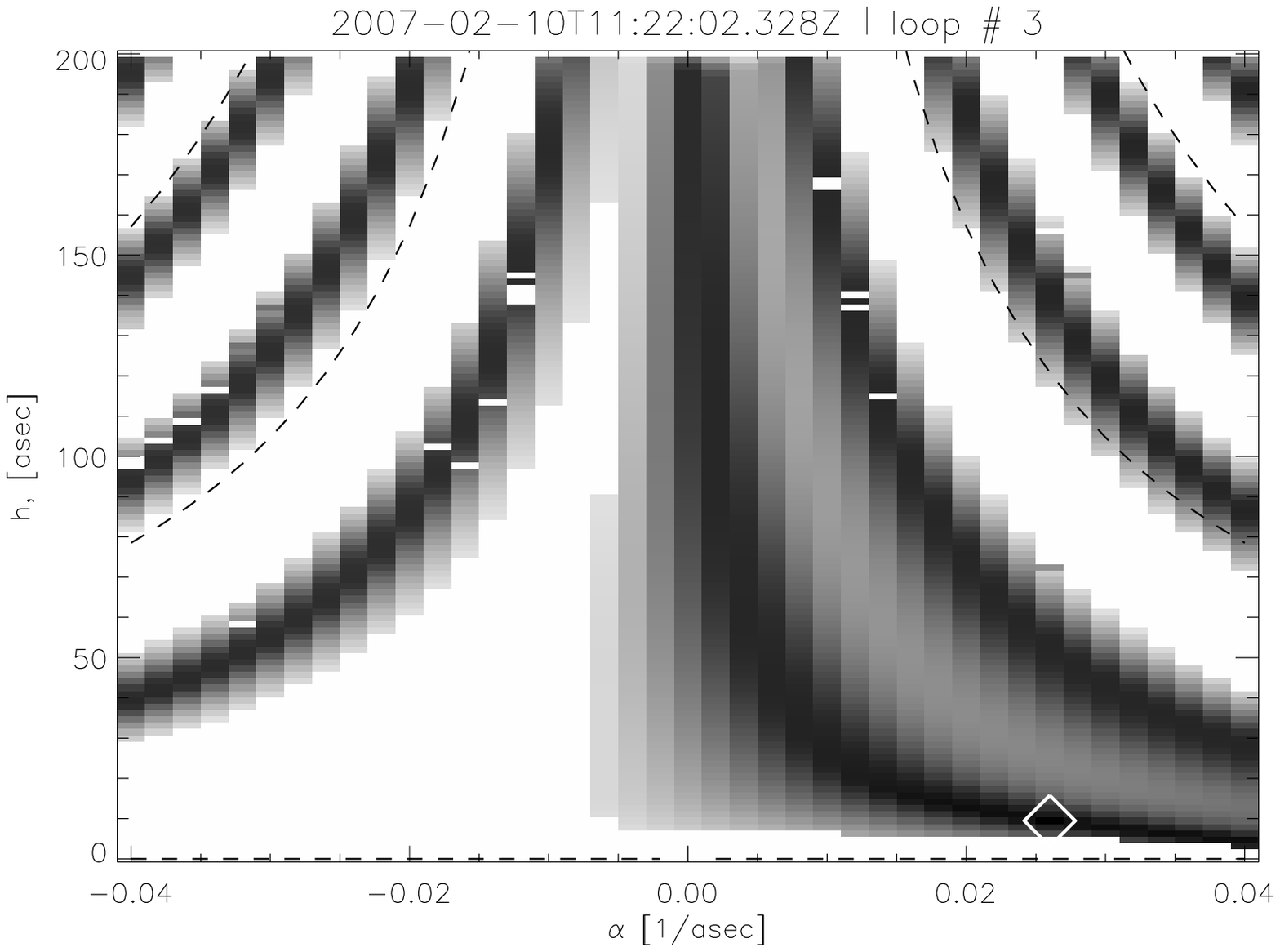} &
  \includegraphics[height=4.5cm]{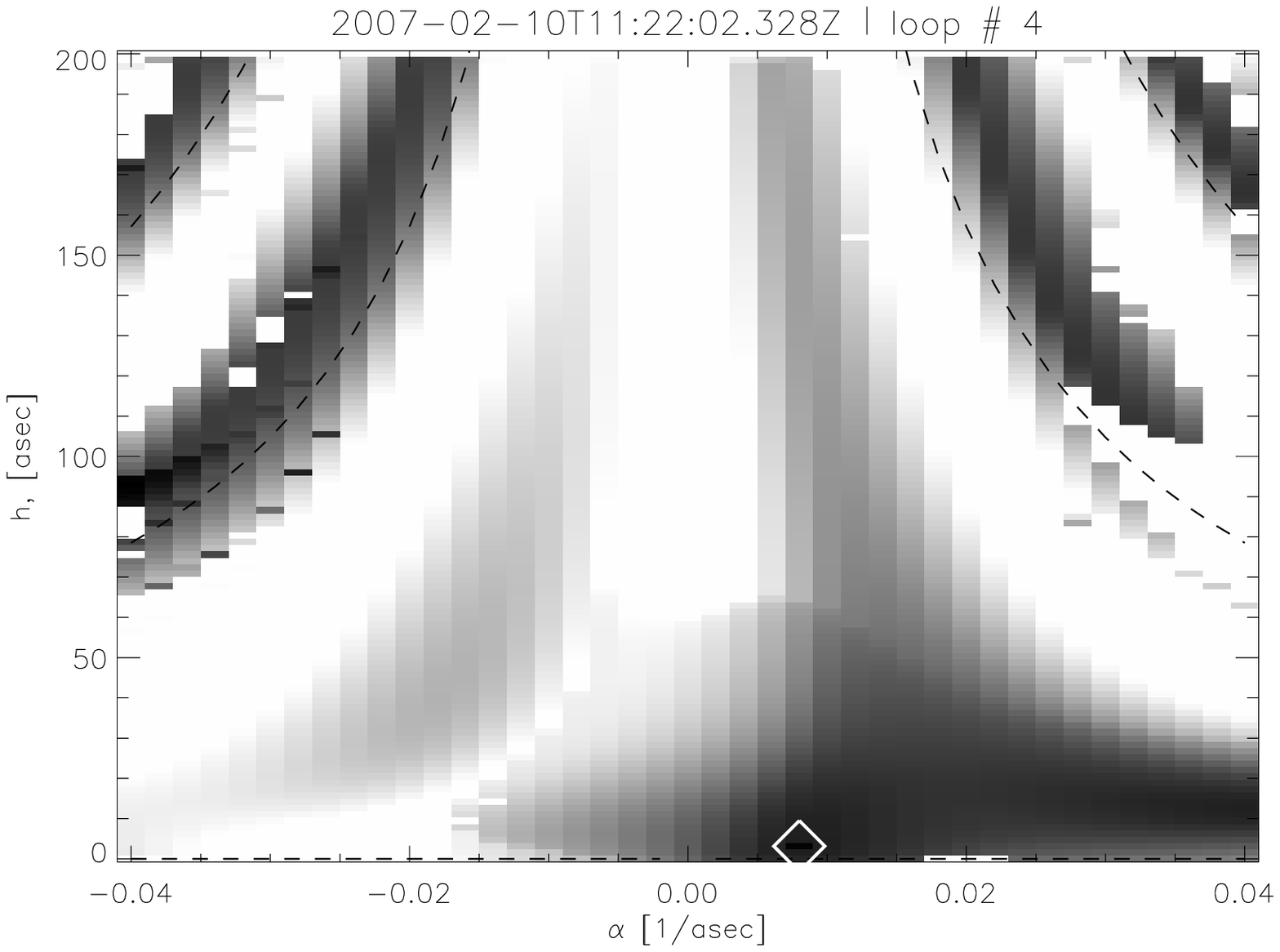} &
  \includegraphics[height=4.5cm]{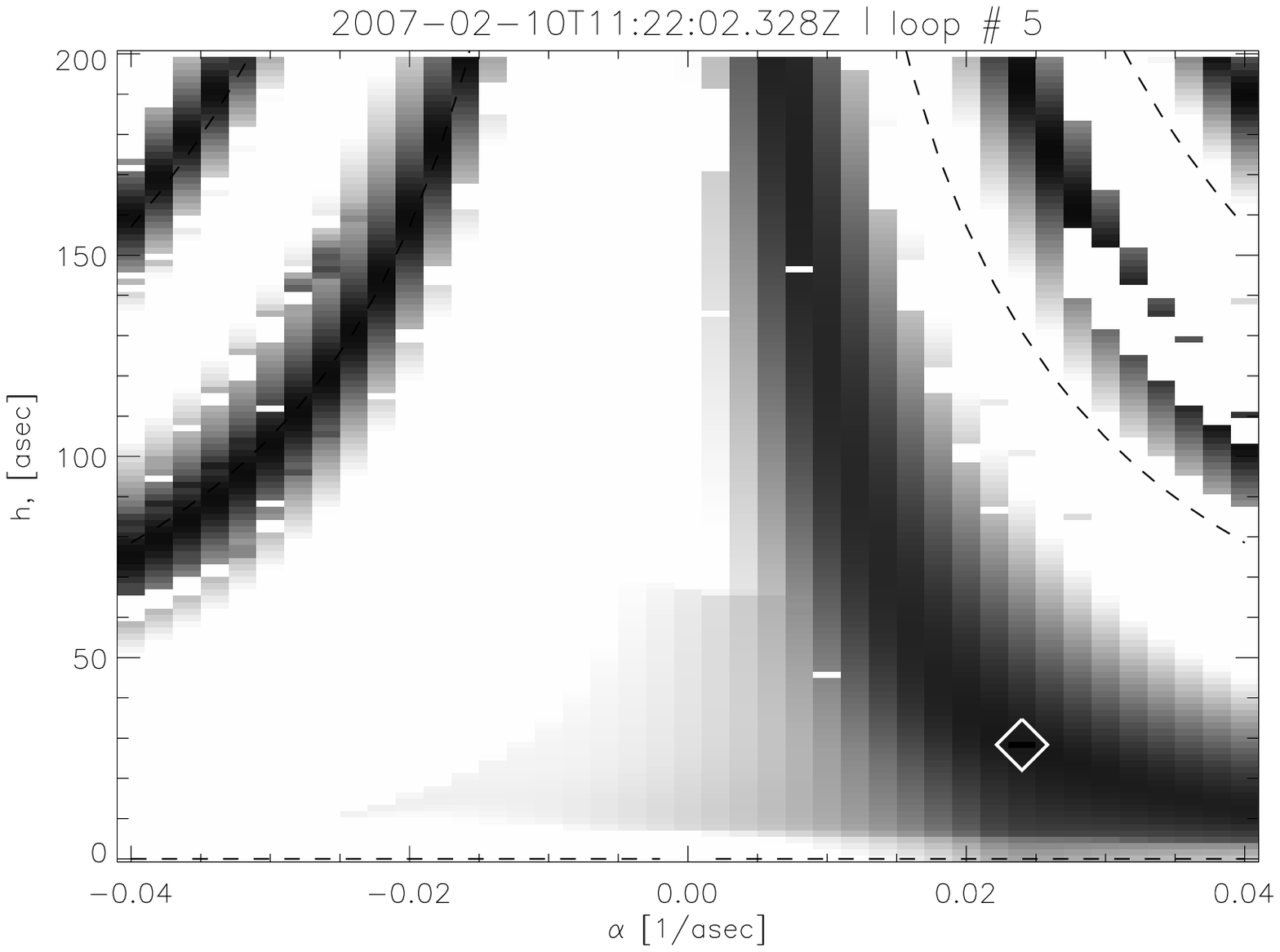} \\
  \includegraphics[height=4.5cm]{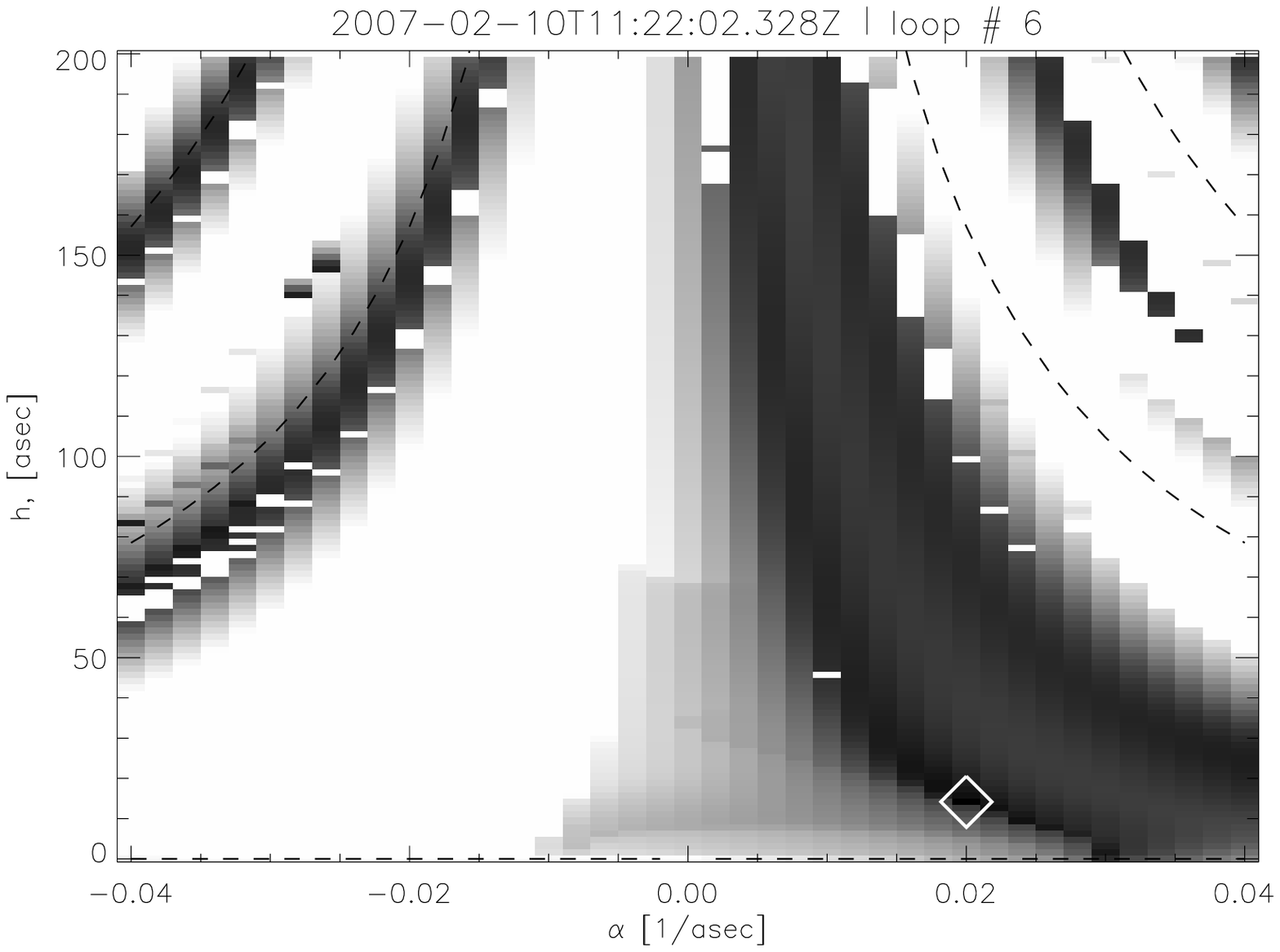} &
  \includegraphics[height=4.5cm]{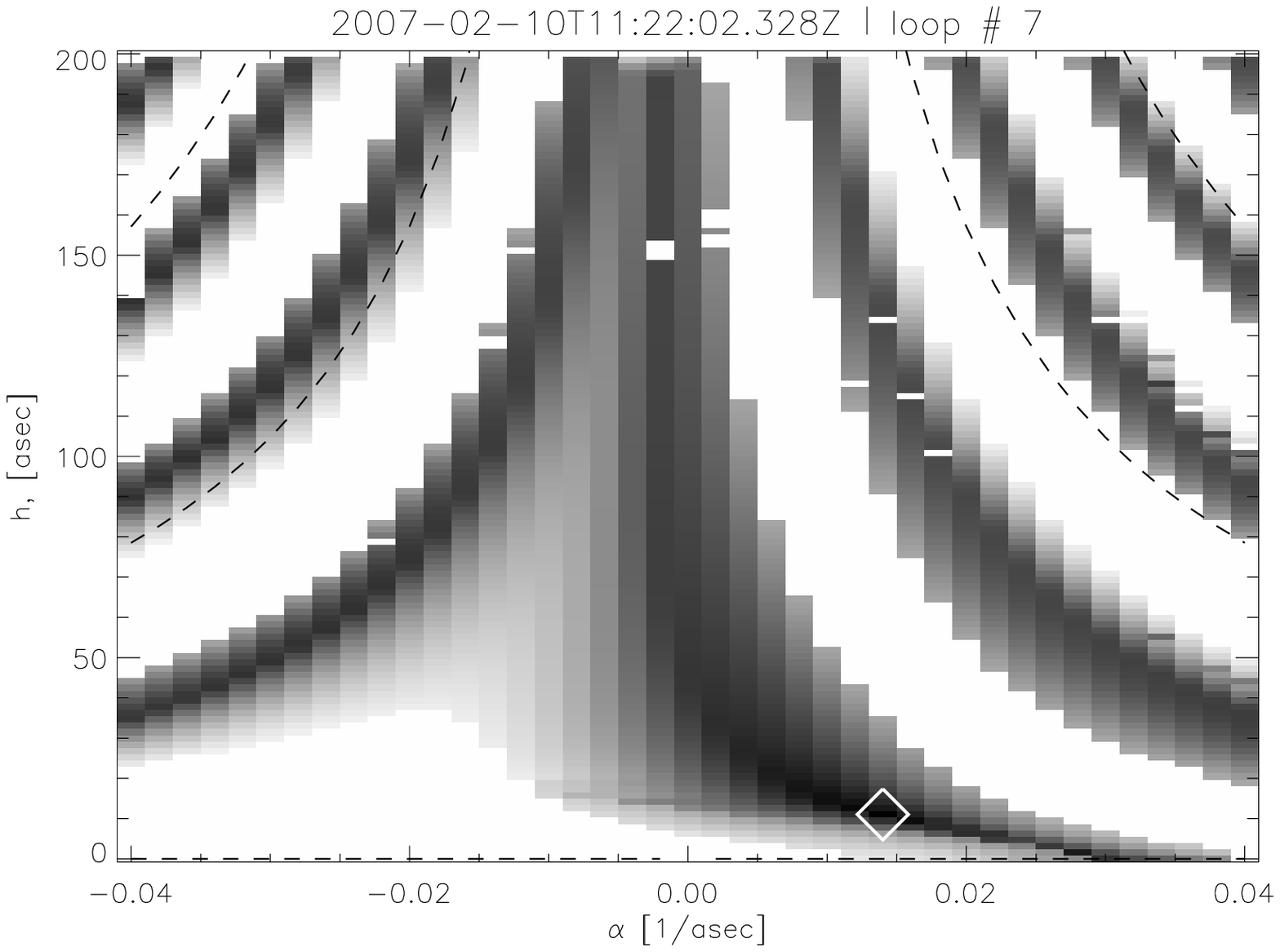} &  
  \includegraphics[height=4.5cm]{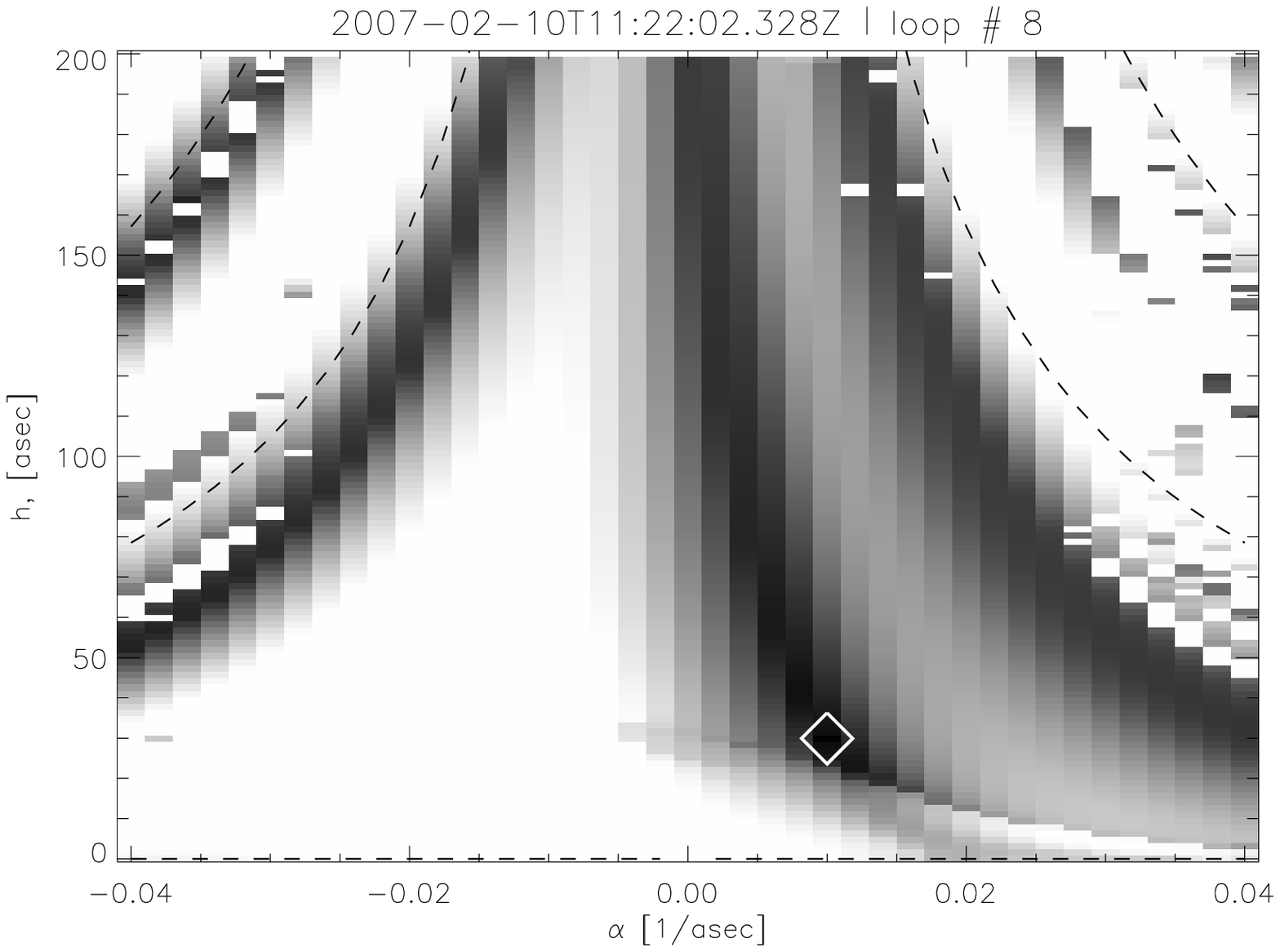} \\  
  \includegraphics[height=4.5cm]{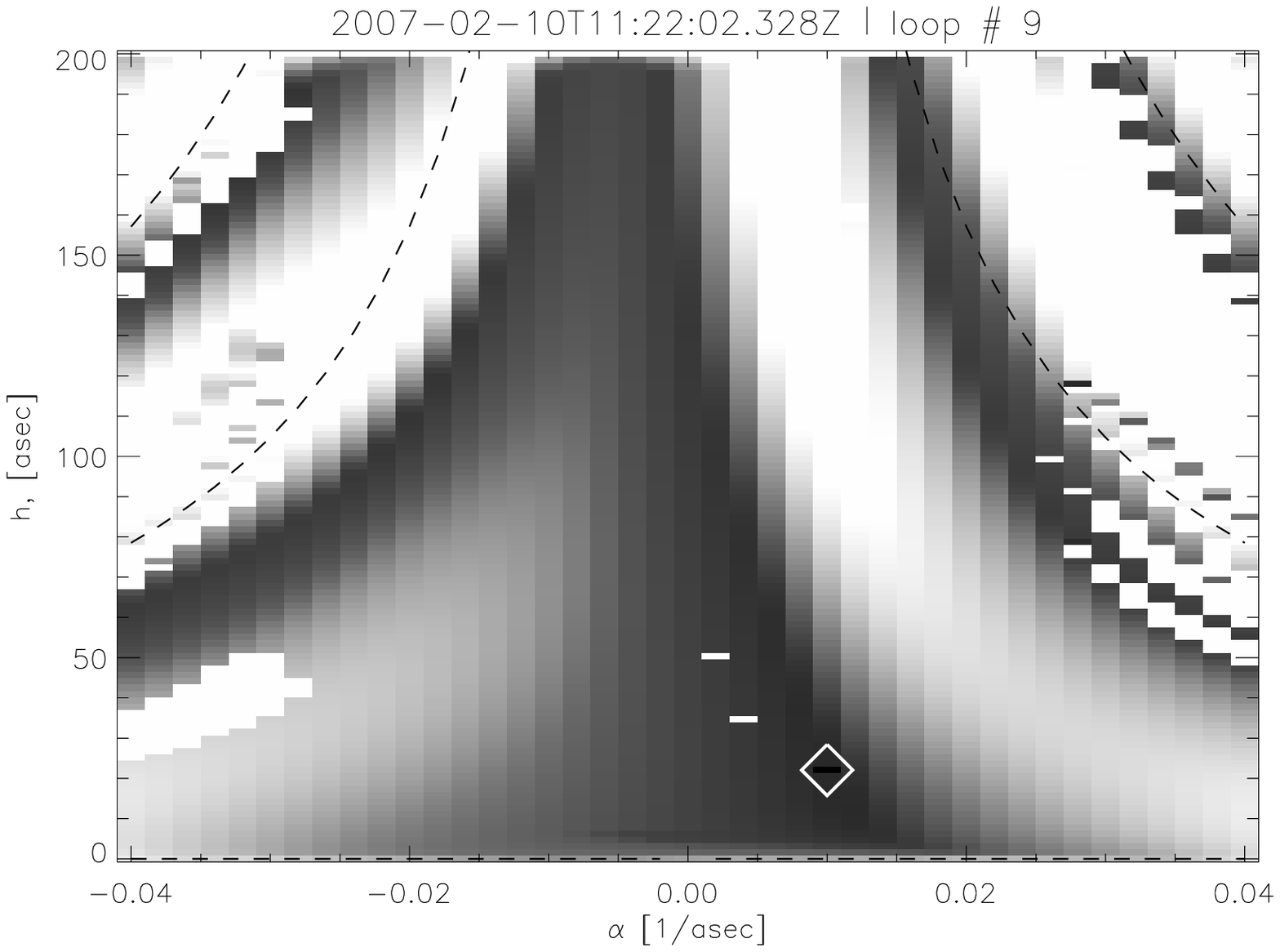} &
  \includegraphics[height=4.5cm]{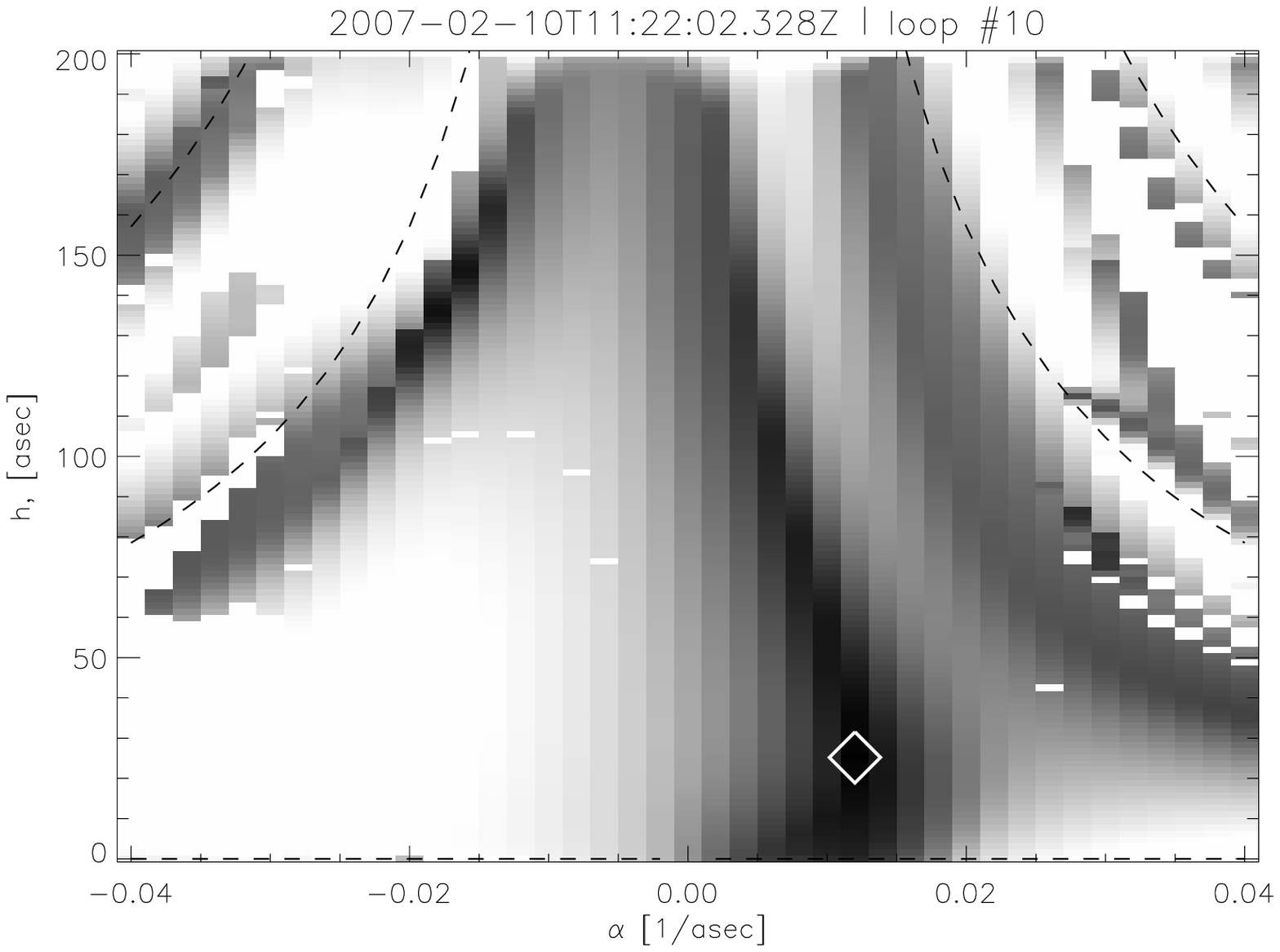} &
  \includegraphics[height=4.5cm]{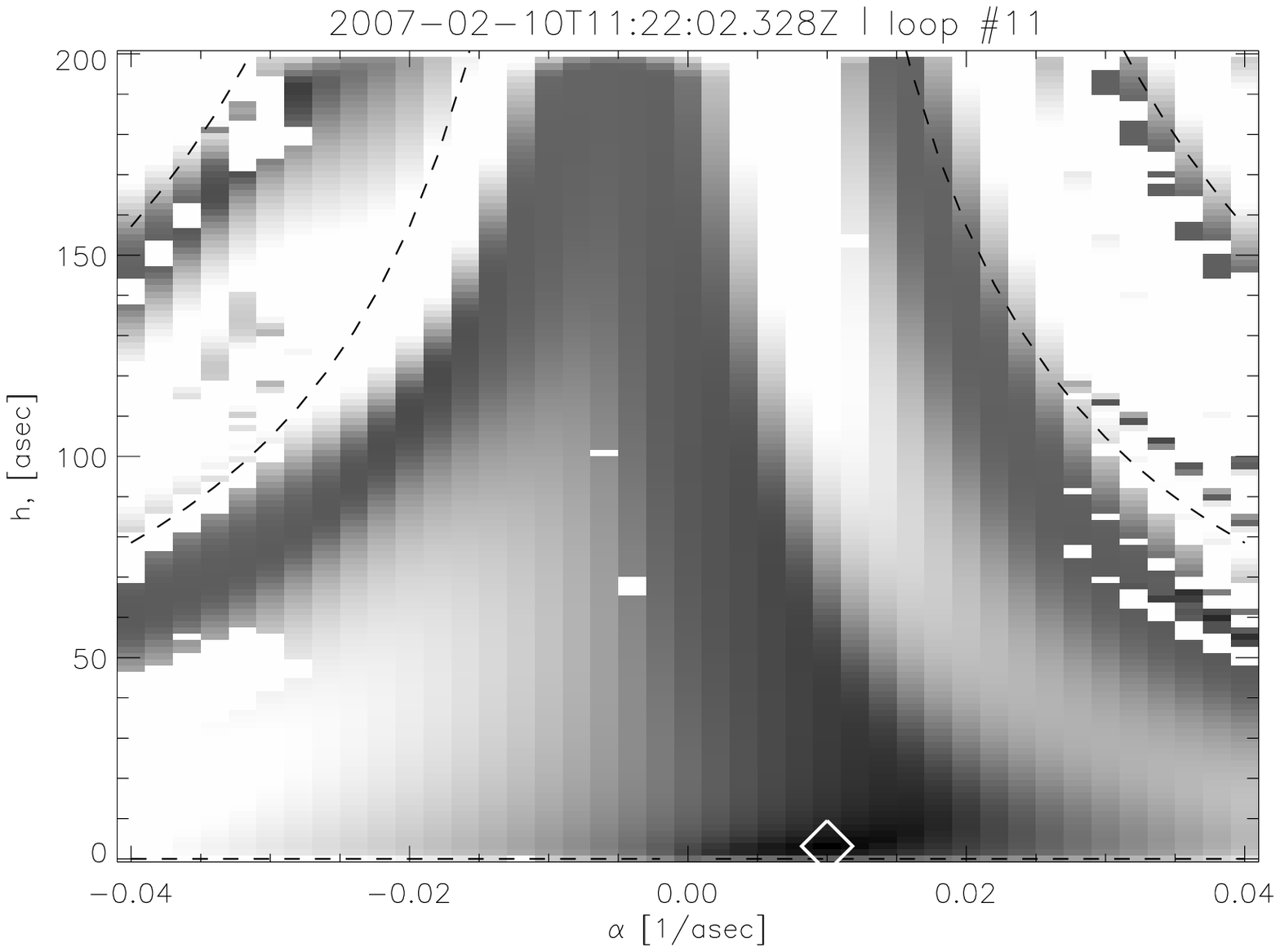} \\
 \end{tabular}
 \end{center} 
 \caption{\small{Parameter spaces for loops shown in Fig.~\ref{real_data}.}}
 \label{real_data_parspaces} 
 \end{figure}

The results are summarized in Table~\ref{real_data_table} and Fig.~\ref{real_data}. Visually it seems that the fit did a good job for all but three loops. The parameter space plots, such as the one shown in Fig.~\ref{real_data_parspaces}, all looked like those for \llfs fields. For all of them we applied the proposed algorithm of selecting local minima on ``non-hyperbolic'' valley.

Disregarding the three ``unsuccessful'' fits, it seems that \als was of the order of $0.010-0.015\mbox{ arcsec}^{-1}$ in the outer region of the sigmoid, and higher in the middle, exceeding $0.02\mbox{ arcsec}^{-1}$. Using the solar radius $983.13''$ and recalling a typical bias of $\alpha_{real}\propto1.23\alpha_{found}$ (see Fig.~\ref{all_on_one_plot}), we estimate $\alpha_{real}$ to be $1.7-2.6\times 10^{-8}\mbox{m}^{-1}$ in the outer regions and over $3.5\times 10^{-8}\mbox{m}^{-1}$ in the core. These value fall in the range typical of active region fields, such as those reported in \citet{Burnette2003}: $\pm4\times 10^{-8}\mbox{m}^{-1}$. The magnetic field strength along all the non-potential field lines was always within 50\% of the strength of the potential field, evaluated along the same path; it was within 25\% in at least half the cases.

Assuming a typical separation distance $L\approx 50''$ (see Fig.~\ref{real_data}) the values of $\alpha_{found}$ was of the order of $0.5-0.75 L^{-1}$ outside of the sigmoid and exceeded $1.0L^{-1}$ inside of the sigmoid. This is within the range of reconstructed \als in the trial \llfs fields (see Fig.~\ref{all_on_one_plot}).

%By visual inspection, the fits appeared to do a relatively good job. The parameter space plots were similar to those obtained in analytic field tests and the range of \als obtained were within the range found to be reliable in such tests. The values of twist were within the typical range reported by other observers. We conclude that the method seems to be applicable to the real solar data. 

\begin{table}
	\centering
\begin{tabular}{ccc}
Loop \# & \al, $\mbox{arcsec}^{-1}$ & $h_{max}$, arcsec \\ 
\hline
 0 & 0.012 & 12.9 \\
 1 & 0.016 &  0.0 \\
 2 & 0.014 & 23.8 \\
 3 & 0.026 &  9.5 \\
 \textit{4} & \textit{0.008} & \textit{35.4} \\
 \textit{5} & \textit{0.024} & \textit{33.9} \\
 6 & 0.020 & 14.3 \\
 7 & 0.014 & 12.7 \\
 8 & 0.010 & 31.5 \\
 \textit{9} & \textit{0.010} & \textit{24.5} \\
10 & 0.012 & 27.4 \\
11 & 0.010 &  3.3 \\
\end{tabular}
\caption{\small{The results of the \al-h fit to solar data. Here $h_{max}$ is maximal height (not height at mid-point, as before in the text). Loops that don't seem to give a good fit to the data are in italic.}}
\label{real_data_table}
\end{table}

%\clearpage
\section{Discussion}\label{sec_discuss}

In this work we have proposed a semi-automated method that, from a given two-dimensional EUV or X-Ray image of coronal loops and a line-of-sight magnetogram, 
reconstructs local twist and a three-dimensional shape of each loop. It tacitly assumes that coronal loops follow magnetic field lines. Our fitting matches the plane of sky projection of each loop to the projection of lines of linear force-free fields, traced from different heights along the line-of-sight and that have different twist in them. The method thus constructs a linear force-free field and one field line in it, that is the best match of an individual loop. 
%REPHRASE?-----------------
The method is similar to the ones proposed by Green et al (2002)\nocite{Green2002} and Lim et al (2007)\nocite{Lim2007}, however, it holds several important advantages. First, it does not require a full length of the loop to be visible for a successful reconstruction. Secondly, it does not require any of the footpoints to be visible. Thirdly, it allows the user to draw a smooth curve (e.g., a B\'ezier spline) interactively on top of the loop, rather than selecting a few points along the loop, thus maximizing the amount of information taken from the coronal image.

We address the question of validity of such a reconstruction, given the fact that the coronal field is probably non-linear force-free and that the superposition of linear force-free fields obtained for each individual loop would not, in general, be a force-free field. To do so, we perform a series of tests on non-linear analytic force-free fields, described in \citet{LowLou1990}, and as model loops we utilize projections of field lines on the photosphere. We compare several results of the method with the original field: 3D shapes of loops, local twist (coronal \al), distribution of twist in the photosphere and the strength of magnetic field. 

We also find that the best-fit line may not be a global minimum of $d(\alpha, h)$, but one in a particular part of parameter space. We developed an algorithm that aids in locating the appropriate region of the parameter space. We followed it manually, but it could be automated. The algorithm seems to improve the results on \llfs fields, but it probably does not describe every possible feature of the parameter space resulting from every single arbitrary magnetic field. The algorithm could be improved and expanded, based on further research involving other non-linear fields and real solar data. 

Based on Table~\ref{table_results} we draw the conclusions (i) that on the trial fields the twist is reconstructed with mean absolute deviation of at most 15\% of the range of photospheric twist, (ii) that heights of the loops are reconstructed with mean absolute deviation of at most 5\% of the range of trial heights, and (iii) that the magnitude of non-potential contribution to photospheric field is reconstructed with mean absolute deviation of at most 10\% of the maximal value. 

As shown in Fig.~\ref{all_on_one_plot}, there is a typical understimation of twist when performing this procedure. Based on the experiments with \llfs fields we conclude, that typically $\alpha_{real}\approx1.23\alpha_{found}$.

%--------------------------

We also demonstrate how this method can be applied to real solar data, by doing the reconstruction based on the data from SOHO/MDI magnetograms and Hinode/XRT X-Ray images. The resulting field lines visually match the observed loops, have reasonable heights and self-consistent amount of twist of the magnitude that agrees with existing measurements of twist in active regions \citep{Burnette2003}. 

\section{Acknowledgements}\label{sec_thankyou}
This work was supported by NASA grants NNX07AI01G and NNX06AB83G.

%\clearpage
\section{Appendix, on Low \& Lou fields.}\label{sec_append}

\citet{LowLou1990} constructed a class of non-linear force-free magnetic fields beginning with an axi-symmetric field. An axi-symmetric, divergence-free magnetic field can be written, in all generality, as 
\begin{equation}
 {\bf B} = \nabla A \times \nabla \phi ~+~ Q \nabla\phi
 = {1\over r \sin\theta}\left({1\over r}{\partial A\over \partial\theta}
 {\bf \hat{r}} - {\partial A\over \partial r}\hat{\theta} + 
 Q\hat{\phi} \right) ~~,
	\label{eq:B}
\end{equation}
where $A(r,\theta)$ is the flux function and the azimuthal component is $B_{\phi}(r,\theta)=Q(r,\theta)/r\sin\theta$. The poloidal components of the force-free condition, $\nabla\times{\bf B}=\alpha{\bf B}$, are satisfied only if $Q$ and $\alpha$ are each functions of the flux function alone
\begin{equation}
 Q(r,\theta) = Q[A(r,\theta)] ~~,~~
 \alpha[A(r,\theta)] = {dQ\over dA} ~~.
\end{equation}
The azimuthal component of the force-free condition 
\begin{equation}
 r^2\sin^2\theta~\nabla\cdot\left({\nabla A\over
 r^2\sin^2\theta}\right) = - \alpha Q {dQ\over dA} =
 -{1\over2}{dQ^2\over dA} ~~,
	\label{eq:GS_eq}
\end{equation}
is known as the Grad-Shafranov equation for flux function $A$.

The Grad-Shafranov equation contains one free function $Q^2(A)$ for which Low and Lou took a particular form. We generalize their choice to 
\begin{equation}
 Q^2(A) = a^2\,|A|^{2+2/n} ~~,
	\label{eq:Q}
\end{equation}
where $a$ and $n$ are free constants. The absolute value signs, absent from the original formulation, are introduced here so that $Q^2$ is real, and non-negative, even where the flux function is negative. Equation (\ref{eq:GS_eq}) can then be made homogeneous in $r$ by proposing a solution
\begin{equation}
 A(r,\theta) = P(\cos\theta)\,r^{-n} ~~,
\end{equation}
for a still-unknown function $P(\mu)$.
Using this in the Grad-Shafranov equation, and defining $\mu=\cos\theta$, leads to the non-linear equation 
\begin{equation}
 (1-\mu^2){d^2P\over d\mu^2} + n(n+1)P = -a^2 
 \left(1+{1\over n}\right)|P|^{2/n}\,P ~~,
	\label{eq:P_eq}
\end{equation}
for the unknown function $P(\mu)$.

Eq.~(\ref{eq:P_eq}) has real solutions for any $a$ and any $n>-2$; $n$ need not be an integer. Boundary conditions, similar to those of \citet{LowLou1990}, are $P'(1)=-10$ and $P(1)=0$, so the solution is regular along the positive $z$ axis. The solution will be regular along the negative $z$ axis only when the solution satisfies the additional
condition $P(-1)=0$. For a given value of $n$ this condition will be satisfied only for certain
choices of the eigenvalue $a$. (Integer choices of $n$ always have one eigenvalue, $a=0$, for which $P(\mu)$ is $\sin^2\theta$ times the Legendre polynomial of order $n$.)

The final magnetic field, defined for $z\ge0$, is constructed by rotating the axi-symmetric field by angle $\Phi$ about the $y$ axis and translating it downward a distance $\ell$. For rotation angles $\Phi\le\pi/2$ and non-vanishing displacement, $\ell>0$, the origin and what had been the
negative $z$ axis lie in $z<0$, outside our domain.  This means
the regularity condition, $P(-1)=0$, is not needed to assure a
regular magnetic field. We therefore make no restriction on $P(-1)$
and consider both $n$ and $a$ to be free parameters.

The function $Q(A)$, required for the final field of eq.\
(\ref{eq:B}), is found from the square root of eq.\
(\ref{eq:Q}). When the solution $P(\mu)$ changes sign there can be
more than one choice of square root for which $Q(A)$ is a
continuous real function.\footnote{There will be $2^m$ distinct choices when
$P(\mu)$ changes sign $m$ times over $-1<\mu<1$.} 
We focus on two such choices, we call
signed and unsigned
\begin{eqnarray}
 Q_U(A) &=& a A|A|^{1/n} ~~,\\
 Q_S(A) &=& a |A|^{1+1/n} ~~,
\end{eqnarray}
both of which satisfy eq.\ (\ref{eq:Q}) when $A(r,\mu)=P(\mu)/r^n$ is real.
Taking the derivative of these functions gives the twist parameters
for the two cases
\begin{eqnarray}
 \alpha_U(A) &=& a \left(1+{1\over n}\right) |A|^{1/n} ~~,\\
 \alpha_S(A) &=& a \left(1+{1\over n}\right) |A|^{1/n} {\rm sgn}(A)~~.
\end{eqnarray}
The unsigned case has a single sense of twist determined by the sign
of the constant $a$; the signed case has both senses provided
$A$ changes sign.

\bibliography{c:/localtexmf/bib/short_abbrevs,c:/localtexmf/bib/full_lib,c:/localtexmf/bib/my_bib}
\end{document}